\documentclass[11pt]{article}

\usepackage{longtable, ltcaption}%
\usepackage{amsmath,amsthm,amssymb}
\usepackage[margin=.7in]{geometry}
\usepackage{hyperref}
\hypersetup{
    colorlinks=true,
    linkcolor=blue,
    citecolor=blue,
    urlcolor=blue,
    filecolor=blue
}
\usepackage{setspace,placeins,booktabs,ctable}
\usepackage{cleveref, graphicx, caption, subcaption, indentfirst, ragged2e, changepage, multirow}
\usepackage[utf8]{inputenc}
\usepackage{bm}
\usepackage{enumitem}
\usepackage{comment}
\usepackage[normalem]{ulem}
\excludecomment{additional-material}
\usepackage{xr}
\allowdisplaybreaks[1]
\usepackage[bibstyle=numeric, citestyle=authoryear-comp, doi=false, url=false, backend=biber, maxbibnames=6, maxcitenames=4, uniquelist=false, uniquename=false, sorting=nyt, natbib=false, style=authoryear-comp, sortcites=false]{biblatex}
\renewbibmacro{in:}{}
\DeclareNameAlias{default}{family-given/given-family}
\newcommand{\citet}{\textcite}
\newcommand{\indicator}[1]{\mathbf{1}\!\left\{#1\right\}}
\newcommand\independent{\protect\mathpalette{\protect\independenT}{\perp}}
    \def\independenT#1#2{\mathrel{\setbox0\hbox{$#1#2$}%
    \copy0\kern-\wd0\mkern4mu\box0}}
\newcommand{\E}{\mathbb{E}}
\renewcommand{\L}{\mathrm{L}}
\renewcommand{\P}{\mathrm{P}}
\newcommand{\F}{\mathrm{F}}
\newcommand{\N}{\mathcal{N}}

\DeclareMathOperator*{\argmin}{arg\,min}
\DeclareMathOperator*{\plim}{plim}

\newcommand{\point}[1]{}

\newtheorem{theorem}{Theorem}

\newtheorem{lemma}{Lemma}

\newtheorem{corollary}{Corollary}
\newtheorem{assumption}{Assumption}

\newtheorem{proposition}{Proposition}

\crefname{figure}{Figure}{Figures}
\crefname{assumption}{Assumption}{Assumptions}
\crefname{inneruassumption}{Assumption}{Assumptions}
\crefname{theorem}{Theorem}{Theorems}
\crefname{assumption}{Assumption}{Assumptions}
\crefname{appendix}{Appendix}{Appendices}

\theoremstyle{definition}
\newtheorem{algorithm}{Algorithm}
\newtheorem{example}{Example}
\newtheorem{remark}{Remark}

\title{\textbf{Distributional Effects with Two-Sided Measurement Error: An Application to Intergenerational Income Mobility}\thanks{We thank Yonghong An, Gary Chamberlain, Hao Dong, Jerry Hausman, Weige Huang, Chengye Jia, Josh Kinsler, Dan Millimet, Ariel Pakes, Xun Tang, and seminar participants at the Harvard/MIT Econometrics Workshop, Monash University, National Tsinghua University, Rice University, Southern Methodist University, UC Irvine, UC Riverside, University of Southern California, University of Surrey, and at the 2018 Triangle Econometrics Conference, the 2019 Tsinghua Econometrics Conference, and the 2019 North American Summer Meetings of the Econometric Society for helpful comments.  An earlier version of this paper circulated under the title ``Nonlinear Approaches to Intergenerational Income Mobility allowing for Measurement Error.''  Code for the approach proposed in our paper is available at \url{https://bcallaway11.github.io/qrme/}.  The U.S. Securities and Exchange Commission disclaims responsibility for any private publication or statement of any SEC employee or Commissioner. This paper expresses the author's views and does not necessarily reflect those of the Commission, the Commissioners, or members of the staff.}}

\author{Brantly Callaway\thanks{John Munro Godfrey, Sr.\ Department of Economics, University of Georgia, B422 Amos Hall, Athens, GA 30605.  Email: \mbox{brantly.callaway@uga.edu}} \and Tong Li\thanks{Department of Economics, Vanderbilt University, VU Station B \#351819, Nashville, TN 37235.  Email: tong.li@vanderbilt.edu} \and Irina Murtazashvili\thanks{Division of Economic and Risk Analysis, U.S. Securities and Exchange Commission, 1617 John F Kennedy Blvd \#520, Philadelphia, PA 19103.  Email: murtazashviliirina@gmail.com} \and Emmanuel S. Tsyawo\footnote{Department of Economics, Finance and Legal Studies, Culverhouse College of Business, University of Alabama. Email: estsyawo@gmail.com} }

\date{\today}

\begin{document}

\maketitle

\abstract{\noindent This paper considers identification and estimation of distributional effect parameters that depend on the joint distribution of an outcome and another variable of interest (``treatment'') in a setting with ``two-sided'' measurement error---that is, where both variables are possibly measured with error.  Examples of these parameters in the context of intergenerational income mobility include transition matrices, rank-rank correlations, and the poverty rate of children as a function of their parents' income, among others.  Building on recent work on quantile regression (QR) with measurement error in the outcome (particularly, \citet{hausman-liu-luo-palmer-2021}), we show that, given (i) two linear QR models separately for the outcome and treatment conditional on other observed covariates and (ii) assumptions about the measurement error for each variable, one can recover the joint distribution of the outcome and the treatment.  Besides these conditions, our approach does not require an instrument, repeated measurements, or distributional assumptions about the measurement error.  Using recent data from the 1997 National Longitudinal Study of Youth, we find that accounting for measurement error notably reduces several estimates of intergenerational mobility parameters.}

\bigskip

\bigskip

\bigskip


\bigskip


\clearpage

\normalsize

\onehalfspacing

\section{Introduction}

Measurement error is a widespread challenge in empirical work in economics.  It is well documented that many important economic variables are often measured with error (e.g., income (\citet{bound-krueger-1991,pischke-1995,bound-brown-mathiowetz-2001}), consumption (\citet{attanasio-battistin-ichimura-2007}), health status (\citet{bound-schoenbaum-stinebrickner-waidmann-1999}), government transfers (\citet{meyer-mok-sullivan-2009}), and participation in social programs (\citet{bollinger-david-1997}), among many others).  Measurement error can arise due to misreporting (e.g., a survey respondent incorrectly reporting their health insurance coverage) or for other reasons.  In the context of our application on intergenerational income mobility, economists have primarily been interested in the effect of parents' \textit{permanent} income on child's \textit{permanent} income.  A key empirical challenge in this literature is that, instead of observing permanent incomes, researchers typically observe annual income (or some other shorter-term measure of income).  The modern literature on intergenerational income mobility (e.g., \citet{solon-1992} and much subsequent work) considers annual income as an error-ridden measure of permanent income and has developed a number of techniques to deal with measured-with-error versions of permanent incomes, particularly in the context of linear models.\footnote{Much of the literature has focused on estimating intergenerational elasticities, which are the coefficient on the log of parents' income in a regression of the log of child's income on the log of parents' income.  This literature demonstrates that accounting for measurement error leads to notable quantitative differences in estimates of intergenerational mobility.
See, for example, \citet{solon-1999,mazumder-2005,haider-solon-2006}, among others.}

In some settings, it is well-established how to deal with measurement error, or it is at least well-known what the consequences of measurement error are.  A leading example is that, in the presence of classical measurement error, a mismeasured right-hand side variable in a linear model leads to attenuation bias, while a mismeasured left-hand side variable does not lead to inconsistency of the parameter estimates in the same setting.  In the current paper, we consider more complicated target parameters that are nonlinear functionals of the joint distribution of a continuous outcome and a continuous treatment variable in a setting where both the outcome and the treatment could be mismeasured.\footnote{As a comment on terminology, we refer to the class of parameters that are possibly nonlinear functionals of the joint distribution as \textit{distributional effects}.  Slightly abusing terminology, we use the term \textit{treatment} to denote the right-hand side variable of primary interest. Under additional conditions, the distributional effects we consider can have a causal interpretation, but this is not a central issue for our main arguments.  We refer to other right-hand side variables as \textit{covariates} and assume that they are not measured with error.\label{fn:terminology}}

It is useful to give some examples of the types of parameters that we consider in the paper.  One example is transition matrices, which summarize transition probabilities from a starting point in the distribution of the treatment variable to an ending point in the distribution of the outcome. In the context of intergenerational mobility, transition matrices provide a way to see how likely children are to move to different parts of the income distribution, conditional on what part of the income distribution their parents were in.  Work on transition matrices includes \citet{jantti-etal-2006,bhattacharya-mazumder-2011,black-devereux-2011,richey-rosburg-2018,millimet-li-roychowdhury-2020}.  Another example is Spearman's Rho---the rank-rank correlation of the outcome and the treatment variable.  This parameter has received considerable recent attention in the intergenerational mobility literature (\citet{chetty-hendren-kline-saez-2014,collins-wanamaker-2017,chetty-hendren-2018,chetty-hendren-jones-porter-2020}).  %
Third, another class of parameters includes those that are functionals of the entire distribution of the outcome as a function of the treatment; these parameters are related to the literature on continuous treatment effects and include specific examples like the fraction of children whose income is below the poverty line as a function of parents' income (\citet{richey-rosburg-2016b,callaway-huang-2020}).  We consider these parameters, along with several more, in substantially more detail in the next section.

What each of the above parameters has in common is that they depend on the joint distribution of the outcome (child's permanent income) and the treatment (parents' permanent income), possibly conditional on covariates. Thus, our main methodological contributions concern identifying joint distributions in the presence of measurement error in both the outcome and the treatment.  The main case that we consider in the paper is the one where a researcher has access to a single measurement of the outcome and of the treatment.  In this case, we impose two main conditions.
First, we impose ``classical'' measurement error conditions; in particular, we impose that measurement errors are additively separable, independent of the true value of the outcome/treatment and other covariates, and independent of each other.\footnote{In the intergenerational mobility literature, the assumption of classical measurement error is typically considered more plausible for some ages of children and parents (typically mid-career ages).  We discuss how to adjust our arguments to the case of ``life-cycle'' measurement error (as in \citet{haider-solon-2006}) rather than classical measurement error; see \Cref{rem:life-cycle} and \Cref{app:life-cycle-measurement-error} in the Supplementary Appendix for more details.} The second main assumption that we make is on auxiliary, preliminary models for the outcome and the treatment conditional on other covariates.  In particular, we assume that both the outcome and the treatment are generated by a quantile regression (QR) model that is linear in parameters across all quantiles.  Besides these two main conditions, we do not require an instrument, repeated measurements, or distributional assumptions about the measurement errors for identification.

The first part of our identification argument concerns separately recovering the marginal distributions of the outcome and the treatment conditional on other covariates in the presence of measurement error.  For this step, our arguments follow from \citet{hausman-liu-luo-palmer-2021}, who show that the structure imposed by the QR model is enough to identify the QR parameters as well as the distribution of the measurement error in the case with classical measurement error in the outcome variable.  Applying this argument to both our outcome and treatment implies that the conditional quantiles and conditional distributions of the outcome and treatment are identified.%
However, identifying these marginal distributions is not enough to identify the distributional effect parameters that we are ultimately interested in, since these parameters depend on the joint distribution of the outcome and the treatment.  From Sklar's Theorem (\citet{sklar-1959}), a well-known result in the literature on copulas, we know that joint distributions can be written as the copula (which captures the dependence between two random variables) of the marginal distributions of the outcome and the treatment.  We show that, under the conditions above, the copula is also identified.  This means that the joint distribution is identified, and hence, all of the distributional effect parameters that we consider are also identified.

For estimation, we propose a three-step approach to estimating the joint distribution of the outcome and the treatment conditional on covariates and then recovering parameters of interest from the joint distribution. In the first step, we estimate the quantile regression parameters and then convert these into estimates of the marginal distributions of the outcome/treatment conditional on covariates.  Our estimation approach here is new.  We implement an EM-type algorithm that iterates back and forth between making draws of the measurement error (conditional on the values of the parameters) and estimating new values of the parameters.  In the second step, we estimate the copula parameters using a simulated maximum likelihood approach that plugs in the first step estimators.  To simplify estimation, we assume that the copula is known up to a finite number of parameters.  This step provides an estimate of the joint distribution.  Third, we can manipulate the estimated joint distribution into any parameters of interest.  We also derive the asymptotic properties of our estimators and show that the empirical bootstrap can be used to conduct inference.

We apply our approach to study intergenerational income mobility using recent data from the 1997 cohort of the National Longitudinal Survey of Youth (NLSY97). This is a rich dataset where we observe much information about families and can link child's income with their parents' income. We employ the 1997 wave of the NLSY97 to obtain information about parents’ income and the 2015 wave of the NLSY to obtain the child’s income. However, there are some notable challenges.  First, we only observe parents' income in a single year.  Second, although in principle we observe child's income in every period where they participate in the survey, children are still relatively young, and, in light of much work in the intergenerational mobility literature, it seems most appropriate to only use their income in more recent waves. Our main results indicate that (i) adjusting for measurement error and (ii) using our copula-based approach relative to quantile regression of child's income directly on parents' income and covariates are both important.  Accounting for measurement error in child's and parents' income tends to reduce various estimates of intergenerational mobility, resulting in estimates that are more similar to those in the literature where multiple observations of income are available relative to estimates coming directly from the observed data that do not account for measurement error.

\subsubsection*{Related Work}

Our work is related to an extensive literature on measurement error in econometrics (see \citet{Hausman01} for an overview of the literature).  Our paper builds on the literature that recovers the distribution of a variable of interest in the presence of measurement error (important early examples include  \citet{carroll-hall-1988,stefanski-carroll-1990,fan-1991,diggle-hall-1993}).  These papers assumed the distribution of the measurement error was known; \citet{li-vuong-1998} relax the distributional assumption on the measurement error when there are repeated measurements available.

Notable advances in measurement error problems have also been made for nonlinear models, particularly for the case of the right-hand side (RHS) mismeasured variable (e.g., \citet{hsiao-1989,li-2000,li-2002,hausman-newey-ichimura-powell-1991,hausman-newey-powell-1995,Schennach04,Schennach08}).  %
Dealing with the left-hand side (LHS) measurement error in a nonlinear context has received much less attention.    This is unfortunate since what we know about the linear setting with either RHS or LHS measurement error does not generally carry over to nonlinear models (some examples include \citet{Hausmanetal98,Cosslett04,Abrevayaetal99,Cavanaghetal98}). %
Furthermore, even less consideration has been given to nonlinear models with \textit{both} RHS and LHS measurement error.  Rare exceptions include \citet{Lewbel96} that studies parametric Engel curves with measurement error on both sides of the equation and \citet{DeNadaiLewbel16} that extends \citet{Schennach07} to allow for both RHS and LHS measurement error in nonparametric models.

Within the measurement error literature, the current paper is most closely related to work on quantile regression with measurement error.  The econometric literature for estimating quantile regressions with RHS measurement error includes \citet{Schennach08,wei-carroll-2009,chesher-2017,firpo-galvao-song-2017}.   Our approach, however, is more closely related to quantile regression with LHS measurement error.  \citet{hausman-liu-luo-palmer-2021} show that the parameters of a linear quantile regression model are identified when there is measurement error in the outcome variable without imposing distributional assumptions on the measurement error or requiring instruments or repeated observations.  Our approach builds on this as we use their identification arguments in our first step to show that the marginal distributions of the outcome and the treatment are identified.  On the other hand, our estimators of the conditional quantiles are different from those proposed in \citet{hausman-liu-luo-palmer-2021}; we propose an EM-type algorithm while their estimators are based on maximum likelihood.  %

Finally, there are a few related papers that examine measurement error in the context of nonlinear approaches to understanding intergenerational mobility.  \citet{millimet-li-roychowdhury-2020} consider transition matrices, particularly in the case where individuals can be misclassified into the wrong cell.  \citet{nybom-stuhler-2017,kitagawa-nybom-stuhler-2018} study rank-rank correlations (i.e., Spearman's Rho) in the presence of measurement error.  \citet{nybom-stuhler-2017} consider measurement error in the ranks of child's income and parents' income rather than in the levels of income. %
The approach in \citet{kitagawa-nybom-stuhler-2018} builds on the small variance approximations of \citet{chesher-1991} to recover rank-rank correlations when there is measurement error in the levels of child's and parents' income.  %
\citet{an-wang-xiao-2020} study a nonparametric version of the intergenerational elasticity in the presence of non-classical measurement error (particularly, life-cycle measurement error).

\subsubsection*{Organization of the paper}

The paper is organized as follows. In \Cref{sec:id}, we introduce a number of parameters that depend on the joint distribution of the outcome and the treatment.  \Cref{sec:3} considers identification of these parameters of interest, allowing for measurement error. In Section \ref{sec:est}, we propose a three-step estimation procedure to estimate the parameters of interest, and, in \Cref{sec:inference}, we study the asymptotic properties of the estimators. Our application on intergenerational income mobility using our approach is in \Cref{sec:emp}. Final remarks are contained in Section \ref{sec:conc}. %

\section{Parameters of Interest} \label{sec:id}

\subsubsection*{Notation}

Let $Y^*$ denote the outcome of interest and $T^*$ denote a continuous treatment variable.  We allow both $Y^*$ and $T^*$ to be mismeasured.  Also, let $X$ denote a $K\times 1$ vector of covariates that are correctly measured.

\bigskip

All of the parameters that we consider are features of the joint distribution $\F_{Y^*T^*\mid X}$.  Later, our identification arguments will center on identifying this joint distribution.  For now, notice that, in the absence of measurement error, this joint distribution is directly identified by the sampling process.  In the presence of measurement error, identifying this distribution is non-trivial.  We take this issue up in \Cref{sec:3} below, but for now, we consider a wide variety of parameters %
that are features of this joint distribution and can, therefore, be obtained if this joint distribution is identified.

We divide the parameters in this section into two categories that we call \textit{Conditional Distribution-type} parameters and \textit{Copula-type} parameters.  In the context of intergenerational mobility, the first category includes parameters such as the fraction of children whose income is below the poverty line as a function of parents' income; we can also consider this parameter after adjusting for differences in the distribution of characteristics (e.g., race and education) across different values of parents' income.  Another parameter in this class is quantiles of child's income conditional on parents' income and other background characteristics (or after adjusting for differences in other background characteristics).  This category also includes parameters from the treatment effects literature (e.g., quantile treatment effects and dose-response functions) as well as counterfactual distributions and distributional policy effects. These types of parameters are of general interest in the case where a researcher is interested in the effect of a continuous treatment on features of the distribution of some outcome.

The copula-type parameters that we consider include transition matrices, measures of upward mobility (i.e., the probability that child's rank in the income distribution is greater than their parents' rank), and rank-rank correlations.  Each of these is a nonlinear measure of intergenerational income mobility and has received considerable interest in recent work on intergenerational mobility (though not much attention has been paid to measurement error for these types of parameters).

\subsection{Conditional Distribution-type Parameters}

The parameters that we consider in this section come from the distribution of $Y^*$ conditional on $T^*$ and $X$, which is given by %
\begin{align} \label{eqn:ycondt}
  \F_{Y^*|T^*X}(y|t,x) &= \P(Y^* \leq y | T^*=t, X=x)
\end{align}
and which will be identified when the joint distribution of $(Y^*,T^*)$ conditional on $X$ is identified.  In our application on intergenerational income mobility, this conditional distribution is the distribution of child's income conditional on parents' income and on other characteristics.  One parameter of particular interest is given by setting $y$ equal to the poverty line, varying $T^*$ (parents' income), and fixing characteristics $X$.  This gives the fraction of children with income below the poverty line as a function of parents' income and holding other characteristics fixed.  Another parameter of interest is the quantiles of $Y^*$ conditional on $T^*$ and $X$.  The quantiles are given by
\begin{align} \label{eqn:qcondt}
  Q_{Y^*|T^*X}(\tau|t,x) = \inf \big\{ y : \F_{Y^*|T^*X}(y|t,x) \geq \tau \big\}
\end{align}
for $\tau \in (0,1)$.  These conditional quantiles are what would be recovered in the absence of measurement error using quantile regression.  Quantile regression has been used in the context of intergenerational mobility in \citet{eide-showalter-1999}, \citet{grawe-2004}, and \citet{Bratbergetal07}.\footnote{The conditional mean of $Y^*$ given $T^*$ can also be recovered from the conditional distribution or the conditional quantiles.
For example, $\E[Y^*|T^*=t, X=x] = \int_0^1 Q_{Y^*|T^*,X}(\tau|t,x) \, d\tau$.  In the context of intergenerational mobility, this parameter is a nonparametric/nonlinear version of the intergenerational elasticity.  Work on nonlinear intergenerational elasticities includes \citet{Behrmanetal90,bratsberg-roed-raaum-naylor-2007,bjorklund-roine-waldenstrom-2012,murtazashvili-2012,murtazashvili-liu-prokhorov-2015,landerso-heckman-2017,callaway-huang-2019,an-wang-xiao-2020,kourtellos-marr-tan-2020}.}

In many applications, it makes sense to integrate out the covariates from the above conditional distributions.  Manipulations of the distribution of covariates (while fixing the conditional distribution $\F_{Y^*|T^*X}$) are called counterfactual distributions.  Work on counterfactual distributions (without measurement error) includes \citet{dinardo-fortin-lemieux-1996,machado-mata-2005,melly-2005,chernozhukov-val-melly-2013,rothe-2010,callaway-huang-2020}.  There are multiple possibilities here, but the most common one is
\begin{align}\label{eqn:counterfactual}
  \F^C_{Y^*|T^*}(y|t) = \int_{\mathcal{X}} \F_{Y^*|T^*X}(y|t,x) \ d\F_X(x)
\end{align}
which adjusts the distribution of covariates at every value of $T^*$ to be given by the overall population distribution of covariates.

\subsection{Copula-type Parameters}

The next set of parameters that we consider depends on the copula of the outcome and the treatment.  The copula is the joint distribution of the ranks of random variables.  That is, $C_{Y^*T^*}(r,s) = \P\big(\F_{Y^*}(Y^*) \leq r, \F_{T^*}(T^*) \leq s\big)$.  Sklar's Theorem, which is one of the most well-known results in the literature on copulas, says that joint distributions can be written as the copula of the marginal distributions, where the copula contains the information about the dependence between the two random variables.\footnote{\citet{joe-1997,nelsen-2007} are leading references on copulas.}  %

Parameters that depend on the copula of child's income and parents' income are common in the intergenerational mobility literature.  Intuitively, one reason why these types of parameters are useful is that they only depend on child and parent ranks in the income distribution and will not be contaminated with information related to how income distributions change across generations (\citet{chetty-hendren-kline-saez-2014,chetty-hendren-kline-saez-turner-2014,nybom-stuhler-2017}).  Main examples include transition matrices, upward mobility parameters, and rank-rank correlations.  We consider each of these in turn next.

\subsubsection*{Transition Matrices}

Transition matrices are one of the main tools to study intergenerational mobility (\citet{jantti-etal-2006,bhattacharya-mazumder-2011,black-devereux-2011,richey-rosburg-2018,millimet-li-roychowdhury-2020}).  However, not much attention has been paid to transition matrices allowing for measurement error (see \citet{millimet-li-roychowdhury-2020} for an exception, though their arguments are substantially different from ours).  Particular cells in transition matrices are given by
\begin{align*}
\theta_{TM}(r_1,r_2,s_1,s_2) = \P\big(r_1 \leq \F_{Y^*}(Y^*) \leq r_2 | s_1 \leq \F_{T^*}(T^*) \leq s_2\big)
\end{align*}
where $(r_1,r_2)$ and $(s_1,s_2)$ all take values from 0 to 1 (in a typical case, these are chosen so that cells in the transition matrix have cutoffs at the quartiles of the income distribution for children and parents).  Notice that transition matrices can be written in terms of the copula of child's income and parents' income.  In particular,
\begin{align}
    \theta_{TM}(r_1,r_2,s_1,s_2) &= \frac{\P\big(r_1 \leq \F_{Y^*}(Y^*) \leq r_2, s_1 \leq \F_{T^*}(T^*) \leq s_2\big)}{\P\big(s_1 \leq \F_{T^*}(T^*) \leq s_2\big)} \nonumber \\[10pt]
    &= \frac{C_{Y^*T^*}(r_2,s_2) - C_{Y^*T^*}(r_1,s_2) - C_{Y^*T^*}(r_2,s_1) + C_{Y^*T^*}(r_1,s_1)}{s_2 - s_1} \label{eqn:tm}
\end{align}

\subsubsection*{Upward Mobility}

Another parameter of interest in the intergenerational mobility literature is the probability that child's income rank is greater than their parents' income rank.  Following \citet{bhattacharya-mazumder-2011}, we define this as a function of parents' income rank,
\begin{align*}
  \theta_{U}(\Delta, s_1, s_2) = \P\big( \F_{Y^*}(Y^*) > \F_{T^*}(T^*) + \Delta | s_1 \leq \F_{T^*}(T^*) \leq s_2\big)
\end{align*}
The leading case here is when $\Delta=0$ (other values of $\Delta$ allow for one to consider the probability of child's income rank exceeding parents' income rank by $\Delta$).  In this case, $\theta_U(0,s_1,s_2)$ is the fraction of children who have a higher rank in the income distribution than their parents' rank in the income distribution, conditional on their parents' rank falling in a particular range.  Notice that $\theta_U$ can be written in terms of the copula of parents' income and child's income.  That is,
\begin{align*}
  \theta_U(\Delta, s_1, s_2) &= \frac{\P\big( \F_{Y^*}(Y^*) > \F_{T^*}(T^*) + \Delta, s_1 \leq \F_{T^*}(T^*) \leq s_2\big)}{s_2 - s_1} \\
  &= \int_{s_1}^{s_2} \int_{\tilde{v} + \Delta}^1 c_{Y^*T^*}(\tilde{u},\tilde{v}) \, d\tilde{u} d\tilde{v} \Big/ (s_2 - s_1)
\end{align*}
where the first equality holds by writing the conditional distribution as the joint distribution divided by the marginal distribution (and because the ranks of parents' income are uniformly distributed), and the second equality holds because the numerator depends on the joint distribution of the ranks of child's income and parents' income (i.e., the copula).\footnote{A related parameter to the upward mobility parameter discussed here is the probability that a child's income itself is greater than their parents' income.  The conditional-on-covariates version of this parameter is given by $\P(Y^* > t | T^*=t, X=x) = 1-\F_{Y^*|T^*X}(t|t,x)$; thus, this parameter can be immediately derived from the conditional distribution in \Cref{eqn:ycondt}, though we note that it should be classified as a conditional distribution-type parameter rather than a copula-type parameter.}

\subsubsection*{Rank-Rank Correlations}

Finally, we consider the rank-rank correlation, which has received much attention in the intergenerational mobility literature (\citet{chetty-hendren-kline-saez-2014,collins-wanamaker-2017,chetty-hendren-2018,chetty-hendren-jones-porter-2020}).  This is the correlation between the rank of child's income and the rank of parents' income.  The primary advantage of the rank-rank correlation over the more traditional intergenerational elasticity is that the rank-rank correlation does not depend on (changes in) the marginal distributions of income over time, which is often a desirable feature of an intergenerational mobility measure.  In the language of the literature on copulas and dependence measures, rank-rank correlations are called Spearman's Rho.  The rank-rank correlation is given by
\begin{align*}
  \rho_S = 12 \int_0^1 \int_0^1 C_{Y^*T^*}(r,s) \, dr \, ds - 3
\end{align*}
which, like the other parameters in this section, depends solely on the copula of child's income and parents' income.

\bigskip

\begin{remark} \label{rem:causal-effects} In \Cref{app:causal-effects} in the Supplementary Appendix, we show that, under the assumption of unconfoundedness, many of the parameters discussed above have a causal interpretation.  For example, $\F^C_{Y^*|T^*}(y|t)$, in \Cref{eqn:counterfactual} above, corresponds to a distributional version of a dose response function, which is a common target parameter in the literature (e.g., \citet{hirano-imbens-2004}, \citet{flores-2007}, \citet{flores-lagunes-gonzalez-neumann-2012}, \citet{galvao-wang-2015}).  Thus, under unconfoundedness, our framework extends work on causal inference with a continuous treatment to settings where the outcome and treatment are measured with error.
\end{remark}

\section{Identification} \label{sec:3}

The previous section suggests that one could obtain all the parameters of interest from the joint distribution of $Y^*$ and $T^*$ conditional on $X$.  In our setting, this is challenging because $Y^*$ and $T^*$ are mismeasured.  We make the following assumptions about the nature of the measurement errors:

\begin{assumption}[Measurement Error]\label{ass:me} \

  (i) $Y = Y^* + U_{Y^*} \qquad \textrm{and} \qquad T = T^* + U_{T^*}$

  (ii) $(U_{Y^*},U_{T^*}) \independent (Y^*,T^*,X)$  %

  (iii) $U_{Y^*} \independent U_{T^*}$ %
\end{assumption}

\Cref{ass:me} is a ``classical'' measurement error assumption.  \Cref{ass:me}(i) says that the researcher observes $Y$ and $T$ which are mismeasured versions of $Y^*$ and $T^*$.  \Cref{ass:me}(ii) says that the measurement error $(U_{Y^*},U_{T^*})$ is independent of $Y^*$, $T^*$, and $X$.  In terms of our application, it says that the joint distribution of the measurement errors does not depend on the true value of child's permanent income, the true value of parents' permanent income, and observed covariates.  %
This type of assumption is very common both in the literature on nonlinear models with measurement error (e.g., \Cref{ass:me}(i) and (ii) are the same as the assumptions in \citet{hausman-liu-luo-palmer-2021} applied to both the outcome conditional on covariates and the treatment conditional on covariates; similar assumptions are also made in \citet{li-vuong-1998,firpo-galvao-song-2017}, among many others) as well as in the literature on intergenerational mobility that allows for measurement error (e.g., the main case considered in \citet{kitagawa-nybom-stuhler-2018} involves classical measurement error though they discuss how their arguments can continue to apply under certain violations of classical measurement error).

\Cref{ass:me}(iii) says that $U_{Y^*}$ and $U_{T^*}$ are mutually independent.  For our application on intergenerational mobility, this assumption says that the measurement error in child's permanent income is independent of the measurement error in parents' permanent income.  This sort of assumption will hold as long as shocks to parents' income are independent of shocks to child's income, which seems plausible because parents' income and child's income are typically observed many years apart.  Related assumptions that measurement error for child's income and parents' income are uncorrelated are common in the literature (see, for example, Assumption 2 and related discussion in \citet{an-wang-xiao-2020}); instead of uncorrelatedness, we require full independence due to focusing on parameters that depend on the entire joint distribution of child's and parents' income.

\begin{remark} \label{rem:life-cycle}
  A leading alternative setup for measurement error in the intergenerational mobility literature is life-cycle measurement error (see \citet{jenkins-1987,haider-solon-2006} as well as \citet{vogel-2006,black-devereux-2011,nybom-stuhler-2016,an-wang-xiao-2020} for additional related discussion).  Here, a typical setup would impose that
  \begin{align*}
    Y_{a_Y} = \lambda^Y_{a_Y} Y^* + U_{Y^*,a_Y} \quad \textrm{and} \quad T_{a_T} = \lambda^T_{a_T} T^* + U_{T^*,a_T}
  \end{align*}
  where $a_Y$ and $a_T$ denote particular ages at which child's income and parents' income could be observed.
  This modification to \Cref{ass:me} allows for things like income at younger ages to be systematically lower than permanent income and for the distribution of the measurement error to change at different ages.  This sort of model is also consistent with the common finding that estimates of intergenerational mobility tend to be higher when child's income is observed when they are relatively young.  We discuss how to adapt our arguments to the case of life-cycle measurement error in detail in \Cref{app:life-cycle-measurement-error} in the Supplementary Appendix.  One of the conditions that is often invoked in the life-cycle measurement error literature is that there exist known ages, typically thought to be starting in the early thirties and perhaps lasting until the late forties (\citet{haider-solon-2006,nybom-stuhler-2016}), when $\lambda^Y_{a_Y}=1$ and $\lambda^T_{a_T}=1$.  Thus, even in the presence of life-cycle measurement error, one simple way to get back to the case with classical measurement error is to use observations from prime-age children and parents only---this is the approach that we take in the application.\footnote{This strategy might not always be available, but, especially for applications like ours that use NLSY data where there is little variation in child's age, this is a viable approach.}%
\end{remark}

\begin{remark} \label{rem:relax-classical-measurement-error}
  Another way to relax the classical measurement error assumption would be to allow for the independence conditions in \Cref{ass:me}(ii) and \Cref{ass:me}(iii) to hold conditional on $X$ (instead of the measurement error also being independent of $X$).  In that case, our arguments exploiting quantile regression do not go through, but it seems likely that one could identify the joint distribution of $(Y^*,T^*)$ conditional on covariates building on existing arguments on recovering distributions of mismeasured variables, particularly in settings where the researcher has access to repeated measurements.  In general, many approaches exist for identifying these distributions even in the presence of complicated forms of non-classical measurement error (see \citet{hu-2017} for a recent review).  In applications where repeated observations are available, it would seem possible to substitute these types of approaches for quantile regression in the first step of our identification arguments.\footnote{One common strategy in the intergenerational mobility literature is to average repeated measures of annual income into a single measure of permanent income and then subsequently ignore measurement error. Generally, in nonlinear settings (such as ours), approaches using averaging would require consistent estimation of permanent income, which, due to the incidental parameters problem, would require the number of time periods to grow to infinity (see, e.g., \citet{kato-galvao-2012,val-weidner-2016}). For these arguments to work, it seems likely that a researcher would already have access to an individual's entire annual earnings history. In this case, there is no measurement error issue at all, as one can simply compute permanent income.  Even in the context of linear models, averaging a few years of earnings may be of limited use in accounting for measurement error, especially when transitory income shocks can be persistent (\citet{mazumder-2005}).}  See \Cref{rem:qr,rem:qr2} for additional discussion on the pros and cons of using quantile regression in the first step.
\end{remark}

\subsection{Identifying Parameters of Interest}

Recall that all the parameters of our interest will be identified if the joint distribution of $Y^*$ and $T^*$ conditional on $X$ is identified.  Besides the assumptions on measurement error above, the main additional condition that we require is that the quantiles of the outcome conditional on the covariates and the quantiles of the treatment conditional on the covariates are linear in parameters across all quantiles.

Next, we present a series of results to identify this joint distribution.  As a first step, recall that by Sklar's Theorem (\citet{sklar-1959}), joint distributions can be written as the copula of their marginals, i.e.,
\begin{align} \label{eqn:sklar}
  \F_{Y^*T^*\mid X}(y,t\mid x) = C_{Y^*T^*\mid X}\big(\F_{Y^*\mid X}(y\mid x),\F_{T^*\mid X}(t\mid x)\mid x\big)
\end{align}
where $C_{Y^*T^*\mid X}$ is the copula of $Y^*$ and $T^*$ conditional on $X$ which captures the dependence between the ranks of $Y^*$ and $T^*$ conditional on $X$.  \Cref{eqn:sklar} is helpful because it splits the identification challenge into two parts: (i) identifying the marginal distributions and (ii) identifying the copula.  This section provides two results showing that (i) the distributions of the outcome/treatment conditional on covariates are identified, and (ii) the conditional copula in \Cref{eqn:sklar} is identified.  Taken together, these results show that the joint distribution is identified and, therefore, that all of our parameters of interest are also identified.

Before stating these results, we make the following additional assumption.

\begin{assumption}[Quantile Regression] \label{ass:qr} \mbox{} %

  (i) $Y^* = Q_{Y^*\mid X}(V_{Y^*}\mid X) = X'\beta_{Y^*}(V_{Y^*})$ with $V_{Y^*} | X \sim U[0,1]$.

  (ii) $T^* = Q_{T^*\mid X}(V_{T^*}\mid X) = X'\beta_{T^*}(V_{T^*})$ with $V_{T^*} | X \sim U[0,1]$.
\end{assumption}

\Cref{ass:qr} imposes linear quantile regression models for $Y^*$ and $T^*$ in terms of $X$.  \Cref{ass:qr} implies that $Y^*$ and $T^*$ arise from models that are both monotonically increasing in a scalar unobservable as well as being known up to the parameters $\beta_{Y^*}(\cdot)$ and $\beta_{T^*}(\cdot)$, respectively.  In the context of our application, the linear QR specification is a substantive modeling assumption.  That being said, it is a natural specification and is in line with QR-based approaches used in the intergenerational mobility literature (e.g., \citet{eide-showalter-1999,grawe-2004,Bratbergetal07}). In addition, one can allow for more flexible parametric specifications for the conditional quantiles by including higher-order terms and interactions among the original covariates. Next, we introduce some additional assumptions for recovering the distributions of the outcome and treatment conditional on covariates in the presence of measurement error.

\begin{assumption}[Additional Assumptions for QR with Measurement Error] \label{ass:additional-qr} \ For $j\in\{Y^*, T^*\}$, we make the following assumptions:

  (i) $\beta_j(\tau)$ is in the space $M[B_1 \times B_2 \times \cdots \times B_K]$ (with $K$ the dimension of $X$) where the functional space $M$ is the collection of all functions $b = (b_1, \ldots, b_K) : [0,1] \rightarrow [B_1 \times \cdots \times B_K]$ with $B_k$ a closed bounded interval of $\mathbb{R}$ for $k=1,\ldots,K$ such that $x'b(\tau) : [0,1] \rightarrow \mathbb{R}$ is monotonically increasing in $\tau$ for all $x \in \textrm{support}(X)$.  Each component $\beta_{jk}(\cdot)$ is twice continuously differentiable with uniformly bounded derivatives.

  (ii) $\E[XX']$ is nonsingular.  There exists at least one covariate $X_1$, such that conditional on the remaining covariates $X_{-1}$, $X_1\mid X_{-1}$ has a continuous distribution with strictly positive density on an open set and the corresponding coefficient $\beta_{j1}(\cdot)$ is strictly monotonic.

  (iii) The measurement error $U_j$ has a continuously differentiable density, mean zero, and satisfies the condition that there exists a constant $C>0$ such that for all $k>0, \E[ | U_j |^k ] < k!\cdot C^k$.
\end{assumption}

\Cref{ass:additional-qr} reproduces Assumptions 1-3 in \citet{hausman-liu-luo-palmer-2021} with slight changes in notation.  Part (i) is mainly a technical condition.  The condition that the mapping $\tau \mapsto x'\beta_j(\tau)$ is monotonically increasing is used to define the admissible class of quantile regression models in \Cref{prop:1} below.  Without such a restriction, rearrangements of quantile indices would yield observationally equivalent models once measurement error is introduced.  Part (ii) is the most substantive part of the assumption.  It requires (a) continuous variation in at least one covariate labeled $X_1$ and (b) a strictly monotone coefficient path for that covariate across quantiles. In the context of our application, a natural candidate for $X_1$ is age. The strict monotonicity requirement encodes a shape restriction on heterogeneous returns to age across the conditional income distribution; e.g., the marginal effect of age should increase with $\tau$.  This sort of condition would hold if those at the higher parts of the conditional income distribution see faster income growth with respect to age than those at lower parts of the conditional distribution.  Part (iii) imposes standard regularity conditions on the measurement error (smooth density, mean zero, and rules out thick tails) that are common in the measurement error literature and ensure the deconvolution is well-posed. All of these conditions are likely to be relatively mild in our application.

\citet{hausman-liu-luo-palmer-2021} show that QR parameters are identified in exactly the sort of setup discussed above.  Notice that identification of $\beta_{Y^*}(\cdot)$ and $\beta_{T^*}(\cdot)$  also implies that the distributions of the outcome/treatment conditional on covariates are also identified.  The next proposition states this result.

\begin{proposition}  \label{prop:1} Under \Cref{ass:me,ass:qr,ass:additional-qr}, $\beta_j(\cdot)$ and $\F_{U_j}$ are uniquely identified within the class of quantile regression models
satisfying \Cref{ass:me,ass:qr,ass:additional-qr} for $j \in \{Y^*,T^*\}$.
\end{proposition}

The proof of \Cref{prop:1} is provided in \Cref{app:proofs}.  With the quantiles identified, one can obtain the marginals simply by inverting the quantiles.  The next part of our identification argument shows that the copula of $Y^*$ and $T^*$ conditional on covariates, $C_{Y^*T^*\mid X}$, is identified.
\begin{theorem}  \label{prop:2} Under \Cref{ass:me,ass:qr,ass:additional-qr}, and under the assumption that the characteristic functions of $U_{Y^*}$ and $U_{T^*}$ are non-vanishing, $C_{Y^*T^*\mid X}$ is identified.
\end{theorem}

The proof of \Cref{prop:2} is provided in \Cref{app:proofs}.  Together, \Cref{prop:1,prop:2} imply that the joint distribution $\F_{Y^*T^*\mid X}$ is identified---\Cref{prop:1} implies that the marginals are identified (because the conditional quantiles can be inverted) and \Cref{prop:2} implies that the conditional copula is identified.  These results also imply that all of our parameters of interest---both the Conditional Distribution-type parameters and the Copula-type parameters---are identified.

\begin{remark} \label{rem:qr}
  The quantile regression structure imposed in \Cref{ass:qr} is crucial for being able to identify the conditional distributions of the outcome and the treatment without requiring repeated measurements of the outcome and treatment.  It does not require the parameters $\beta_{Y^*}(\cdot)$ and $\beta_{T^*}(\cdot)$ to have a structural/causal interpretation, but it does require that both models of the conditional quantiles are correctly specified.  There is an interesting connection between this requirement and early work on identifying the distribution of a variable of interest in the presence of measurement error (e.g., \citet{carroll-hall-1988,stefanski-carroll-1990,fan-1991,diggle-hall-1993});  this literature assumed the distribution of the measurement error was known (e.g., that the measurement error follows a normal distribution with known variance), which enabled recovering the distribution of the random variable of interest without extra requirements such as repeated measurements.  Instead of requiring the measurement error distribution to be known, \Cref{ass:qr} places structure on the conditional distributions of the outcome and treatment while leaving the measurement error distribution unrestricted up to some regularity conditions.
\end{remark}

  \begin{remark} \label{rem:qr2}
    \Cref{ass:qr} is relatively less important for showing the result in \Cref{prop:2}; for example, given that the conditional distributions of the outcome and treatment are identified and that the distributions of the measurement errors in the outcome and treatment equation are identified, \Cref{prop:2} does not otherwise rely on \Cref{ass:qr}.  This suggests that our arguments in the second step are likely to go through under alternative first step assumptions.  In particular, in applications where a researcher has access to repeated measurements of the outcome and treatment, it might be attractive to impose less structure than is provided by the quantile regression setup in \Cref{ass:qr}.
\end{remark}

\begin{remark}
  A leading alternative to our approach would be to use quantile regression of the outcome conditional on the treatment and covariates.  We are unaware of work that allows for measurement error in both the outcome and the treatment in a quantile regression setup.  Conditional on handling measurement error, this setup would deliver the Conditional Distribution-type parameters that we consider; however, it would not immediately deliver the Copula-type parameters that we consider.
\end{remark}

\section{Estimation} \label{sec:est}

This section considers estimating the parameters of interest from \Cref{sec:id} in the presence of measurement error based on our identification results from \Cref{sec:3}.  We make some simplifying assumptions next that make estimation simpler; in general, we have tried to strike a balance between generality and being able to implement our model in a practical way in realistic applications, though noting that one could proceed in a less restrictive way here at the cost of more complicated estimation.  In \Cref{app:simulations}, we assess the performance of our estimation strategy using Monte Carlo simulations.

\begin{assumption}[Random Sampling] \label{ass:sampling}
  $\{Y_i,T_i,X_i\}_{i=1}^n$ are $i.i.d.$ draws from the joint distribution $\F_{YTX}$.
\end{assumption}

\begin{assumption}[Distribution of Measurement Error]\label{ass:unobs} \ %
  $U_{Y^*} \sim \F_{U_{Y^*}}(\cdot|\sigma_{Y^*})$ and $U_{T^*} \sim \F_{U_{T^*}}(\cdot|\sigma_{T^*})$ where $\sigma_{Y^*}$ and $\sigma_{T^*}$ are finite-dimensional parameters.
\end{assumption}

\begin{comment}
    \begin{assumption}[Copula] \label{ass:cop} \ $C_{Y^*T^*\mid X}(\cdot,\cdot\mid x)$ is the same across all values of $x \in \textrm{support}(X)$.  Under this condition, let $C_{Y^*T^*\mid X}(\cdot, \cdot)$ denote the conditional copula.  In addition, $C_{Y^*T^*\mid X}(\cdot, \cdot) = C_{Y^*T^*\mid X}(\cdot, \cdot ;\delta)$; i.e., the conditional copula is known up to the finite-dimensional parameters $\delta$. Moreover, $ C_{Y^*T^*\mid X}(r, s ;\delta) $  is twice continuously differentiable in $(r,s) \in (0,1)^2 $. The conditional copula density $ c_{Y^*T^*\mid X}(r, s ;\delta) := \frac{\partial^2 }{\partial r \partial s} C_{Y^*T^*\mid X}(r, s ;\delta)  $ is twice continuously differentiable in $\delta$ and uniformly bounded in its arguments.
\end{assumption}
\end{comment}

\begin{assumption}[Copula]\label{ass:cop}
The conditional copula $C_{Y^*T^*\mid X}(\cdot,\cdot\mid x)$ is invariant in $x\in\mathrm{support}(X)$ and is denoted by $C_{Y^*T^*\mid X}(r,s;\delta)$ for some finite-dimensional parameter $\delta$. It is twice continuously differentiable in $(r,s)\in(0,1)^2$, and its density
\[
c_{Y^*T^*\mid X}(r,s;\delta)
:=\frac{\partial^2}{\partial r\,\partial s} C_{Y^*T^*\mid X}(r,s;\delta)
\]
is strictly positive, uniformly bounded in its arguments, and twice continuously differentiable in $\delta$.
\end{assumption}

\begin{assumption}[Splines] \label{ass:splines} Let $L$ denote a (fixed) number of knots and consider the sequence of equally spaced $\tau$, $0 < \tau_1 < \tau_2 < \cdots < \tau_L < 1$.  For any $\tau \in [\tau_1, \tau_L]$ and $j \in \{Y^*,T^*\}$,
  \begin{align*}
    \beta_j(\tau) = \beta_j(\tau_l) + (\tau - \tau_l) \frac{\beta_j(\tau_u) - \beta_j(\tau_l)}{\tau_u - \tau_l}
  \end{align*}
  where $\tau_l = \sup\{ u \in \{\tau_1, \ldots, \tau_L\} : u \leq \tau\}$ and $\tau_u = \inf\{ u \in \{\tau_1, \ldots, \tau_L\} : u \geq \tau\}$; i.e., $\tau_l$ is the closest smaller value of $\tau$ for which $\beta_j(\tau)$ has been estimated and $\tau_u$ is the closest larger value of $\tau$ for which $\beta_j(\tau)$ has been estimated. For $ \tau < \tau_1 $ and $ \tau \geq \tau_L $:
\begin{equation*}%
    \beta_{Y^*}(\tau) = \begin{cases}
        \beta_{Y^*}(\tau_1) + \frac{\log(\tau/\tau_1)}{1-\tau_1}e_1, & \tau < \tau_1 \\
        \beta_{Y^*}(\tau_L) - \frac{\log((1-\tau)/(1-\tau_L))}{\tau_L}e_1, & \tau \geq \tau_L
    \end{cases}
\end{equation*}where $e_1$ denotes the $K\times 1$ standard basis vector with 1 as its first element and zeroes everywhere else.
\end{assumption}

\Cref{ass:sampling} says that we have access to a sample of $n$ $i.i.d.$ draws of $Y$, $T$, and $X$.  \Cref{ass:unobs} says that the distribution of the measurement errors $U_{Y^*}$ and $U_{T^*}$ is known up to a finite number of parameters.  In practice, we will assume that $U_{Y^*}$ and $U_{T^*}$ are both mixtures of normal distributions.  This is similar to what \citet{hausman-liu-luo-palmer-2021} do in their application, and mixtures of normals are likely to have good approximating properties for continuously distributed measurement error.\footnote{A side-effect of assuming that the distributions of measurement error are known up to a finite number of parameters is that the smoothness of the measurement error---e.g., ordinary-smooth vs.~super-smooth measurement error---does not affect the rate of convergence of our estimator or our inference procedure.  This is in contrast with  \citet{hausman-liu-luo-palmer-2021}, whose main estimator does not specify a parametric model for the measurement error, and, hence, the smoothness of the measurement error affects inference and rates of convergence.\label{fn:smoothness}}   %
The first part of \Cref{ass:cop} says that the conditional copula does not vary across different values of the covariates (note that this does not imply that it is equal to the unconditional copula).  This sort of condition is frequently invoked in the literature on conditional copulas, where it is often referred to as the ``simplifying assumption'' as it greatly simplifies estimating conditional copulas. Despite being a common assumption, it is a strong assumption (see, for example, the discussion in  \citet{derumigny-fermanian-2017} as well as \citet{veraverbeke-omelka-gijbels-2011,gijbels-veraverbeke-omelka-2011}); \citet{nagler-2025} provides a recent systematic discussion of the plausibility of this assumption both in terms of models that rationalize it and in terms of empirical evidence on its plausibility across several different applications.  The second part of \Cref{ass:cop} says that the conditional copula of $Y^*$ and $T^*$ is known up to a finite number of parameters.  In principle, the conditional copula could be nonparametrically estimated (\Cref{prop:2} implies that it is identified, though it does not provide a directly estimable expression).  While possible, proceeding this way is likely to be quite complicated in practice (see, for example, \citet{li-vuong-1998, li-2002, veraverbeke-omelka-gijbels-2011, firpo-galvao-song-2017}).  Instead, we proceed by considering the case where $C_{Y^*T^*\mid X}$ is specified parametrically.
\Cref{ass:splines} allows us to recover the process $\beta_j(\cdot)$ from estimating a finite number of $\beta_j(\tau)$, i.e., estimate them over a finite grid of $\tau$ and interpolate the process from the particular set of $\beta_j(\tau)$ that we estimate (see \citet{wei-carroll-2009} and \citet{arellano-bonhomme-2016} for similar assumptions in related contexts).\footnote{The parameters $\rho_S$ and $\theta_U(\Delta, s_1, s_2)$ depend on integrals over the distributional tails. The second part of \Cref{ass:splines} complements the first part by specifying a spline model for the extreme intervals, following \citet[Appendix D]{arellano-bonhomme-2016} and \citet{geraci-2007-quantile}.}

\subsection{Step 1: Estimating Conditional Quantiles}

The first step is to estimate $Q_{Y^*\mid X}(\tau\mid x) = x'\beta_{Y^*}(\tau)$ and $Q_{T^*\mid X}(\tau\mid x)=x'\beta_{T^*}(\tau)$.  In this section, we outline the estimation procedure for $Q_{Y^*\mid X}$ and note that the arguments for $Q_{T^*\mid X}$ hold analogously.

First, notice that if $U_{Y^*}$ were observed, then we could estimate the QR parameters by quantile regression of $(Y-U_{Y^*})$ on $X$.
Next, let $\psi_{\tau}(w) = \tau - \indicator{w < 0}$ which is the derivative of the check function $\rho_\tau(w) = (\tau - \indicator{w < 0})w$.  If $Y^*$ were observed, the first-order condition for estimating the QR parameters is given by
\begin{align*}
  0 &= \E\big[X \psi_\tau\big(Y^*-X'\beta_{Y^*}(\tau)\big) \big] = \E\big[X \psi_{\tau}\big(Y-U_{Y^*} - X'\beta_{Y^*}(\tau)\big)\big]
\end{align*}
This is an infeasible moment condition because $Y^*$ (and $U_{Y^*}$) is unobserved.  However, notice that it can be rewritten as
\begin{align} \label{eqn:est1}
    0 = \E\left[ X \int_{\mathcal{U}} \psi_\tau\big(Y-u-X'\beta_{Y^*}(\tau)\big) f_{U_{Y^*}|Y,X}(u|Y,X) \ du \right]
\end{align}
Further, notice that from Bayes' Theorem,
\begin{align}
  f_{U_{Y^*}|Y,X}(u|y,x) = \frac{f_{Y|U_{Y^*},X}(y|u,x)f_{U_{Y^*}\mid X}(u\mid x)}{f_{Y\mid X}(y\mid x)} \label{eqn:me-bayes}
\end{align}
and that
\begin{align*}
  f_{Y|U_{Y^*},X}(y|u,x) = f_{Y^*\mid X}(y-u\mid x) \quad \textrm{and} \quad f_{U_{Y^*}\mid X}(u\mid x) = f_{U_{Y^*}}(u)
\end{align*}
which hold by \Cref{ass:me}(i)\footnote{To see the first part, notice that $\P(Y \leq y | X, U_{Y^*}=u) = \P(Y^* \leq y-u\mid X)$ which holds by \Cref{ass:me}(ii) and then the result follows immediately.} and \Cref{ass:me}(ii).  Moreover, given $\beta_{Y^*}(\cdot)$, the conditional cdf of $Y^*$ can be recovered as $ \displaystyle \F_{Y^*\mid X}(y\mid x) = \int_0^1 \indicator{x'\beta_{Y^*}(\tau) \leq y} \, d\tau$.  In estimation, this transformation ensures that the conditional cdf will be increasing in $y$ (\citet{chernozhukov-val-galichon-2010}). Similarly, the conditional density of $Y^*$ can be recovered as $f_{Y^*\mid X}(y\mid x) = \big[x'\beta^\partial_{Y^*}\big(\F_{Y^*\mid X}(y\mid x)\big)\big]^{-1}$ where $\beta_{Y^*}^{\partial}(\tau):= \frac{\partial \beta_{Y^*}(\tau) }{\partial \tau}$ (see, for example, \citet{rothe-wied-2020}).  \Cref{eqn:est1} thus suggests a two-step estimation procedure.  In the first step, make draws from $f_{U_{Y^*}|Y,X}$; in the second step, run QR using the observed $Y$, $X$, and the simulated measurement error.  In practice, we use the following estimation scheme, iterating back and forth between making draws of the measurement error and estimating the parameters using QR and maximum likelihood.

\begin{algorithm}\label{alg:Est} \mbox

  \bigskip

  \hspace{-10pt} \textit{Initialize:}

  \begin{adjustwidth}{20pt}{}
  \begin{enumerate}
      \item Set the counter $l=0$.  Set initial values of the parameters $\{\widehat{\beta}^{(l)}_{Y^*}, \hat{\sigma}^{(l)}_{Y^*} \}$.\footnote{EM algorithms are not guaranteed to converge to the global optimum and may instead converge to a local optimum depending on the choice of starting values (\citet{wu-1983convergence}).  The most common strategy for guarding against this is to run the algorithm from multiple random starting values and select the solution with the highest log-likelihood; see, e.g., the discussion in \citet{balakrishnan-2017statistical}. \label{fn:algo-initialization}}  %
      Set a tolerance level $\varepsilon = o\big(n^{-1/2}\big)$.
  \end{enumerate}
  \end{adjustwidth}

  \bigskip

  \hspace{-10pt} \textit{Iterate:}

  \begin{adjustwidth}{20pt}{}
  \begin{enumerate}
    \setcounter{enumi}{1}
    \item Set $l=l+1$.  For $i=1,\ldots,n$, make $S$ draws from $\widehat{f}_{U_{Y^*}|Y,X}(u|Y_i,X_i) \propto \widehat{f}_{Y^*\mid X}(Y_i - u\mid X_i; \widehat{\beta}^{(l-1)}_{Y^*}) \widehat{f}_{U_{Y^*}}(u|\hat{\sigma}_{Y^*}^{(l-1)})$.  %

    \item Using the $nS$ pseudo-observations from Step 2 above, estimate $\widehat{\beta}_{Y^*}^{(l)}$ using quantile regression of $Y_i - U_{is}$ as the outcomes and $X_i$ as the covariates.  Estimate $\hat{\sigma}_{Y^*}^{(l)}$ using the same $nS$ observations.

    \item If $\left\|\left(\widehat{\beta}_{Y^*}^{(l)}, \hat{\sigma}_{Y^*}^{(l)}\right) - \left(\widehat{\beta}_{Y^*}^{(l-1)}, \hat{\sigma}_{Y^*}^{(l-1)}\right) \right\| < \varepsilon$, return $\left(\widehat{\beta}_{Y^*}^{(l)}, \hat{\sigma}_{Y^*}^{(l)}\right)$, else continue iterating.
  \end{enumerate}
  \end{adjustwidth}
\end{algorithm}

\subsection{Step 2: Estimating the copula parameter}
Under \Cref{ass:cop}, the copula $C_{Y^*T^*\mid X}$ is known up to a finite number of parameters.  In this section, we propose to estimate the copula parameters by simulated maximum likelihood.

Building up to an estimation routine based on maximum likelihood, first notice that the observed distribution of mismeasured variables $Y$ and $T$ is given by
\begin{align} \label{eqn:est-cop1}
    \P(Y \leq \bar{y}, T \leq \bar{t} | X=x) &= \P(Y^* + U_{Y^*} \leq \bar{y}, T^* + U_{T^*} \leq \bar{t} | X=x) \nonumber \\
    &= \int \indicator{y + u \leq \bar{y}, t + v \leq \bar{t}} f_{Y^*T^*\mid X}(y,t\mid x) f_{U_{Y^*}}(u) f_{U_{T^*}}(v) \, dy \, dt \, du \, dv \nonumber \\
    &= \int C_{Y^*T^*\mid X}\big( \F_{Y^*\mid X}(\bar{y} - u|x), \F_{T^*\mid X}(\bar{t} - v|x) \big) f_{U_{Y^*}}(u) f_{U_{T^*}}(v) \, du \, dv
\end{align}
where the second equality holds by \Cref{ass:me}, and the third holds (after some manipulation) by \Cref{prop:2} and \Cref{ass:cop}. %
Note that \Cref{prop:1} implies that $\F_{Y^*\mid X}$, $\F_{T^*\mid X}$, $f_{U_{Y^*}}$, and $f_{U_{T^*}}$ are identified; they are each known up to a finite number of parameters by \Cref{ass:unobs,ass:cop,ass:splines}.  What remains is estimating the copula parameters.

\Cref{eqn:est-cop1} suggests estimating the copula parameters by maximum likelihood, and it implies
\begin{align*}
  f_{YT\mid X}(\bar{y},\bar{t}\mid x) &= \int c_{Y^*T^*\mid X}\big(\F_{Y^*\mid X}(\bar{y}-u\mid x), \F_{T^*\mid X}(\bar{t}-v \mid x);\delta\big) \\
  & \hspace{50pt} \times f_{Y^*\mid X}(\bar{y} - u \mid x) f_{T^*\mid X}(\bar{t} - v \mid x) f_{U_{Y^*}}(u) f_{U_{T^*}}(v) \, du \, dv \\
  &= \E\Big[c_{Y^*T^*\mid X}\big(\F_{Y^*\mid X}(\bar{y}- U_{Y^*}\mid x), \F_{T^*\mid X}(\bar{t}- U_{T^*}\mid x);\delta\big) f_{Y^*\mid X}(\bar{y} - U_{Y^*} \mid x) f_{T^*\mid X}(\bar{t} - U_{T^*} \mid x) \Big]
\end{align*}
where $c_{Y^*T^*\mid X}$ is the copula pdf, and the expectation is over the measurement error variables $U_{Y^*}$ and $U_{T^*}$. The copula parameter can be estimated by maximizing the log-likelihood function.  Let
\begin{align*}
  \widehat{L}_i^{(S)}(\delta) &= \sum_{s=1}^S c_{Y^*T^*\mid X}\big(\widehat{\F}_{Y^*\mid X}(Y_i- U_{Y^*}^{(s)}\mid X_i), \widehat{\F}_{T^*\mid X}(T_i- U_{T^*}^{(s)}\mid X_i); \delta\big)  \\
  &\hspace{50pt} \times \widehat{f}_{Y^*\mid X}(Y_i - U_{Y^*}^{(s)} | X_i) \widehat{f}_{T^*\mid X}(T_i- U_{T^*}^{(s)} | X_i)
\end{align*}
where $U_{Y^*}^{(s)}$ and $U_{T^*}^{(s)}$ are independent draws from the estimated measurement error distributions $\widehat{f}_{U_{Y^*}}(\cdot;\widehat{\sigma}_{Y^*})$ and $\widehat{f}_{U_{T^*}}(\cdot;\widehat{\sigma}_{T^*})$, respectively.
Then, we estimate the copula parameter by
\begin{align*}
  \widehat{\delta} = \arg \max_{\delta} \sum_{i=1}^n \log\big(\widehat{L}_i^{(S)}(\delta)\big).
\end{align*}

\subsection{Step 3: Estimating Parameters of Interest}

\subsubsection{Conditional Distribution-type Parameters}

First, we focus on estimating the Conditional Distribution-type parameters such as $\F_{Y^*|T^*X}$ and $Q_{Y^*|T^*X}$.  \citet{bouye-salmon-2009} study the relationship between copulas and quantile regression in the bivariate case (see also \citet{chen-fan-2006}).  The bivariate case is the relevant case for the setup in the current paper, as the results for quantile regression and a bivariate copula go through in our case, with everything holding conditional on $X$.  The results of \citet[p.~726]{bouye-salmon-2009} imply that
\begin{align}\label{eqn:Fc}
  \F_{Y^*|T^*X}(y|t,x) = C_{2\mid X}\big(\F_{Y^*\mid X}(y\mid x), \F_{T^*\mid X}(t\mid x)\big)
\end{align}
where $C_{2\mid X}(u_1, u_2) = \frac{ \partial C_{Y^*T^*\mid X}(u_1,u_2)}{\partial u_2}$ which provides an alternative expression for $\F_{Y^*|T^*X}$ (here we also use \Cref{ass:cop} which says that the conditional copula does not vary with $x$). Under the conditions that $C_{2\mid X}$ is invertible in its first argument, the quantiles can be directly obtained by
\begin{align} \label{eqn:Qc}
  Q_{Y^*|T^*X}(\tau|t,x) = Q_{Y^*\mid X}\left(C_{2;1\mid X}^{-1}\big(\tau; \F_{T^*\mid X}(t\mid x)\big) \ \Big| \  x\right)\end{align}
where $C_{2;1\mid X}^{-1}(\cdot;\cdot)$ is the inverse of $C_{2\mid X}$ with respect to its first argument.

\begin{example}[Archimedean Family of Copulas]
  An Archimedean copula is given by
\begin{align*}
  C(u_1,u_2) = \psi^{-1}\big(\psi(u_1) + \psi(u_2)\big)
\end{align*}
where $\psi$ is called a generator function (see \citet{nelsen-2007} for more details).  The results of \citet[p. 729]{bouye-salmon-2009} imply that
\begin{align*}
  \F_{Y^*|T^*X}(y|t,x) = \frac{\psi'\big(\F_{T^*\mid X}(t\mid x)\big)}{\psi'\big(\psi^{-1}\big[\psi\big(\F_{T^*\mid X}(t\mid x)\big) + \psi\big(\F_{Y^*\mid X}(y\mid x)\big)\big]\big)}
\end{align*}
and
\begin{align*}
  Q_{Y^*|T^*X}(\tau|t,x) = Q_{Y^*\mid X}\left( \psi^{-1} \left[ \psi\left( \psi'^{-1}\left( \frac{1}{\tau} \psi'\big(\F_{T^*\mid X}(t\mid x)\big)\right)\right) - \psi\big(\F_{T^*\mid X}(t\mid x)\big)  \right] \ \Bigg| \ x\right)
\end{align*}
\end{example}

\begin{example}
The Clayton copula is a particular Archimedean family with $C(u_1,u_2;\delta) = (u_1^{-\delta} + u_2^{-\delta} -1)^{-1/\delta}$ where we restrict $\delta > 0$; the generator function is given by $\psi(s; \delta) = \delta^{-1}(s^{-\delta} - 1)$, and one can show that
\begin{align*}
  \F_{Y^*|T^*X}(y|t,x) = \Big[1 + \F_{T^*\mid X}(t\mid x)^\delta\big(\F_{Y^*\mid X}(y\mid x)^{-\delta} - 1\big)\Big]^{-(1+\delta)/\delta}
\end{align*}
and
\begin{align*}
  Q_{Y^*|T^*X}(\tau|t,x) = Q_{Y^*\mid X}\left(\left[(\tau^{-\delta/(1+\delta)} - 1)\F_{T^*\mid X}(t\mid x)^{-\delta} + 1\right]^{-1/\delta} \ \Bigg| \ x \right)
\end{align*}
\end{example}

The Clayton copula has the nice property of having a one-to-one correspondence with Kendall's Tau.

\begin{example}[Gaussian Copula]
The Gaussian copula is given by $C(u_1,u_2;\rho) = \Phi_2\big(\Phi^{-1}(u_1), \Phi^{-1}(u_2); \rho\big)$ where $\Phi$ is the cdf of a standard normal random variable and $\Phi_2$ is the cdf of a pair of random variables that are jointly normally distributed with mean 0, variance 1, and correlation coefficient $\rho$.  Then,
\begin{align*}
  \F_{Y^*|T^*X}(y|t,x) = \Phi \left( \frac{ \Phi^{-1}\big(\F_{Y^*\mid X}(y\mid x)\big) - \rho \Phi^{-1}\big(\F_{T^*\mid X}(t\mid x)\big)}{ \sqrt{1-\rho^2} }\right)
\end{align*}
and
\begin{align*}
  Q_{Y^*|T^*X}(\tau|t,x) = Q_{Y^*\mid X}\left( \Phi( \rho \Phi^{-1}\big(\F_{T^*\mid X}(t\mid x)\big) + \sqrt{1-\rho^2} \Phi^{-1}(\tau) \  \Big| \ x\right)
\end{align*}
\end{example}
See \citet{bouye-salmon-2009} for more discussion of particular parametric families of copulas and quantile regression.

\subsubsection{Copula-type Parameters}

Next, we consider estimating our Copula-type parameters.  One thing to notice is that the copula that we estimated was the \textit{conditional} copula, but the copula parameters that we focus on are functionals of the unconditional copula.  Thus, the key step to estimating the Copula-type parameters is to back out the unconditional copula from our estimates of the conditional copula and conditional distributions of $Y^*$ and $T^*$.  In particular, notice that
\begin{align*}
\widehat{C}_{Y^*T^*}(r,s) :=
\frac{1}{n}\sum_{i=1}^n C_{Y^*T^*\mid X}\big(\widehat{\F}_{Y^*\mid X}(\widehat{\F}_{Y^*}^{-1}(r)\mid X_i),\widehat{\F}_{T^*\mid X}(\widehat{\F}_{T^*}^{-1}(s)\mid X_i);\, \widehat{\delta}\big)
\end{align*}
is an estimator of the unconditional copula.  Then, to estimate particular cells in the transition matrix, one can simply plug into \Cref{eqn:tm}; i.e.,
\begin{align*}
  \widehat{\theta}_{TM}(r_1,r_2,s_1,s_2) = \frac{\widehat{C}_{Y^*T^*}(r_2,s_2) - \widehat{C}_{Y^*T^*}(r_1,s_2) - \widehat{C}_{Y^*T^*}(r_2,s_1) + \widehat{C}_{Y^*T^*}(r_1,s_1)}{s_2 - s_1}
\end{align*}
Similarly, a plug-in estimator via numerical integration of the upward mobility parameter is given by
\begin{align*}
    \widehat{\theta}_U(\Delta, s_1, s_2): = \int_{0}^{1}\!\!\int_{0}^{1}\frac{\indicator{r>s+\Delta}\,\indicator{s_1\le s\le s_2}}{(s_2 - s_1)}\,\frac{\widehat{f}_{Y^* T^*}\!\big(\widehat{\F}_{Y^*}^{-1}(r),\,\widehat{\F}_{T^*}^{-1}(s)\big)}
{\widehat{f}_{Y^*}\!\big(\widehat{\F}_{Y^*}^{-1}(r)\big)\,\widehat{f}_{T^*}\!\big(\widehat{\F}_{T^*}^{-1}(s)\big)}
\,dr\,ds.
\end{align*}where
\begin{align*}
    \widehat{f}_{Y^* T^*}\!\big(y,\,t\big) = \frac{1}{n}\sum_{i=1}^n c_{Y^*T^*\mid X}\big(\widehat{\F}_{Y^*\mid X}(y\mid X_i), \widehat{\F}_{T^*\mid X}(t\mid X_i);\, \widehat{\delta}\big)\widehat{f}_{Y^*\mid X}(y\mid X_i)\widehat{f}_{T^*\mid X}(t\mid X_i)
\end{align*}is an estimator of the unconditional joint pdf.

\section{Inference} \label{sec:inference}

In this section, we develop the asymptotic properties of our estimators. We focus on showing asymptotic normality and bootstrap validity of the estimators of our main target parameters.  The key ingredient for providing these results is to establish asymptotic normality of our first step estimators of the QR and measurement error parameters.  \Cref{alg:Est} delivers estimates of the QR and measurement error parameters \textit{at convergence}.\footnote{The consistency and asymptotic normality results below apply to the global optimum of the sample criterion.  In practice, since EM algorithms may converge to a local rather than global optimum depending on the choice of starting values, one must take care to select starting values that yield the highest observed log-likelihood across multiple initializations; see the implementation details in \Cref{sec:emp}. \label{fn:assume-global-max}}  Thus, the estimated parameters solve the sample first-order conditions up to an asymptotically negligible term, and we can use proof techniques similar to those for establishing the limiting distribution of M-estimators.  In particular, after establishing the consistency of our first step estimators, we linearize the score around the population parameters and apply the Central Limit Theorem (CLT). The main technical challenge lies in handling the nuisance conditional density arising from the measurement-error component. Specifically,
\begin{align*}
f_{U_{Y^*}\mid Y,X}(u\mid y,x;\sigma_{Y^*})
:= \frac{\left[ x'\beta_{Y^*}^{\partial}\left(\int_0^1 \indicator{ x'\beta_{Y^*}(\tilde{\tau}) \leq (y-u) } \, d\tilde{\tau}\right) \right]^{-1} f_{U_{Y^*}}(u ; \sigma_{Y^*})}{\int_{\mathcal{U}} f_{Y \mid U_{Y^*}, X}(y \mid \tilde{u}, x)\,
   f_{U_{Y^*}}(\tilde{u} ; \sigma_{Y^*})\, d\tilde{u}}\,,
\end{align*}
where this expression follows from \Cref{eqn:me-bayes} after invoking \Cref{ass:unobs}.  This expression makes explicit the dependence on $\sigma_{Y^*}$, the quantile-slope function $\beta_{Y^*}^{\partial}(\tau):=\partial\beta_{Y^*}(\tau)/\partial\tau$, as well as the entire quantile coefficient process $ \big\{\beta_{Y^*}(\tau): \tau\in(0,1) \big\}$. Consequently, conventional asymptotic arguments that treat each index $\tau $ separately are not applicable. Note that, for the discussion of results for our first step estimators, we present results for $(Y^*,X',U_{Y^*})$, but corresponding results apply for $(T^*,X',U_{T^*})$ using analogous arguments. Finally, given asymptotic normality of our estimator of the QR parameters, we conclude this section by proving asymptotic normality and bootstrap validity for all of our main target parameters.

\subsection{Consistency of First Step Estimators of QR and Measurement Error Parameters}

To start with, we show the consistency of our first step estimators of the quantile regression parameters and measurement error parameters.  This part relies on the identification result in \Cref{prop:1}.  We make the following assumption.

\begin{assumption}[Compact parameter space and uniform interior]\label{ass:compact_par_space}
Let $\mathcal T:=[\tau_1, \, \tau_L], \, 0 < \tau_1 < \tau_L < 1 $, $\Gamma_\beta\subset\mathbb R^{K}$, and $ \Gamma_\delta \subset \mathbb{R}^{P} $ be compact with nonempty interior,  and
$\Gamma_\sigma=[\underline{\sigma},\overline{\sigma}]$ with
$0\le \underline{\sigma}<\overline{\sigma}<\infty$.
Define $\Gamma:=\Gamma_\beta\times\Gamma_\sigma\times \Gamma_\delta $.
For all $\tau\in\mathcal T$,
\[
\big(\beta_{Y^*}(\tau)',\ \sigma_{Y^*}, \delta' \big)' \in \operatorname{int}(\Gamma),
\quad\text{and}\quad
\inf_{\tau\in\mathcal T}\operatorname{dist}\!\big(\,(\beta_{Y^*}(\tau)',\sigma_{Y^*},\delta')',\ \partial\Gamma\,\big)\ \ge\ \kappa
\]
for some $\kappa>0$, where $\operatorname{dist}(\cdot,\partial\Gamma)$ is the Euclidean distance to the boundary of $\Gamma$.
\end{assumption}
\noindent \Cref{ass:compact_par_space} is a standard regularity condition on the parameter space commonly imposed in the analysis of M-estimators.  Importantly, the product space $\Gamma $ does not depend on $\tau$, and we focus on $\mathcal T := [\tau_1, \, \tau_L] $, a compact subset of $(0,1)$.

For the proof of consistency, we first proceed by providing a representation of the population criterion corresponding to \Cref{alg:Est}.  The population criterion for a fixed quantile index $\tau \in \mathcal{T}$ is given by
\begin{align*}
    Q_\tau(\beta,\sigma) &= \mathbb{E} \left[ \int_{\mathcal{U}} \rho_\tau\left( Y - u - X'\beta \right) f_{U_{Y^*} | Y, X}(u \mid Y, X;\sigma) \, du \right]
\end{align*}where $ \displaystyle (\beta_{Y^*}(\tau)',\sigma_{Y^*})' := \argmin_{(\beta',\sigma)' \in \Gamma } Q_\tau(\beta,\sigma) $.

The next crucial ingredient for the consistency result is uniform convergence in probability of the sample criterion. Define the empirical (Monte–Carlo) conditional density built from the $S$ draws $ U_{is}, \, s=1,\ldots,S $ for each
$i$ by
\[
\widehat f^{(S)}_{U_{Y^*}\mid Y,X}(u \mid Y_i, X_i; \widehat\sigma_{Y^*})
:= \frac{1}{S}\sum_{s=1}^S \delta_{U_{is}}(u)
\] where $\delta_{U_{is}}(u)$ is the Dirac point mass at $u$. Consider the following sample criterion:
\begin{align*}
    Q_{n,\tau}(\beta,\sigma) &=\frac{1}{n} \sum_{i=1}^n \left[\int_{\mathcal{U}}\rho_\tau\left( Y_i - u - X_i'\beta \right) \widehat{f}_{U_{Y^*} | Y, X}^{(S)}(u \mid Y_i, X_i; \sigma ) du \right]
\end{align*}with the estimator in \Cref{alg:Est} defined by
$ \big(\widehat{\beta}_{Y^*}(\tau)',\widehat{\sigma}_{Y^*}\big)':= \displaystyle \argmin_{(\beta',\sigma)' \in \Gamma } Q_{n,\tau}(\beta,\sigma), \, \tau\in \{\tau_1,\ldots,\tau_L \} $.

For each $i\in\{1,\dots,n\}$ and $\sigma\in\Gamma_\sigma$, let $K_{i,\sigma}$ be the transition kernel of the Markov chain \( \{U_{i1}, \ldots, U_{iS} \}(\sigma) \) on $\mathcal U$, conditional on $(Y_i,X_i)$, with invariant law $\mathcal{F}_{U_{Y^*} \mid Y,X}(\cdot\mid Y_i,X_i;\sigma)$ where $\mathcal{F}_{U_{Y^*} \mid Y,X}\big((-\infty,u]\mid Y_i,X_i;\sigma\big) := \F_{U_{Y^*} \mid Y,X}(u\mid Y_i,X_i;\sigma) $. The following assumption is imposed on the convergence of the Markov chain uniformly in $ \sigma$.
\begin{assumption}[MCMC Chain]\label{ass:MCMC_conv}
There exists $q:\mathcal U\to[1,\infty)$, $M<\infty$, and $\rho\in(0,1)$ such that, for all $u\in\mathcal U$, $s\ge1$,
 $$
 \sup_{\sigma\in\Gamma_\sigma}
 \big\|\,K_{i,\sigma}^s(u,\cdot)
 - \mathcal{F}_{U\mid Y,X}(\cdot\mid Y_i,X_i;\sigma)\,\big\|_{\mathrm{TV}}
 \;\le\; M\,q(u)\,\rho^s
 $$
\noindent where \( \displaystyle \sup_{\sigma\in\Gamma_\sigma}\mathbb E[q(U_{i0}(\sigma))\mid Y_i,X_i]<\infty \ almost \ surely \ (a.s.)\) and $ \| \cdot \|_{\mathrm{TV}} $ denotes the total variation norm.
\end{assumption}
\noindent \Cref{ass:MCMC_conv} enables the use of a conditional strong law of large numbers, ensuring that the Monte Carlo approximation error is $O_p(S^{-1/2})$. Sufficient conditions for \Cref{ass:MCMC_conv} are provided, for example, by \citet{rosenthal-1995-minorization}. Notably, \Cref{ass:MCMC_conv} implies that the Markov chain $\{U_{i1},\ldots,U_{iS}\}(\sigma)$ for each $i$ is $\beta$-mixing (see \citet[p.89]{doukhan-1995-mixing}), which in turn guarantees that the Monte Carlo approximation error is $O_p(S^{-1/2})$; see, e.g., \citet[Theorem 10]{prakasa2009conditional}. We also make the following assumption.

\begin{assumption}[Dominance]\label{ass:dominance_Xf}
    There exists a positive constant $C<\infty$ such that almost surely ($a.s.$), \quad (a) $ \E[\lVert X \rVert^2 ] + \E[\lVert Y \rVert^2 ] \leq C $; (b) \( \displaystyle \sup_{\sigma \in \Gamma_\sigma } f_{U_{Y^*} | Y, X}(u \mid Y, X;\sigma) \leq C \); (c) \( \displaystyle \sup_{\sigma \in \Gamma_\sigma } \int_{\mathcal{U}} u^2 f_{U_{Y^*} | Y, X} (u \mid Y, X; \sigma ) \leq C \); (d) $ f_{Y^*\mid X}(y\mid X) \in (0,\infty) $ and continuous in $y \in \mathcal{Y} $; (e) $ f_{U_{Y^*} | Y, X}(u \mid Y, X;\sigma) $ is continuously differentiable with respect to $ \sigma $, and \( \displaystyle \E\Big[\Big(\int_{\mathcal{U}}\big|u f_{U_{Y^*} | Y, X}^{\partial} (u \mid Y_i, X_i; \sigma_{Y^*} ) \big| \, du \Big)^2 \Big] \leq C. \)
\end{assumption}

\Cref{ass:dominance_Xf} is a standard regularity condition that imposes boundedness and smoothness conditions on the outcomes, covariates, and measurement error.

The following lemma provides a uniform weak law of large numbers result necessary for establishing the consistency of the estimator in \Cref{alg:Est}.
\begin{lemma}\label{lem:UWLLN}
    Suppose \Cref{ass:sampling,ass:MCMC_conv,ass:compact_par_space,ass:dominance_Xf,ass:unobs} hold, then as $ n\rightarrow \infty $ and $n/S \rightarrow 0 $,
    \[
    \sup_{(\tau, \sigma)' \in \mathcal{T}\times \Gamma_\sigma }\Big|Q_{n,\tau}\big(\beta(\tau),\sigma\big) - Q_\tau\big(\beta(\tau),\sigma\big)\Big| \xrightarrow{p} 0.
    \]
\end{lemma}
The proof of \Cref{lem:UWLLN} is provided in the Supplementary Appendix.  Next, the following theorem provides the consistency result of $ \big(\widehat{\beta}_{Y^*}(\tau)',\widehat{\sigma}_{Y^*}\big) $ uniformly in $\mathcal{T}$.
\begin{theorem}\label{theorem:consistency}
    Suppose \Cref{ass:me,ass:qr,ass:additional-qr,ass:sampling,ass:MCMC_conv,ass:dominance_Xf,ass:unobs,ass:splines,ass:compact_par_space} hold, then as $ n\rightarrow \infty $ and $n/S \rightarrow 0 $,
    \[
    \sup_{\tau \in \mathcal{T}} \big\lVert \big(\widehat{\beta}_{Y^*}(\tau)',\widehat{\sigma}_{Y^*}\big)' - \big(\beta_{Y^*}(\tau)',\sigma_{Y^*}\big)' \big\rVert = o_p(1).
    \]
\end{theorem}
The proof of \Cref{theorem:consistency} is provided in \Cref{app:inference-proofs}.

\subsection{Asymptotic Normality}

This section establishes joint asymptotic linearity and normality for the stacked $ (LK + 1)\times 1 $ parameter vector
\(
\big(\,(\widehat{\bm\beta}_{Y^*}^{\,L}-{\bm\beta}_{Y^*}^{\,L})',\ \widehat{\sigma}_{Y^*}-\sigma_{Y^*}\big)'
\),
where \({\bm\beta}_{Y^*}^{\,L}:=(\beta_{Y^*}(\tau_1)',\ldots,\beta_{Y^*}(\tau_L)')'\).
This result provides the foundation for deriving the limiting distributions of the target parameters introduced in \Cref{sec:id}.

\begin{theorem}\label{theorem:anorm_beta}
    Suppose \Cref{ass:me,ass:qr,ass:additional-qr,ass:sampling,ass:unobs,ass:splines,ass:MCMC_conv,ass:dominance_Xf,ass:compact_par_space} hold. Further, assume the $ L(K+1) \times (LK +1) $ matrix $\mathbb{H}_{Y^*}^L$ in \eqref{eqn:H_L} is full column rank,  then as $ (n/S) \rightarrow 0 $, (a) $ \Big( \big(\widehat{\bm\beta}_{Y^*}^L - {\bm\beta}_{Y^*}^L \big)', \ \widehat{\sigma}_{Y^*} - \sigma_{Y^*} \Big)' $ has the following asymptotically linear representation:
     \begin{align*}
   \sqrt{n} \begin{bmatrix}
        \widehat{\bm\beta}_{Y^*}^L - {\bm\beta}_{Y^*}^L\\
        \widehat{\sigma}_{Y^*} - \sigma_{Y^*}
    \end{bmatrix} &=
    \big({\mathbb{H}_{Y^*}^L}'\mathbb{H}_{Y^*}^L\big)^{-1}{\mathbb{H}_{Y^*}^L}'
    \sqrt{n}\mathbb{C}_{n,Y^*}^L + o_p(1)
\end{align*}
and,
(b) \[
\sqrt{n} \begin{bmatrix}
        \widehat{\bm\beta}_{Y^*}^L - {\bm\beta}_{Y^*}^L\\
        \widehat{\sigma}_{Y^*} - \sigma_{Y^*}
    \end{bmatrix}
    \xrightarrow{d} \mathcal{N}\big( 0, \,  \Omega_{Y^*}^L \big)
\]where \( \E\big[ \mathbb{C}_{n,Y^*}^L \big] = 0 \) and $ \displaystyle \Omega_{Y^*}^L:= \big({\mathbb{H}_{Y^*}^L}'\mathbb{H}_{Y^*}^L\big)^{-1}{\mathbb{H}_{Y^*}^L}' \Big(\lim_{n\rightarrow \infty} n\mathrm{Var}\big[ \mathbb{C}_{n,Y^*}^L \big] \Big) \mathbb{H}_{Y^*}^L \big({\mathbb{H}_{Y^*}^L}'\mathbb{H}_{Y^*}^L\big)^{-1} $.
\end{theorem}
\noindent The proof of \Cref{theorem:anorm_beta} is provided in \Cref{app:inference-proofs}.  The condition that $ (n/S) \rightarrow 0 $ requires $S$ to go to infinity faster than $n$. In practice, this can be achieved by setting $S = an^{1+d}, \, a>0, d>0 $.

Next, we state a corollary to this result that says that our estimator of $\F_{Y^*|X}(y|x)$, which is directly estimable given our estimates of the QR parameters and is an important input for estimating our main target parameters, is also asymptotically normal.

\begin{corollary} \label{cor:anorm_Ystar_given_X}
    Suppose \Cref{ass:me,ass:qr,ass:additional-qr,ass:sampling,ass:unobs,ass:splines,ass:MCMC_conv,ass:dominance_Xf,ass:compact_par_space} hold, then
    \begin{align*}
        \sqrt{n}\big( \widehat{\F}_{Y^* \mid X}(y \mid x) - \F_{Y^* \mid X}(y \mid x) \big) \xrightarrow{d} \mathcal{N}\big(0,\, \sigma(\F_{Y^* \mid X}(y \mid x))\big),
    \end{align*}
    where $\sigma(\F_{Y^* \mid X}(y \mid x))$ is explicitly defined in \Cref{app:proof-cor1}.
\end{corollary}
The proof of \Cref{cor:anorm_Ystar_given_X} is provided in \Cref{app:inference-proofs}.  The next proposition shows that our estimator of the parameters of the conditional copula is also asymptotically normal.
\begin{proposition} \label{prop:anorm_copula}
    Suppose \Cref{ass:me,ass:qr,ass:additional-qr,ass:sampling,ass:splines,ass:dominance_Xf,ass:MCMC_conv,ass:unobs,ass:compact_par_space,ass:cop} hold, then
    \begin{align*}
        \sqrt{n}\big( \widehat{\delta} - \delta\big) \xrightarrow{d} \mathcal{N}\big(0,\, \Sigma(\delta)\big),
    \end{align*}
    where $\Sigma(\delta)$ is explicitly defined in \Cref{app:anorm_copula}.
\end{proposition}

Finally, our main result in this section is to show that our estimators of the main target parameters discussed above are all asymptotically normal and that the empirical bootstrap can be used to approximate their limiting distribution. Specifically, each bootstrap iteration draws $n$ observations with replacement from the original sample and re-runs the complete three-step estimation procedure---including the EM algorithm for the QR and measurement error parameters, the simulated maximum likelihood step for the copula parameters, and the computation of the target parameters---on the resampled dataset.
\begin{theorem}\label{theorem:anorm_pars}
    Suppose \Cref{ass:me,ass:qr,ass:additional-qr,ass:sampling,ass:splines,ass:dominance_Xf,ass:MCMC_conv,ass:unobs,ass:compact_par_space,ass:cop} hold, then
    \begin{align*}
       (a)& \quad \sqrt{n}\big( \widehat{\F}_{Y^*T^*\mid X}(y,t\mid x) - \F_{Y^*T^*\mid X}(y,t\mid x) \big) \xrightarrow{d} \mathcal{N}\big(0,\, \sigma(\F_{Y^*T^*\mid X}(y,t\mid x))\big); \\
       (b)& \quad \sqrt{n}\big( \widehat{\F}_{Y^*|T^*X}(y|t,x) - \F_{Y^*|T^*X}(y|t,x) \big) \xrightarrow{d} \mathcal{N}\big(0,\, \sigma(\F_{Y^*|T^*X}(y|t,x))\big); \\
       (c)& \quad \sqrt{n}\big( \widehat{Q}_{Y^*|T^*X}(\tau \mid t,x) - Q_{Y^*|T^*X}(\tau \mid t,x) \big) \xrightarrow{d} \mathcal{N}\big(0,\, \sigma(Q_{Y^*|T^*X}(\tau \mid t,x))\big); \\
       (d)& \quad \sqrt{n}\big( \widehat{\theta}_{TM}(r_1,r_2,s_1,s_2) - \theta_{TM}(r_1,r_2,s_1,s_2) \big) \xrightarrow{d} \mathcal{N}\big(0,\, \sigma(\theta_{TM}(r_1,r_2,s_1,s_2))\big); \\
       (e)& \quad \sqrt{n}(\widehat{\rho}_S - \rho_S ) \xrightarrow{d} \mathcal{N}\big(0,\, \sigma(\rho_S)\big);  \text{ and}\\
       (f)& \quad \sqrt{n}\big( \widehat{\theta}_{U}(\Delta,s_1,s_2) - \theta_{U}(\Delta,s_1,s_2) \big) \xrightarrow{d} \mathcal{N}\big(0,\, \sigma(\theta_{U}(\Delta,s_1,s_2))\big).
    \end{align*}
    where each of the asymptotic variances is defined in the proof of \Cref{theorem:anorm_pars}. In addition, the empirical bootstrap consistently estimates the limiting distribution of each estimator.
\end{theorem}
We provide proofs of part (a) of \Cref{theorem:anorm_pars} and bootstrap validity in \Cref{app:inference-proofs}, and we provide the proofs of parts (b)-(f) in Appendix \ref{app:additional-proofs} in the Supplementary Appendix.

\section{Intergenerational Income Mobility allowing for Measurement Error}
\label{sec:emp}

In this section, we use the methods developed in the previous sections to estimate a variety of distributional effect parameters in an application about intergenerational mobility, allowing for measurement error in both parents' permanent income and child's permanent income, and assess how much allowing for measurement error affects the resulting estimates.

\subsection*{Data}

The data that we use comes from the 1997 National Longitudinal Survey of Youth (NLSY97). The NLSY97 represents the 1997 cohort of the National Longitudinal Survey that contains detailed panel data on individuals who were 12-16 years old at the start of 1997.  Following much work on intergenerational income mobility, we focus on a subset of the data involving fathers and sons. We use the logarithm of son's total labor income in 2014 (which is self-reported in the 2015 survey wave) as the outcome variable. In 2014, survey respondents were between 29 and 33 years old.\footnote{This is essentially right at the start of the age range where the literature has generally observed that annual incomes are not systematically different from permanent incomes (\citet{haider-solon-2006,nybom-stuhler-2016}).  Thus, for the results below, we do not make any adjustments related to life-cycle measurement error as discussed in \Cref{app:life-cycle-measurement-error} in the Supplementary Appendix. \label{fn:application-life-cycle-measurement-error}} We use the logarithm of father's total labor income in 1996 (in 2014 dollars) as the treatment variable.  We drop all father-son pairs in which either the father or the son has zero or missing earnings.

The main advantages of using the NLSY97 relative to other datasets, such as the Panel Study of Income Dynamics (PSID), are (i) the NLSY97 sample sizes tend to be larger, and there are more individuals who are of a similar age, and (ii) the NLSY97 tends to have more individual-level covariates that can be included in our analysis.  The main complication is that we do not observe permanent incomes.  For fathers, we only observe a single year of income from the first year of the survey.  For sons, we only use income reported for the year 2014 and discard information about income in subsequent years.  Having one observation of father's and son's incomes fits into the methodological framework considered in the paper.

\renewcommand{\arraystretch}{.7}
\setlength{\tabcolsep}{15pt}
\newcolumntype{.}{D{.}{.}{-1}}
\ctable[caption={Summary Statistics},label=tab:ss,pos=!tbp,]{lcccc}{\tnote[]{\vspace{1pt} \textit{Notes:}  The table provides summary statistics by father's income quartiles in each column.  Son's income and father's income are reported in thousands of dollars.  Standard deviations of each variable are reported in parentheses.

    \vspace{2pt} \textit{Sources:}  NLSY97, as described in text.}}{\FL
\multicolumn{1}{l}{}&\multicolumn{1}{c}{Q1}&\multicolumn{1}{c}{Q2}&\multicolumn{1}{c}{Q3}&\multicolumn{1}{c}{Q4}\ML
Son's Income&46.17&50.82&59.18&66.34\NN
&(33.77)&(34.91)&(42.2)&(46.89)\NN[5pt]
Father's Income&19.46&43.2&64.22&112.28\NN
&(9.64)&(5.66)&(6.48)&(51.19)\NN[5pt]
\% White&0.53&0.67&0.74&0.84\NN
&(0.50)&(0.47)&(0.44)&(0.37)\NN[5pt]
\% Black&0.14&0.13&0.13&0.06\NN
&(0.35)&(0.34)&(0.33)&(0.24)\NN[5pt]
\% Hispanic&0.32&0.19&0.12&0.09\NN
&(0.47)&(0.40)&(0.33)&(0.28)\NN[5pt]
Son's Age 1997&14.15&14.16&14.27&14.27\NN
&(1.43)&(1.48)&(1.51)&(1.47)\NN[5pt]
Father's Age 1997&40.61&40.91&41.97&42.79\NN
&(7.30)&(5.87)&(5.44)&(4.98)\NN[5pt]
Father's Educ.&11.38&13.03&13.92&15.32\NN
&(3.45)&(2.46)&(2.45)&(2.76)\LL
}
\renewcommand{\arraystretch}{1}
\setlength{\tabcolsep}{6pt}

Besides the income variables for fathers and sons, we also observe race, ethnicity, the ages of both fathers and sons, and the father's education (in years). We drop observations that are missing values for any of these. This results in a dataset with 1,096 observations. Summary statistics for our dataset are available in \Cref{tab:ss}. Each column provides average values of each variable by quartile of father's income. Like almost all work on intergenerational mobility, we see that son's income is increasing in father's income and that sons from low-income families tend to have higher incomes than their fathers, while sons from high-income families tend to have lower incomes than their fathers. The estimated intergenerational elasticity (IGE) is 0.054 in our data. This is broadly similar to early studies of intergenerational income mobility that ignored measurement error in income variables. However, the estimated value of the IGE using our data is small relative to more recent IGE estimates that account for measurement error in income in a linear regression framework.
In terms of other covariates, the percentage of sons that are white, father's age, and father's education are all increasing with father's income. The percentage of sons who are black and the percentage of sons who are Hispanic both decrease with fathers' income. The son's age in 1997 does not appear to be correlated with father's income.

\subsection*{Implementation Details}

In this section, the only covariates we include are son's age and father's age; this specification is common in the intergenerational mobility literature.
In \Cref{app:additional-results} in the Supplementary Appendix, we provide additional results that condition on more covariates.  For our first step quantile regression estimates (that allow for measurement error), we estimate $\beta_j(\tau)$ over $L=25$ equally spaced values of $\tau$ from 0.02 to 0.98.  We initialize \Cref{alg:Est} by setting $\widehat{\beta}_{j}^{(0)}(\tau)$ equal to the estimates of the QR parameters ignoring measurement error.  We specify the measurement error for both son's permanent income and father's permanent income as a mixture of two normal distributions.    We initialize the mixture parameters for equation $j$ in a data-driven way.  We set the starting mixture probabilities to be $1/M$ each, with $M$ denoting the number of mixture components;  component means to be equally-spaced values in $\big[-\tfrac{1}{4}\hat{\sigma}_j, \tfrac{1}{4}\hat{\sigma}_j\big]$ centered at zero, where $\hat{\sigma}_j$ is the sample standard deviation of the dependent variable in equation $j$; and component standard deviations so that the mixture variance equals $\hat{\sigma}_j^2/4$.  Across alternative trial starting values and with multiple restarts, we found that this initialization led to reliable convergence of the algorithm to the global optimum.  At each iteration, given estimates of $\left(\widehat{\beta}^{(l-1)}_j(\cdot), \hat{\sigma}^{(l-1)}_j\right)$ from the previous iteration, for each observation, we run a Metropolis–Hastings algorithm for 400 steps, discarding the first 20 as burn-in, and use the remaining ones as measurement error draws. The proposal distribution is a Gaussian random walk whose standard deviation is initialized to the standard deviation of the measurement error mixture implied by the starting parameters and updated after each M-step to track the current measurement error scale.  As discussed in \Cref{alg:Est}, we update the QR parameters using quantile regression accounting for the measurement error draws.  We update the parameters for the measurement error itself by fitting a two-component mixture of normals on the measurement error draws. For this step, we use the R \texttt{mixtools} package (\citet{benaglia2010mixtools}), which implements a standard (and relatively computationally cheap) EM-algorithm for estimating mixture models (see also \citet{mclachlan-lee-rathnayake-2019}). Our first step algorithm terminates when the maximum scaled parameter change, $\max_k |\hat{\theta}_k^{(l)} - \hat{\theta}_k^{(l-1)}| / (1 + |\hat{\theta}_k^{(l-1)}|)$, falls below $0.01$, where the maximum is taken over all QR coefficients and measurement error parameters.  For estimating the conditional copula in the second step of our estimation procedure, we specify that the conditional copula is a Gaussian copula and estimate its parameter.  This specification performed well in terms of AIC and BIC relative to a candidate set of copula families and measurement error distributions; in Supplementary Appendix \ref{app:additional-results}, we discuss model selection in more detail and provide analogous results using a Frank copula and Laplace measurement error instead.  To estimate the copula parameter, we use 1000 simulated measurement error draws in our simulated maximum likelihood procedure.  We report standard errors and confidence intervals below that come from using the empirical bootstrap with 200 iterations, where each iteration draws a sample of size $n$ with replacement from the original data and re-runs the full three-step estimation procedure on the resampled dataset.

\subsection*{Copula-type Parameters}

To start with, we consider copula-type parameters: rank-rank correlations, transition matrices, and upward mobility measures.  Although our estimation approach involves estimating quantile regressions conditional on covariates (and accounting for measurement error), the copula-type parameters that we consider are unconditional and, therefore, can be directly compared to rank-rank correlations, transition matrices, or upward mobility measures computed directly from the observed data.  We report both of these types of results in this section, which provides a straightforward way to compare our estimates that allow for measurement error with estimates that ignore measurement error.

We start with rank-rank correlations.  Using the raw data (and ignoring measurement error), our estimate of the rank-rank correlation is 0.209 (s.e.=0.031).  By contrast, our estimate of the rank-rank correlation that allows for measurement error is 0.353 (s.e.=0.062).  The difference between these estimates is large---our estimates that adjust for measurement error suggest substantially less intergenerational mobility (due to the rank-rank correlations being higher) than is implied by the estimates that use the observed data directly.\footnote{Among the distributional effect parameters that we consider, the effect of measurement error on rank-rank correlations has been considered in more depth in the literature relative to other parameters.  First, using administrative data from Sweden with a long earnings history, \citet{nybom-stuhler-2017} provide results (see, in particular, Panel c of their Figure 1) comparing estimates of intergenerational rank-rank correlations for children at different ages to rank-rank correlations using the entire earnings history (and thus accounting for measurement error).  At age 30, which is comparable to our estimates reported above that ignore measurement error, they estimate a rank-rank correlation of about 0.18.  When they use the full earnings history, which is infeasible for us but comparable to our estimates that account for measurement error, they estimate a rank-rank correlation of about 0.26.  This is fairly similar to our estimate, perhaps reflecting somewhat lower intergenerational mobility in the United States than in Sweden.  Second, \citet{chetty-hendren-kline-saez-2014} use administrative data from the United States for parent-child pairs that are 29-32 in 2011 and 2012, which is very similar to the ages and years that we consider.  Their main rank-rank correlation estimate is 0.341, which is quite similar to ours.  \label{fn:empirical-rank-rank-correlations} %
}

\begin{table}[t]
  \centering
  \caption{Transition Matrix}
  \label{tab:tmat}
  \begin{subtable}[t]{.45 \linewidth}
  \begin{tabular}[t]{rrrrrr}
    \toprule
    \multicolumn{1}{c}{} & \multicolumn{1}{c}{} & \multicolumn{4}{c}{Father's Income Quartile} \\
    \cmidrule(l{3pt}r{3pt}){3-6}
    &  & 1 & 2 & 3 & 4\\
    \midrule
    & 4 & 0.113 & 0.190 & 0.272 & 0.425\\

    &  & (0.022) & (0.012) & (0.005) & (0.031)\\

    & 3 & 0.201 & 0.256 & 0.276 & 0.268\\

    &  & (0.012) & (0.003) & (0.008) & (0.006)\\

    & 2 & 0.279 & 0.276 & 0.255 & 0.190\\

    &  & (0.005) & (0.008) & (0.003) & (0.013)\\

    & 1 & 0.408 & 0.278 & 0.196 & 0.117\\

    \multirow{-8}{*}{\rotatebox[origin=c]{90}{Son's Income Quartile}} &  & (0.031) & (0.005) & (0.012) & (0.022)\\
    \bottomrule
  \end{tabular}
  \caption{Measurement Error}
  \end{subtable}
  \begin{subtable}[t]{.45\linewidth}
    \begin{tabular}[t]{rrrrrr}
      \toprule
      \multicolumn{1}{c}{} & \multicolumn{1}{c}{} & \multicolumn{4}{c}{Father's Income Quartile} \\
      \cmidrule(l{3pt}r{3pt}){3-6}
      &  & 1 & 2 & 3 & 4\\
      \midrule
      & 4 & 0.161 & 0.234 & 0.256 & 0.376\\

      &  & (0.023) & (0.025) & (0.030) & (0.029)\\

      & 3 & 0.197 & 0.215 & 0.303 & 0.266\\

      &  & (0.026) & (0.027) & (0.028) & (0.029)\\

      & 2 & 0.329 & 0.230 & 0.314 & 0.175\\

      &  & (0.030) & (0.032) & (0.037) & (0.026)\\

      & 1 & 0.303 & 0.277 & 0.175 & 0.190\\

      \multirow{-8}{*}{\rotatebox[origin=c]{90}{Son's Income Quartile}} &  & (0.028) & (0.030) & (0.025) & (0.024)\\
      \bottomrule
    \end{tabular}
    \caption{Observed}
  \end{subtable}
  \begin{justify}
    {\footnotesize \noindent \textit{Notes:}  The table provides estimates of transition matrices either allowing for measurement error using the techniques developed in the paper (Panel (a)) or coming directly from the observed data (Panel (b)).  The columns are organized by quartiles of father's income; e.g., columns labeled ``1'' use data from fathers whose income is in the first quartile. Some columns in Panel (b) do not sum to exactly one due to ties in income data.  Similarly, rows are organized by quartiles of son's income.  Standard errors are computed using the bootstrap.

    \noindent \textit{Sources:}  NLSY97, as described in text. }
  \end{justify}
\end{table}

The results on transition matrices are presented in \Cref{tab:tmat}.  First, ignoring measurement error (as in Panel (b)), the estimated transition matrix indicates a large degree of intergenerational income mobility.  While sons generally are somewhat more likely to stay in the same income quartile as their father, large movements in the income distribution do not appear to be uncommon.  For example, for a son whose father was in the lowest quartile of the income distribution, our estimate shows that the probability that the son is in the lowest quartile of the income distribution is 30\%, but that the probability that the son moves to the top quartile of the income distribution is 16\%.  Similarly, for fathers in the top quartile of the income distribution, we estimate the probability that their son is in the top quartile of the income distribution to be 38\% and that the probability that their son is in the bottom quartile of the income distribution to be 19\%.  Allowing for measurement error (as in Panel (a)) indicates substantially less intergenerational mobility.  For example, for fathers in the lowest quartile of the permanent income distribution, we estimate that 41\% of their sons stay in the lowest quartile of the permanent income distribution and only 11\% move to the top quartile of the permanent income distribution.  For fathers in the top quartile of the permanent income distribution, we estimate that 43\% of their sons stay in the top quartile while only 12\% move to the bottom quartile.

\setlength{\tabcolsep}{10pt}
\ctable[pos=t,caption={Upward Mobility},label=tab:um]{lllll}{\tnote[]{\textit{Notes:}  The table provides estimates of upward mobility allowing for measurement error using the techniques developed in the paper (top panel) or coming directly from the observed data (bottom panel).  The columns are organized by quartiles of father's income; e.g., columns labeled ``1'' use data from fathers whose income is in the first quartile.  Standard errors are computed using the bootstrap.

    \textit{Sources:}  NLSY97, as described in text.}}{\FL
\multicolumn{1}{l}{\bfseries }&\multicolumn{4}{c}{Father's Income Quartile}\NN
\cline{2-5}
\multicolumn{1}{l}{}&\multicolumn{1}{c}{1}&\multicolumn{1}{c}{2}&\multicolumn{1}{c}{3}&\multicolumn{1}{c}{4}\ML
{Measurement Error}&&&&\NN
~~&0.792&0.586&0.409&0.202\NN
~~&(0.016)&(0.007)&(0.008)&(0.016)\ML
{Observed}&&&&\NN
~~&0.830&0.573&0.387&0.167\NN
~~&(0.009)&(0.005)&(0.005)&(0.009)\LL
}
\setlength{\tabcolsep}{6pt}

Finally, we present upward mobility estimates in \Cref{tab:um}.  These are estimates of the fraction of sons whose rank in the income distribution exceeds the rank of their fathers.  We present these results by quartile of father's income as well as separately by whether we use our approach that allows for measurement error or just use the observed data directly.  Compared to using the observed data directly, the upward mobility estimates allowing for measurement error are lower when the father is in the first quartile of the permanent income distribution and higher when the father is in the second, third, or fourth quartile of the permanent income distribution.  As for the other parameters in this section, these results suggest that allowing for measurement error reduces estimates of intergenerational income mobility.

\subsection*{Conditional Distribution-type Results}

The next set of results that we present corresponds to the Conditional Distribution-type parameters.  %
In this section, we primarily compare estimates using our approach that allows for measurement error to estimates of conditional distribution-type parameters that ignore measurement error and come directly from quantile regression estimates of son's income on father's income, son's age, and father's age.

\begin{figure}[t]
  \caption{Quantiles of Son's Income as a Function of Father's Income}
  \label{fig:qytx}
  \begin{subfigure}[t]{.5\textwidth}
    \includegraphics[width=\textwidth]{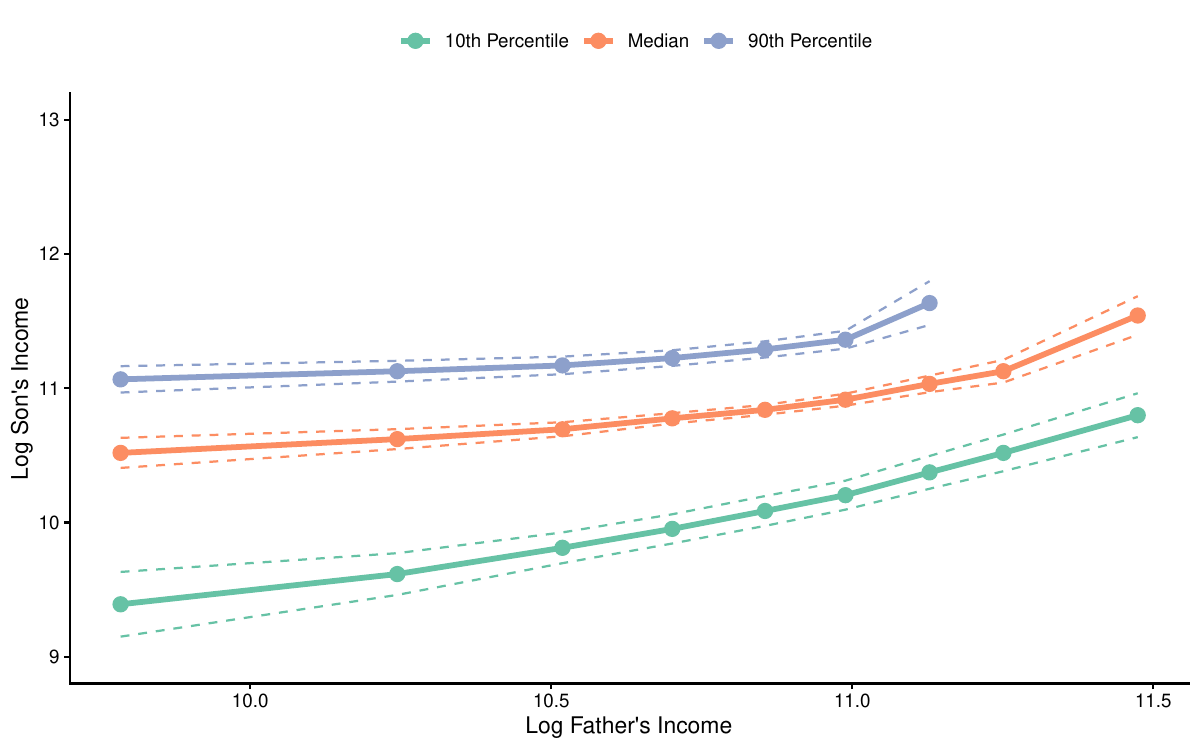}
    \caption{Measurement Error}
  \end{subfigure}
  \begin{subfigure}[t]{.5\textwidth}
    \includegraphics[width=\textwidth]{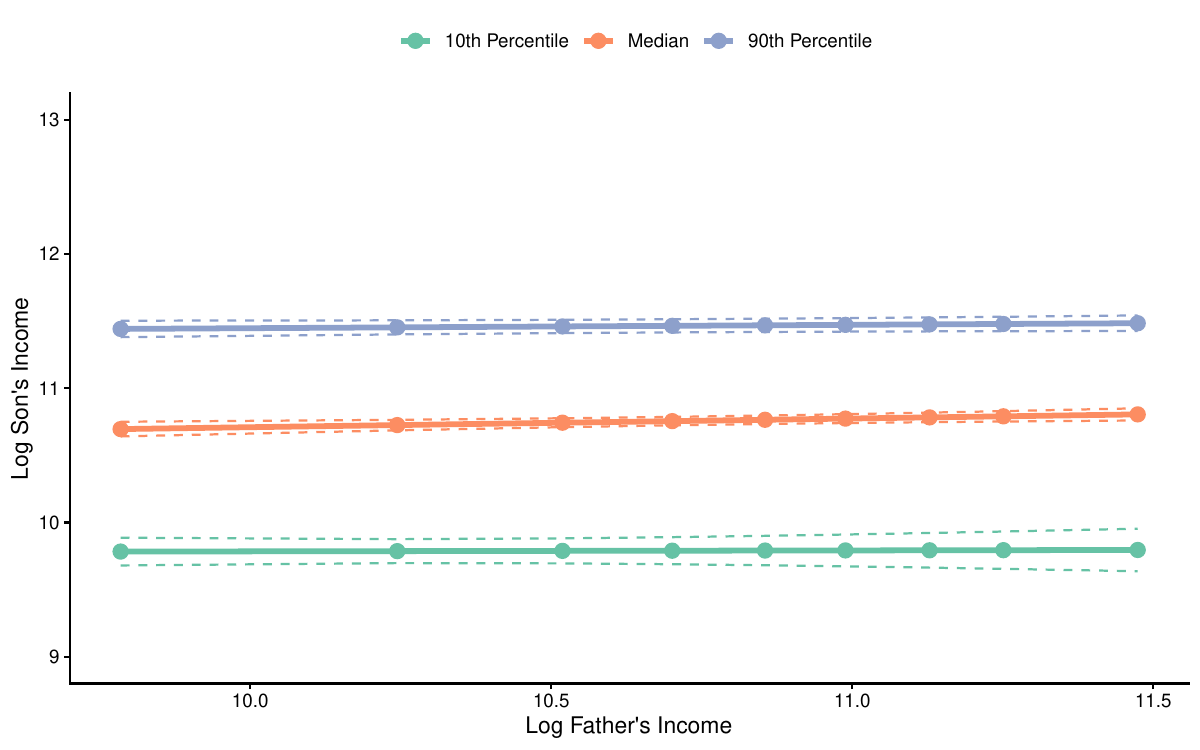}
    \caption{QR No Measurement Error}
  \end{subfigure}
  \begin{justify}
    { \footnotesize \textit{Notes:} The figure provides estimates of quantiles of son's income as a function of father's income and conditional on son's age and father's age being equal to their averages in the sample.  The estimates in Panel (a) come from the approach suggested in the current paper that allows for measurement error.  The estimates in Panel (b) ignore measurement error and come from quantile regression of observed son's income on observed father's income and covariates.  The three lines in each panel are estimates of the 10th percentile, median, and 90th percentile of son's income, which are estimated at the 10th, 20th, \ldots, and 90th percentiles of observed father's income.  In Panel (a), the 90th percentile is not shown at the 80th and 90th percentiles of father's income; at those values, computing the ME-corrected 90th percentile requires evaluating the conditional quantile function of $Y^*$ given $X$ at quantile levels above 0.98, where extrapolation beyond the estimated range is unreliable. Standard errors are computed using the bootstrap.}
  \end{justify}
\end{figure}

\Cref{fig:qytx} contains estimates of conditional quantiles (the 10th percentile, median, and 90th percentile) of son's income as a function of father's income, where son's age and father's age (both in 1997) are set at their average values in our sample (14 and 42, respectively).\footnote{We estimate these conditional quantiles by inverting the sample analog of \Cref{eqn:Fc}, which ensures monotonicity of the conditional quantiles.\label{fn:monotonic-conditional-quantiles}}  Estimates of these conditional quantiles are reported at the 10th, 20th, \ldots, and 90th percentiles of observed father's income.  These are unequally spaced values of father's income, but they range from about \$18,000 to just over \$96,000.  Panel (a) contains estimates using our approach that allows for measurement error, and Panel (b) contains estimates coming from quantile regression of son's income on father's income and covariates ignoring measurement error.

We start by discussing the estimates that ignore measurement error in Panel (b).  By construction, the quantile regression estimates in Panel (b) are linear.  Ignoring measurement error, the slopes of the quantile regression estimates are very flat, indicating little effect of father's income on son's income.  For interpreting the results, we mainly discuss results conditional on father's income being in the 10th percentile (which is roughly equal to the poverty line) and in the 90th percentile.  Using quantile regression and ignoring measurement error, the 10th percentile of son's income conditional on having father's income at the 10th percentile (and conditional on having average values for father's age and son's age) is estimated to be \$17,700, the median is estimated to be \$44,200, and the 90th percentile is estimated to be \$93,200.  For sons whose father is in the 90th percentile of the income distribution, the distribution of their income is very similar---we estimate that the 10th percentile (conditional on having average values for father's age and son's age) is \$17,900, the median is \$49,300, and the 90th percentile is \$97,200.\footnote{For $\tau \in \{0.1,0.5,0.9\}$, from our quantile regression estimates of the log of son's income on the log of father's income and father's and son's age,  the estimated effect of father's income is small (ranging from 0.005 to 0.067) but positive across all values of $\tau$.  The estimated effect is only statistically significant for $\tau=0.5$.  This immediately translates to the small estimated effects of father's income in Panel (b) of \Cref{fig:qytx}.}

Moving to results that allow for measurement error (Panel (a) of \Cref{fig:qytx}), there are substantial differences.  The slopes of all three quantile curves are steeper, indicating a much stronger relationship between father's and son's permanent income than is apparent from using quantile regression on the raw data.  For sons whose father's permanent income is at the 10th percentile, correcting for measurement error reduces estimated income across the distribution: the 10th percentile falls from \$17,700 to \$12,000, the median falls from \$44,200 to \$37,000, and the 90th percentile falls from \$93,200 to \$64,000.  There are also large differences at the high end of father's income.  For sons whose father's permanent income is at the 90th percentile, the 10th percentile of their income distribution rises from \$17,900 to \$49,100, and the median rises from \$49,300 to  \$103,000, more than double the estimate that ignores measurement error.

\begin{figure}[t]
  \caption{Poverty Rate}
  \begin{subfigure}[t]{.5\textwidth}
    \includegraphics[width=\textwidth]{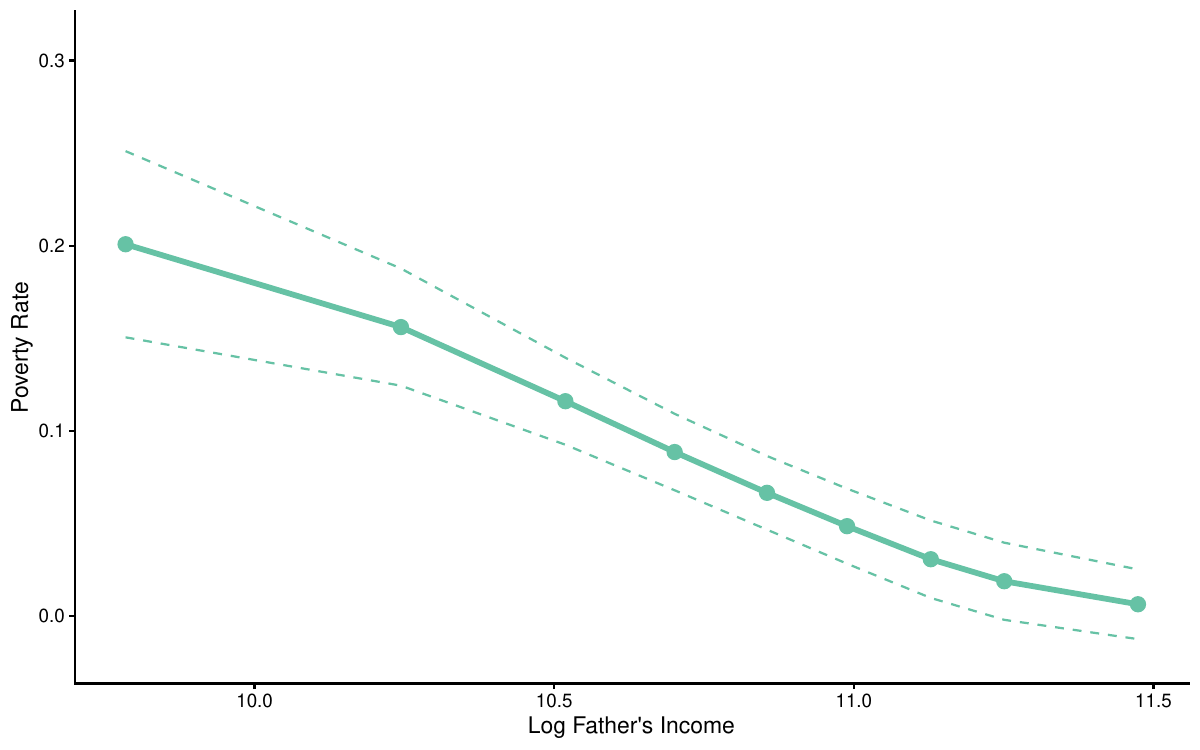}
    \caption{Measurement Error}
  \end{subfigure}
  \begin{subfigure}[t]{.5\textwidth}
    \includegraphics[width=\textwidth]{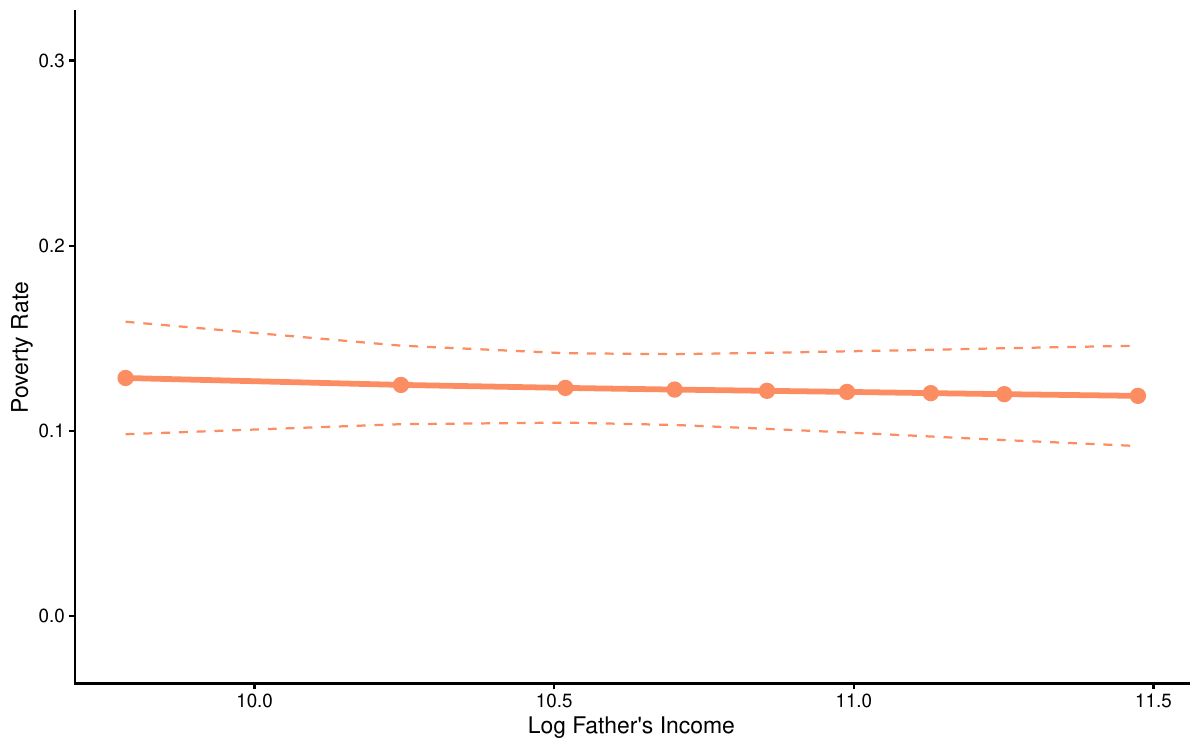}
    \caption{QR No Measurement Error}
  \end{subfigure}
  \begin{justify}
    { \footnotesize \textit{Notes:} The figure provides estimates of son's poverty rate as a function of father's income and conditional on son's age and father's age being equal to their averages in the sample.  The estimates in Panel (a) come from the approach suggested in the current paper that allows for measurement error.  The estimates in Panel (b) ignore measurement error and come from inverting quantile regressions of observed son's income on observed father's income and covariates.  The poverty line is set at (the logarithm of) \$20,000.  Standard errors are computed using the bootstrap.}
  \end{justify}
\end{figure}

Next, we consider estimates of the fraction of sons whose income is below the poverty line (which we set at \$20,000) as a function of father's income and conditional on father's and son's ages being set at their average in our sample.  Using quantile regression (and ignoring measurement error), we estimate that 12.9\% of sons whose father's income is at the 10th percentile have income that is below the poverty line.  For sons whose father's income was at the 90th percentile, we estimate that 11.9\% have income that is below the poverty line.  %
Allowing for measurement error again has a large effect and indicates substantially less intergenerational mobility.  For sons whose father's permanent income is at the 10th percentile, we estimate that 20.1\% have permanent income below the poverty line, but for sons whose father's permanent income is at the 90th percentile, we estimate that only 0.6\% have permanent income below the poverty line.

\bigskip

Our results in this section indicate that accounting for measurement error generally reduces estimates of various ways to summarize intergenerational income mobility, which is in line with the literature in many different contexts (e.g., \citet{solon-1992,mazumder-2005,haider-solon-2006,nybom-stuhler-2016,an-wang-xiao-2020}).  These results especially matter in the tails of the distribution where the sensitivity to measurement error is higher, which is also in line with the literature (e.g., \citet{bhattacharya-mazumder-2011,nybom-stuhler-2017}).

\section{Conclusion}
\label{sec:conc}

In this paper, we developed a new method to obtain the joint distribution of an outcome and a continuous treatment conditional on covariates when the outcome and the treatment are potentially measured with error. Our main innovation and departure from the existing literature was to model both the outcome and the treatment as a function of the covariates; that is, changing the problem from measurement error on the left and right to two cases of measurement error on the left. Then, we showed that the copula of the outcome and treatment was identified in this setup,  which, in turn, implies that a large number of distributional effect parameters of interest are identified.

We applied our approach to study intergenerational income mobility. We use this as a leading example for our approach because researchers studying intergenerational income mobility typically only have noisy measurements of an individual's permanent income.  In addition, we used recent data from the NLSY where only one observation of child's and parents' income was available, which means that many existing approaches to accounting for measurement error are not applicable.  We found that accounting for measurement error led to substantially lower estimates of a variety of measures of intergenerational mobility relative to implementing common alternative approaches while ignoring measurement error.

\FloatBarrier

\onehalfspace

{
\begingroup
\setstretch{.95}
\renewcommand*{\bibfont}{\small}
\setlength\bibitemsep{5pt}
\printbibliography
\endgroup
}

\appendix\crefalias{section}{appendix}\renewcommand{\theequation}{A\arabic{equation}}\setcounter{equation}{0}
\setcounter{assumption}{0}
\renewcommand{\theassumption}{A\arabic{assumption}}

\section{Proofs of Results in Section \ref{sec:3} } \label{app:proofs}

\subsubsection*{Proof of \Cref{prop:1}}

\begin{proof}
  The result holds immediately from \citet[Theorem 1]{hausman-liu-luo-palmer-2021}.
\end{proof}

\subsubsection*{Proof of \Cref{prop:2}}

\begin{proof}
  Let $\phi_X(t_1,t_2) = \E\big[\exp(it_1 Y + it_2 T)\mid X\big]$ denote the characteristic function of the observed $(Y,T)$ conditional on $X$.  Let $\phi_X^*(t_1,t_2) = \E\big[\exp(it_1Y^* + it_2T^*)\mid X\big]$ denote the characteristic function of $(Y^*,T^*)$ conditional on $X$ and let $\phi_{U_{Y^*}}$ and $\phi_{U_{T^*}}$ denote the characteristic functions of $U_{Y^*}$ and $U_{T^*}$, respectively.  Then,
  \begin{align*}
    \phi_X(t_1,t_2) = \phi_X^*(t_1,t_2)\phi_{U_{Y^*}}(t_1)\phi_{U_{T^*}}(t_2)
  \end{align*}
  which holds by \Cref{ass:me}.  Further, $\phi_{U_{Y^*}}$ and $\phi_{U_{T^*}}$ are identified due to the result in \Cref{prop:1}.  This implies
  \begin{align*}
    \phi_X^*(t_1,t_2) = \frac{\phi_X(t_1,t_2)}{\phi_{U_{Y^*}}(t_1)\phi_{U_{T^*}}(t_2)}
  \end{align*}
  which holds because the characteristic functions of $U_{Y^*}$ and $U_{T^*}$ are non-vanishing.  This proves that the joint distribution of $(Y^*,T^*)\mid X$ is identified, which implies that the copula is identified.
\end{proof}

\section{Proofs of Results in Section \ref{sec:inference} } \label{app:inference-proofs}

\subsection{Proof of Theorem \ref{theorem:consistency} }

\begin{proof}
Under the conditions of \Cref{prop:1}, namely \Cref{ass:me,ass:qr,ass:additional-qr} and \Cref{lem:UWLLN}, the proof of this part follows from a straightforward modification of existing results, e.g., \citet[Proof of Theorem 22.1]{hansen-2022}. The details are therefore omitted.
\end{proof}

\subsection{Proof of Theorem \ref{theorem:anorm_beta}}

\begin{proof}

\noindent \textbf{Part (a):}

The $ (k+1)\times 1 $ population first-order conditions with respect to \( \big(\beta_{Y^*}(\tau)',\,\sigma_{Y^*}\big)' \) are given by
\begin{align}
    \mathbb{E} \left[X \cdot \int_{\mathcal{U}} \psi_\tau\left( Y - u - X'\beta_{Y^*}(\tau) \right)
f_{U_{Y^*} | Y, X}(u \mid Y, X; \sigma_{Y^*}) \, du \right] &= 0 \label{eqn:FOC_beta} \\
\mathbb{E} \Big[\int_{\mathcal{U}}
\rho_\tau\left( Y - u - X'\beta_{Y^*}(\tau) \right) \cdot f_{U_{Y^*} | Y, X}^{\partial}(u \mid Y, X; \sigma_{Y^*})   du \Big] &= 0 \label{eqn:FOC_sig}
\end{align}where $ \displaystyle f_{U_{Y^*} | Y, X}^{\partial}(u \mid Y, X;\sigma_{Y^*}) := \frac{\partial }{\partial \sigma_{Y^*}}f_{U_{Y^*} | Y, X}(u \mid Y, X; \sigma_{Y^*}) $. \eqref{eqn:FOC_beta} can be alternatively expressed as
\begin{align}
0 =& \mathbb{E} \left[ X \int_{\mathcal{U}} \psi_\tau\left( Y - u - X'\beta_{Y^*}(\tau) \right) f_{U_{Y^*} | Y, X}(u \mid Y, X) \, du \right] \nonumber \\
=& (\tau-1)\mathbb{E}[X] +  \E\left[ X\int_{\mathcal{U}}\indicator{u\leq Y - X'\beta_{Y^*}(\tau)} f_{U_{Y^*} | Y, X}(u \mid Y, X) \, du \right] \nonumber \\
=& (\tau - 1)\mathbb{E}[X] + \mathbb{E} \left[ X \cdot \F_{U_{Y^*} \mid Y,X}\left(Y - X'\beta_{Y^*}(\tau) \mid Y, X\right) \right].
\tag{\ref{eqn:FOC_beta}'}
\end{align}

The following decomposition of the sample moment condition corresponding to \eqref{eqn:FOC_beta}, for $\tau \in \mathcal{T} $ satisfies

\begingroup
\small

\begin{align}
    & o_p(n^{-1/2}) = \frac{1}{nS} \sum_{i=1}^n \sum_{s=1}^S \left[ X_i \psi_\tau\left( Y_i - U_{is} - X_i'\widehat{\beta}_{Y^*}(\tau) \right) \right] \label{eqn:sample-qr-score}\\
    &=: \frac{1}{n} \sum_{i=1}^n X_i \int_{\mathcal{U}} \psi_\tau\left( Y_i - u - X_i'\widehat{\beta}_{Y^*}(\tau) \right)  \widehat{f}_{U_{Y^*} | Y, X}^{(S)}(u \mid Y_i, X_i; \widehat{\sigma}_{Y^*} ) \, du \nonumber \\
    &= \frac{1}{n} \sum_{i=1}^n X_i \underbrace{\Big\{  \int_{\mathcal{U}} \left[  \psi_\tau\left( Y_i - u - X_i'\widehat{\beta}_{Y^*}(\tau) \right) \left(\widehat{f}_{U_{Y^*} | Y, X}^{(S)}(u \mid Y_i, X_i; \widehat{\sigma}_{Y^*} )  - \widehat{f}_{U_{Y^*} | Y, X}(u \mid Y_i, X_i; \widehat{\sigma}_{Y^*} ) \right)  \right] \, du \Big\} }_{R_{i,MC}=O_p(S^{-1/2})} \nonumber \\
    &+ \underbrace{\frac{1}{n} \sum_{i=1}^n X_i \int_{\mathcal{U}} \left[ \psi_\tau\left( Y_i - u - X_i'\widehat{\beta}_{Y^*}(\tau) \right) \left( \widehat{f}_{U_{Y^*} | Y, X}(u \mid Y_i, X_i; \widehat{\sigma}_{Y^*} ) - \widehat{f}_{U_{Y^*} | Y, X}(u \mid Y_i, X_i; \sigma_{Y^*} ) \right) \, du \right]}_{B_{1n}} \nonumber \\
    &+ \underbrace{\frac{1}{n} \sum_{i=1}^n X_i \int_{\mathcal{U}} \left[ \psi_\tau\big( Y_i - u - X_i'\widehat{\beta}_{Y^*}(\tau) \big) \underbrace{\Big(\widehat{f}_{U_{Y^*} | Y, X}(u \mid Y_i, X_i; \sigma_{Y^*} ) - f_{U_{Y^*} | Y, X}(u \mid Y_i, X_i; \sigma_{Y^*} ) \Big) }_{R_{i,MC}} \, du \right]}_{B_{2n}} \nonumber \\
    &+ \underbrace{\frac{1}{n} \sum_{i=1}^n X_i \int_{\mathcal{U}} \left[ \psi_\tau\big( Y_i - u - X_i'\widehat{\beta}_{Y^*}(\tau) \big) f_{U_{Y^*} | Y, X}(u \mid Y_i, X_i; \sigma_{Y^*} ) \, du \right]}_{B_{3n}} \nonumber
\end{align}

\endgroup \noindent where $ \displaystyle \widehat{f}_{U_{Y^*} | Y, X}(u \mid Y_i, X_i; \widehat{\sigma}_{Y^*} ) := \frac{\left[ X_i'\widehat{\beta}_{Y^*}^{\partial}\left(\widehat{\tau}_y(Y_i-u)\right) \right]^{-1}}{\int_{\mathcal{U}} f_{Y \mid U_{Y^*}, X}(Y_i \mid \tilde{u}, X_i)\,
   f_{U_{Y^*} \mid X}(\tilde{u} ; \widehat{\sigma}_{Y^*})\, d\tilde{u}} f_{U_{Y^*}}(u ; \widehat{\sigma}_{Y^*}) $ and $\displaystyle \widehat{\tau}_y(Y_i-u) := \int_{0}^{1} \indicator{ X_i'\widehat{\beta}_{Y^*}(\tilde{\tau}) \leq (Y_i-u) } \, d\tilde{\tau} $ thanks to \Cref{ass:splines}.  The normalizing property of the MCMC algorithm ensures $ \widehat{f}_{U_{Y^*} | Y, X}(u \mid Y_i, X_i; \sigma ) $ is a proper pdf, i.e., integrates to one, uniformly in $ \Gamma_{\sigma} $.

Thanks to \Cref{ass:MCMC_conv} and that $| \psi_\tau(\cdot)| \leq 1 $, the Monte Carlo approximation error $ R_{i,MC} = O_p(S^{-1/2}) = o_p(n^{-1/2}) $ for each $i \in \{1,\ldots,n \} $ since $n/S=o(1)$ following the argument used in the proof of \Cref{lem:UWLLN} analogously.

Thanks to the differentiability of $f_{U_{Y^*}}(u ; \sigma)$ with respect to $\sigma$ (\Cref{ass:dominance_Xf}(e)), it follows from the Mean-Value Theorem (MVT) that
\begin{align*}
    \widehat{f}_{U_{Y^*} | Y, X}(u \mid Y_i, X_i; \widehat{\sigma}_{Y^*} ) - \widehat{f}_{U_{Y^*} | Y, X}(u \mid Y_i, X_i; \sigma_{Y^*} ) = \widehat{f}_{U_{Y^*} | Y, X}^{\partial}(u \mid Y_i, X_i; \bar{\sigma}_{Y^*} ) (\widehat{\sigma}_{Y^*} - \sigma_{Y^*})
\end{align*}for some $ \bar{\sigma}_{Y^*} $ that satisfies $|\bar{\sigma}_{Y^*} - \sigma_{Y^*}|\leq |\widehat{\sigma}_{Y^*} - \sigma_{Y^*}| $. It follows from the above that
\begin{align*}
    B_{1n} = \frac{1}{n} \sum_{i=1}^n X_i \int_{\mathcal{U}} \left[ \psi_\tau\left( Y_i - u - X_i'\widehat{\beta}_{Y^*}(\tau) \right) \widehat{f}_{U_{Y^*} | Y, X}^{\partial}(u \mid Y_i, X_i; \bar{\sigma}_{Y^*} ) \, du\right](\widehat{\sigma}_{Y^*} - \sigma_{Y^*}).
\end{align*}

From \Cref{lem:R_AB}, the Monte Carlo approximation error has the following representation:
\begin{align*}
    R_{i,MC}: &= \big( \widehat{f}_{U_{Y^*} | Y, X}(u \mid Y_i, X_i; \sigma_{Y^*} ) - f_{U_{Y^*} | Y, X}(u \mid Y_i, X_i; \sigma_{Y^*} ) \big)\\
    &= \sum_{\ell=1}^{L}\widehat{\mathbb{R}}_{\ell,f_{U_{Y^*} | Y, X}}(u,Y_i,X_i;\sigma_{Y^*})'\,
\big(\widehat{\beta}_{Y^*}(\tau_{\ell})-\beta_{Y^*}(\tau_{\ell})\big).
\end{align*}Thus,
\begin{align*}
    B_{2n} =& \sum_{\ell = 1}^{L} \underbrace{\frac{1}{n} \sum_{i=1}^n\Big\{ \int_{\mathcal{U}} \left[ \psi_\tau\big( Y_i - u - X_i'\widehat{\beta}_{Y^*}(\tau) \big)  X_i\widehat{\mathbb{R}}_{\ell,f_{U_{Y^*} | Y, X}}(u,Y_i,X_i;\sigma_{Y^*})'\, du \right] \Big\} }_{ \mathbb{R}_{\beta,\ell,n}(\tau) } \big(\widehat{\beta}_{Y^*}(\tau_{\ell})-\beta_{Y^*}(\tau_{\ell})\big)\\
    =&: \sum_{\ell = 1}^{L}\mathbb{R}_{\beta,\ell,n}(\tau)\big( \widehat{\beta}_{Y^*}(\tau_\ell) - \beta_{Y^*}(\tau_\ell) \big).
\end{align*}

In addition to \Cref{ass:splines}, $ f_{U_{Y^*} | Y, X}(u \mid Y_i, X_i; \sigma ) $ is continuous in $u$, thus the resulting CDF $ \F_{U_{Y^*} | Y, X}(u\mid Y_i, X_i; \sigma_{Y^*}) $ is continuously differentiable in $u$. By the MVT, $ \F_{U_{Y^*} | Y, X}(Y_i - X_i'\widehat{\beta}_{Y^*}(\tau) \mid Y_i, X_i) = \F_{U_{Y^*} | Y, X}(Y_i - X_i'\beta_{Y^*}(\tau) \mid Y_i, X_i) - f_{U_{Y^*} | Y, X}(Y_i - X_i'\bar{\beta}_{Y^*}(\tau) \mid Y_i, X_i)X_i'\big(\widehat{\beta}_{Y^*}(\tau) - \beta_{Y^*}(\tau)\big) $. It follows from the representation in (\ref{eqn:FOC_beta}') that
\begin{align*}
    B_{3n}&=\frac{1}{n} \sum_{i=1}^n X_i \int_{\mathcal{U}} \left[ \psi_\tau\left( Y_i - u - X_i'\widehat{\beta}_{Y^*}(\tau) \right) f_{U_{Y^*} | Y, X}(u \mid Y_i, X_i; \sigma_{Y^*} ) \, du\right] \\
    &= (\tau-1)\frac{1}{n} \sum_{i=1}^n X_i + \frac{1}{n} \sum_{i=1}^n X_i  \int_{\mathcal{U}} \left[ \indicator{u \leq Y_i - X_i'\widehat{\beta}_{Y^*}(\tau) } f_{U_{Y^*} | Y, X}(u \mid Y_i, X_i; \sigma_{Y^*} ) \, du\right]\\
    &= (\tau-1)\frac{1}{n} \sum_{i=1}^n X_i + \frac{1}{n} \sum_{i=1}^n X_i \F_{U_{Y^*} | Y, X}(Y_i - X_i'\widehat{\beta}_{Y^*}(\tau) \mid Y_i, X_i)\\
    &= (\tau-1)\frac{1}{n} \sum_{i=1}^n X_i + \frac{1}{n} \sum_{i=1}^n X_i \F_{U_{Y^*} | Y, X}(Y_i - X_i'\beta_{Y^*}(\tau) \mid Y_i, X_i)\\
    &- \Big\{\frac{1}{n} \sum_{i=1}^n X_iX_i' f_{U_{Y^*} | Y, X}(Y_i - X_i'\bar{\beta}_{Y^*}(\tau) \mid Y_i, X_i)\Big\}\big(\widehat{\beta}_{Y^*}(\tau) - \beta_{Y^*}(\tau)\big).
\end{align*}

Putting terms together,
\begin{equation}\label{eqn:score_beta}
\begin{split}
    &o_p(n^{-1/2}) \\
    =& \underbrace{- \Big\{\frac{1}{n} \sum_{i=1}^n X_i \widehat{f}_{U_{Y^*} | Y, X}(Y_i - X_i'\bar{\beta}_{Y^*}(\tau) \mid Y_i, X_i) X_i' \Big\}}_{\mathbb{A}_{\beta,n}(\tau)}\big(\widehat{\beta}_{Y^*}(\tau) - \beta_{Y^*}(\tau)\big)  \\
    & + \sum_{\ell = 1}^{L}\mathbb{R}_{\beta,\ell,n}(\tau)\big( \widehat{\beta}_{Y^*}(\tau_\ell) - \beta_{Y^*}(\tau_\ell) \big) \\
    & + \underbrace{\frac{1}{n} \sum_{i=1}^n X_i \int_{\mathcal{U}} \left[ \psi_\tau\left( Y_i - u - X_i'\widehat{\beta}_{Y^*}(\tau) \right) \widehat{f}_{U_{Y^*} | Y, X}^{\partial}(u \mid Y_i, X_i; \bar{\sigma}_{Y^*} ) \, du\right]}_{\mathbb{B}_{\beta,n}(\tau)}(\widehat{\sigma}_{Y^*} - \sigma_{Y^*})  \\
    &+ \underbrace{(\tau-1)\frac{1}{n} \sum_{i=1}^n X_i + \frac{1}{n} \sum_{i=1}^n X_i \F_{U_{Y^*} | Y, X}(Y_i - X_i'\beta_{Y^*}(\tau) \mid Y_i, X_i)}_{\mathbb{C}_{\beta,n}(\tau)} \\
    =&: \mathbb{A}_{\beta,n}(\tau)\big(\widehat{\beta}_{Y^*}(\tau) - \beta_{Y^*}(\tau)\big) + \sum_{\ell = 1}^{L}\mathbb{R}_{\beta,\ell,n}(\tau)\big( \widehat{\beta}_{Y^*}(\tau_\ell) - \beta_{Y^*}(\tau_\ell) \big)+ \mathbb{B}_{\beta,n}(\tau)(\widehat{\sigma}_{Y^*} - \sigma_{Y^*}) + \mathbb{C}_{\beta,n}(\tau).
\end{split}
\end{equation}

Recalling that $ \rho_{\tau}(w) $ is differentiable in $w$, the following decomposition of the sample moment corresponding to \eqref{eqn:FOC_sig}, for $\tau \in (0,1)$ holds:
\begin{align}
&o_p(n^{-1/2}) + O_p(S^{-1/2}) \nonumber \\
&= \frac{1}{n} \sum_{i=1}^n \Big[\int_{\mathcal{U}}
\rho_\tau\big( Y_i - u - X_i'\widehat{\beta}_{Y^*}(\tau) \big) \widehat{f}_{U_{Y^*} | Y, X}^\partial(u \mid Y_i, X_i; \widehat{\sigma}_{Y^*} )  du \Big] \label{eqn:sample-me-score} \\
&= \frac{1}{n} \sum_{i=1}^n \int_{\mathcal{U}}\Big(
\rho_\tau\big( Y_i - u - X_i'\widehat{\beta}_{Y^*}(\tau) \big) - \rho_\tau\left( Y_i - u - X_i'\beta_{Y^*}(\tau) \right) \Big) \widehat{f}_{U_{Y^*} | Y, X}^\partial(u \mid Y_i, X_i; \widehat{\sigma}_{Y^*} ) \, du \nonumber \\
&+ \frac{1}{n} \sum_{i=1}^n \int_{\mathcal{U}} \rho_\tau\left( Y_i - u - X_i'\beta_{Y^*}(\tau) \right) \Big( \widehat{f}_{U_{Y^*} | Y, X}^\partial(u \mid Y_i, X_i; \widehat{\sigma}_{Y^*} ) - \widehat{f}_{U_{Y^*} | Y, X}^\partial(u \mid Y_i, X_i; \sigma_{Y^*} ) \Big) \, du \nonumber \\
&+ \frac{1}{n} \sum_{i=1}^n \frac{\partial}{\partial \sigma}\int_{\mathcal{U}} \rho_\tau\left( Y_i - u - X_i'\beta_{Y^*}(\tau) \right) \underbrace{\Big(\widehat{f}_{U_{Y^*} | Y, X}(u \mid Y_i, X_i; \sigma_{Y^*} ) - f_{U_{Y^*} | Y, X}(u \mid Y_i, X_i; \sigma_{Y^*} ) \Big) }_{R_{i,MC}} \, du \nonumber \\
&+ \frac{1}{n} \sum_{i=1}^n \int_{\mathcal{U}} \rho_\tau\left( Y_i - u - X_i'\beta_{Y^*}(\tau) \right) f_{U_{Y^*} | Y, X}^{\partial} (u \mid Y_i, X_i; \sigma_{Y^*} ) \, du \nonumber \\
&=: \underbrace{ -\Big\{\frac{1}{n} \sum_{i=1}^n \int_{\mathcal{U}}\Big[
\psi_\tau\left( Y_i - u - X_i'\bar{\beta}_{Y^*}(\tau) \right) \widehat{f}_{U_{Y^*} | Y, X}^\partial(u \mid Y_i, X_i; \widehat{\sigma}_{Y^*} ) \, du\Big]X_i'\Big\}}_{\mathbb{A}_{\sigma,n}(\tau)}\big(\widehat{\beta}_{Y^*}(\tau) - \beta_{Y^*}(\tau)\big) \nonumber \\
&+ \underbrace{\Big\{\frac{1}{n} \sum_{i=1}^n \int_{\mathcal{U}} \rho_\tau\left( Y_i - u - X_i'\beta_{Y^*}(\tau) \right)  \widehat{f}_{U_{Y^*} | Y, X}^{\partial\partial}(u \mid Y_i, X_i; \bar{\sigma}_{Y^*} ) \, du \Big\}}_{\mathbb{B}_{\sigma,n}(\tau)} (\widehat{\sigma}_{Y^*} - \sigma_{Y^*}) \nonumber \\
&+ \sum_{\ell = 1}^{L}\mathbb{R}_{\sigma,\ell,n}(\tau)\big( \widehat{\beta}_{Y^*}(\tau_\ell) - \beta_{Y^*}(\tau_\ell) \big) \nonumber \\
&+ \underbrace{\frac{1}{n} \sum_{i=1}^n \int_{\mathcal{U}} \rho_\tau\left( Y_i - u - X_i'\beta_{Y^*}(\tau) \right) f_{U_{Y^*} | Y, X}^{\partial} (u \mid Y_i, X_i; \sigma_{Y^*} ) \, du }_{\mathbb{C}_{\sigma,n}(\tau)} \nonumber \\
&=: \mathbb{A}_{\sigma,n}(\tau)\big(\widehat{\beta}_{Y^*}(\tau) - \beta_{Y^*}(\tau)\big) + \sum_{\ell = 1}^{L}\mathbb{R}_{\sigma,\ell,n}(\tau)\big( \widehat{\beta}_{Y^*}(\tau_\ell) - \beta_{Y^*}(\tau_\ell) \big)  + \mathbb{B}_{\sigma,n}(\tau)(\widehat{\sigma}_{Y^*} - \sigma_{Y^*}) + \mathbb{C}_{\sigma,n}(\tau). \label{eqn:score_sig}
\end{align}
where under the conditions of \Cref{lem:R_AB},
\begin{align*}
    &\frac{1}{n} \sum_{i=1}^n \frac{\partial}{\partial \sigma}\int_{\mathcal{U}} \rho_\tau\left( Y_i - u - X_i'\beta_{Y^*}(\tau) \right) \Big(\widehat{f}_{U_{Y^*} | Y, X}(u \mid Y_i, X_i; \sigma_{Y^*} ) - f_{U_{Y^*} | Y, X}(u \mid Y_i, X_i; \sigma_{Y^*} ) \Big)\, du\\
    &= \sum_{\ell = 1}^{L} \underbrace{\frac{1}{n} \sum_{i=1}^n \Big\{ \int_{\mathcal{U}} \Big[\rho_\tau\left( Y_i - u - X_i'\beta_{Y^*}(\tau) \right) \frac{\partial}{\partial \sigma} \widehat{\mathbb{R}}_{\ell,f_{U_{Y^*} | Y, X}}(u,Y_i,X_i;\sigma_{Y^*})'\, du\Big]\Big\}}_{ \mathbb{R}_{\sigma,\ell,n}(\tau) } \times\big( \widehat{\beta}_{Y^*}(\tau_\ell) - \beta_{Y^*}(\tau_\ell) \big)\\
    &=: \sum_{\ell = 1}^{L}\mathbb{R}_{\sigma,\ell,n}(\tau)\big( \widehat{\beta}_{Y^*}(\tau_\ell) - \beta_{Y^*}(\tau_\ell) \big).
\end{align*}

\begin{comment}
    \begin{align}
    &o_p(n^{-1/2}) \nonumber \\
    &= -\Big\{\frac{1}{n} \sum_{i=1}^n \int_{\mathcal{U}}\Big[
\psi_\tau\left( Y_i - u - X_i'\bar{\beta}_{Y^*}(\tau) \right) \widehat{f}_{U_{Y^*} | Y, X}^\partial(u \mid Y_i, X_i; \widehat{\sigma}_{Y^*} ) \, du\Big]X_i'\Big\}\big(\widehat{\beta}_{Y^*}(\tau) - \beta_{Y^*}(\tau)\big) \nonumber \\
&+ \Big\{\frac{1}{n} \sum_{i=1}^n \int_{\mathcal{U}} \rho_\tau\left( Y_i - u - X_i'\beta_{Y^*}(\tau) \right)  \widehat{f}_{U_{Y^*} | Y, X}^{\partial\partial}(u \mid Y_i, X_i; \bar{\sigma}_{Y^*} ) \, du \Big\} (\widehat{\sigma}_{Y^*} - \sigma_{Y^*}) \nonumber \\
&+ \frac{1}{n} \sum_{i=1}^n \Big\{\int_{\mathcal{U}} \rho_\tau\left( Y_i - u - X_i'\beta_{Y^*}(\tau) \right) \Big(\E\big[\widehat{f}_{U_{Y^*} | Y, X}^{\partial} (u \mid Y_i, X_i; \sigma_{Y^*} ) \mid Y_i, X_i\big] - f_{U_{Y^*} | Y, X}^{\partial} (u \mid Y_i, X_i; \sigma_{Y^*} ) \Big) \, du \nonumber \\
&- \mathrm{Bias}_{n,f^\partial} \Big\} \nonumber \\
&+ \frac{1}{n} \sum_{i=1}^n \int_{\mathcal{U}} \rho_\tau\left( Y_i - u - X_i'\beta_{Y^*}(\tau) \right) f_{U_{Y^*} | Y, X}^{\partial} (u \mid Y_i, X_i; \sigma_{Y^*} ) \, du + \mathrm{Bias}_{n,f^\partial} + O_p(n^{-1}) \nonumber \\
&=: \mathbb{A}_{2n}\big(\widehat{\beta}_{Y^*}(\tau) - \beta_{Y^*}(\tau)\big) + \mathbb{B}_{2n}(\widehat{\sigma}_{Y^*} - \sigma_{Y^*}) + \mathbb{C}_{2n}  + \mathrm{Bias}_{n,f^\partial} + O_p(n^{-1}). \label{eqn:score_sig}
\end{align}
\end{comment}

Define
\begin{align*}
    \mathbb{A}_{n}(\tau):= \underbrace{\begin{bmatrix}
        \mathbb{A}_{\beta,n}(\tau) \\
        \mathbb{A}_{\sigma,n}(\tau)
    \end{bmatrix}}_{(K+1) \times K }, \
    \mathbb{R}_{\ell,n}(\tau):= \underbrace{\begin{bmatrix}
        \mathbb{R}_{\beta,\ell,n}(\tau) \\
        \mathbb{R}_{\sigma,\ell,n}(\tau)
    \end{bmatrix}}_{(K+1) \times K }, \
    \mathbb{B}_{n}(\tau):= \underbrace{\begin{bmatrix}
        \mathbb{B}_{\beta,n}(\tau) \\
        \mathbb{B}_{\sigma,n}(\tau)
    \end{bmatrix}}_{(K+1) \times 1 }, \text{ and } \mathbb{C}_{n}(\tau): =
    -\underbrace{\begin{bmatrix}
        \mathbb{C}_{\beta,n}(\tau)\\
        \mathbb{C}_{\sigma,n}(\tau)
    \end{bmatrix}}_{(K+1) \times 1 }.
\end{align*} Then, stacking both expanded moments \eqref{eqn:score_beta} and \eqref{eqn:score_sig} vertically obtains
\begin{align*}
    \mathbb{A}_{n}(\tau)\big(\widehat{\beta}_{Y^*}(\tau) - \beta_{Y^*}(\tau)\big) + \sum_{\ell = 1}^{L}\mathbb{R}_{\ell,n}(\tau)\big( \widehat{\beta}_{Y^*}(\tau_\ell) - \beta_{Y^*}(\tau_\ell) \big)+ \mathbb{B}_{n}(\tau)(\widehat{\sigma}_{Y^*} - \sigma_{Y^*}) =  \mathbb{C}_{n}(\tau) + o_p(n^{-1/2}),
\end{align*}which, evaluated at each $\tau = \tau_{\ell'} \in \{ \tau_1,\ldots,\tau_L\} $ (from \Cref{ass:splines}) gives,

\begin{align*}
    &\big(\mathbb{A}_{n}(\tau_{\ell'}) + \mathbb{R}_{\ell',n}(\tau_{\ell'})\big)\big( \widehat{\beta}_{Y^*}(\tau_{\ell'}) - \beta_{Y^*}(\tau_{\ell'}) \big) \\
    &+ \sum_{\ell \neq \ell'}^{L}\mathbb{R}_{\ell,n}(\tau_{\ell'})\big( \widehat{\beta}_{Y^*}(\tau_\ell) - \beta_{Y^*}(\tau_\ell) \big)+ \mathbb{B}_{n}(\tau_{\ell'})(\widehat{\sigma}_{Y^*} - \sigma_{Y^*}) =  \mathbb{C}_{n}(\tau_{\ell'}) + o_p(n^{-1/2}).
\end{align*}

\noindent For example, the above evaluated at $\ell' = 1$ has the matrix form
\begin{align*}
    \underbrace{\Bigg[
\ \mathbb{A}_{n}(\tau_{1})+\mathbb{R}_{1,n}(\tau_{1})
\ \Big|\
\mathbb{R}_{2,n}(\tau_{1})
\ \Big|\ \cdots \Big|\
\mathbb{R}_{L,n}(\tau_{1})
\ \Big|\
\mathbb{B}_{n}(\tau_{1})
\ \Bigg]}_{(K+1) \times (LK + 1) }
\underbrace{\begin{bmatrix}
\widehat{\beta}_{Y^*}(\tau_{1})-\beta_{Y^*}(\tau_{1})\\
\widehat{\beta}_{Y^*}(\tau_{2})-\beta_{Y^*}(\tau_{2})\\
\vdots\\
\widehat{\beta}_{Y^*}(\tau_{L})-\beta_{Y^*}(\tau_{L})\\
\widehat{\sigma}_{Y^*}-\sigma_{Y^*}
\end{bmatrix}}_{(LK+1) \times 1 }
=
\underbrace{\mathbb{C}_{n}(\tau_{1})}_{(K+1) \times 1 } + o_p(n^{-1/2}).
\end{align*}
\noindent Next, stacking vertically across $ \tau \in \{ \tau_1,\ldots,\tau_L \} $,

\begingroup
\footnotesize

\begin{align}\label{eqn:moments_stacked}
\underbrace{\begin{bmatrix}
\mathbb{A}_{n}(\tau_{1})+\mathbb{R}_{1,n}(\tau_{1}) & \mathbb{R}_{2,n}(\tau_{1}) & \cdots & \mathbb{R}_{L,n}(\tau_{1}) & \mathbb{B}_{n}(\tau_{1})\\
\mathbb{R}_{1,n}(\tau_{2}) & \mathbb{A}_{n}(\tau_{2})+\mathbb{R}_{2,n}(\tau_{2}) & \cdots & \mathbb{R}_{L,n}(\tau_{2}) & \mathbb{B}_{n}(\tau_{2})\\
\vdots & \vdots & \ddots & \vdots & \vdots\\
\mathbb{R}_{1,n}(\tau_{L}) & \mathbb{R}_{2,n}(\tau_{L}) & \cdots & \mathbb{A}_{n}(\tau_{L})+\mathbb{R}_{L,n}(\tau_{L}) & \mathbb{B}_{n}(\tau_{L})
\end{bmatrix}}_{L(K+1) \times (LK+1) }
\underbrace{\begin{bmatrix}
\widehat{\beta}_{Y^*}(\tau_{1})-\beta_{Y^*}(\tau_{1})\\
\widehat{\beta}_{Y^*}(\tau_{2})-\beta_{Y^*}(\tau_{2})\\
\vdots\\
\widehat{\beta}_{Y^*}(\tau_{L})-\beta_{Y^*}(\tau_{L})\\
\widehat{\sigma}_{Y^*}-\sigma_{Y^*}
\end{bmatrix}}_{(LK+1)\times 1}
=
\underbrace{\begin{bmatrix}
\mathbb{C}_{n}(\tau_{1})\\
\mathbb{C}_{n}(\tau_{2})\\
\vdots\\
\mathbb{C}_{n}(\tau_{L})
\end{bmatrix}}_{L(K+1)\times 1} + o_p(n^{-1/2}).
\end{align}
\endgroup
\noindent Define
\begin{align}\label{eqn:H_L}
   \mathbb{H}_{Y^*}^L := \plim_{n\rightarrow \infty} \begin{bmatrix}
\mathbb{A}_{n}(\tau_{1})+\mathbb{R}_{1,n}(\tau_{1}) & \mathbb{R}_{2,n}(\tau_{1}) & \cdots & \mathbb{R}_{L,n}(\tau_{1}) & \mathbb{B}_{n}(\tau_{1})\\
\mathbb{R}_{1,n}(\tau_{2}) & \mathbb{A}_{n}(\tau_{2})+\mathbb{R}_{2,n}(\tau_{2}) & \cdots & \mathbb{R}_{L,n}(\tau_{2}) & \mathbb{B}_{n}(\tau_{2})\\
\vdots & \vdots & \ddots & \vdots & \vdots\\
\mathbb{R}_{1,n}(\tau_{L}) & \mathbb{R}_{2,n}(\tau_{L}) & \cdots & \mathbb{A}_{n}(\tau_{L})+\mathbb{R}_{L,n}(\tau_{L}) & \mathbb{B}_{n}(\tau_{L})
\end{bmatrix},
\end{align}

\begin{align*}
    \big(\widehat{\bm\beta}_{Y^*}^L - {\bm\beta}_{Y^*}^L\big) := \begin{bmatrix}
\widehat{\beta}_{Y^*}(\tau_{1})-\beta_{Y^*}(\tau_{1})\\
\widehat{\beta}_{Y^*}(\tau_{2})-\beta_{Y^*}(\tau_{2})\\
\vdots\\
\widehat{\beta}_{Y^*}(\tau_{L})-\beta_{Y^*}(\tau_{L})
\end{bmatrix}, \quad \text{ and } \quad
\mathbb{C}_{n,Y^*}^L:= \begin{bmatrix}
\mathbb{C}_{n}(\tau_{1})\\
\mathbb{C}_{n}(\tau_{2})\\
\vdots\\
\mathbb{C}_{n}(\tau_{L})
\end{bmatrix}.
\end{align*}
\noindent Then under the full column rank condition on $\mathbb{H}_{Y^*}^L$ in the theorem, ${\mathbb{H}_{Y^*}^L}'\mathbb{H}_{Y^*}^L$ is invertible. In addition to the conditions of \Cref{theorem:consistency}, one obtains the asymptotically linear representation
\begin{align*}
   \sqrt{n} \begin{bmatrix}
        \widehat{\bm\beta}_{Y^*}^L - {\bm\beta}_{Y^*}^L\\
        \widehat{\sigma}_{Y^*} - \sigma_{Y^*}
    \end{bmatrix} &=
    \big({\mathbb{H}_{Y^*}^L}'\mathbb{H}_{Y^*}^L\big)^{-1}{\mathbb{H}_{Y^*}^L}'
    \sqrt{n}\mathbb{C}_{n,Y^*}^L + o_p(1)
\end{align*}

\noindent \textbf{Part (b):}

By (\ref{eqn:FOC_beta}') and \eqref{eqn:FOC_sig},
\begin{align*}
    \mathbb{C}_{n}(\tau) = \frac{1}{n} \sum_{i=1}^n \begin{bmatrix}
    X_i \big( (\tau-1) + \F_{U_{Y^*} | Y, X}(Y_i - X_i'\beta_{Y^*}(\tau) \mid Y_i, X_i) \big) \\
     \int_{\mathcal{U}} \rho_\tau\left( Y_i - u - X_i'\beta_{Y^*}(\tau) \right) f_{U_{Y^*} | Y, X}^{\partial} (u \mid Y_i, X_i; \sigma_{Y^*} ) \, du
\end{bmatrix} =: \frac{1}{n} \sum_{i=1}^n \mathbb{C}_{in}(\tau)
\end{align*}
is mean-zero for any $ \tau \in (0,1) $. This implies $ \E[\mathbb{C}_{n,Y^*}^L] = 0 $. Further, under \Cref{ass:sampling} and \Cref{ass:dominance_Xf}(a),
\begin{align*}
    \E\big[\big\lVert X_i \big( (\tau-1) + \F_{U_{Y^*} | Y, X}(Y_i - X_i'\beta_{Y^*}(\tau) \mid Y_i, X_i) \big)\big\rVert^2\big] \leq \E[\lVert X \rVert^2 ] \leq C.
\end{align*} Define $\mathcal{E}_{i\tau}^*:= Y_i - X_i'\beta_{Y^*}(\tau)$. Note that $ \rho_\tau(w) \leq |w| $ and $\rho_\tau(w)$ is 1-Lipschitz. By the triangle inequality,

\begin{align*}
    &\Big|\int_{\mathcal{U}} \rho_\tau\left( Y_i - u - X_i'\beta_{Y^*}(\tau) \right) f_{U_{Y^*} | Y, X}^{\partial} (u \mid Y_i, X_i; \sigma_{Y^*} ) \, du \Big| \\
    &= \Big|\int_{\mathcal{U}} \left(\rho_\tau\left( \mathcal{E}_{i\tau}^* - u\right) - \rho_\tau( \mathcal{E}_{i\tau}^* ) + \rho_\tau( \mathcal{E}_{i\tau}^* ) \right) f_{U_{Y^*} | Y, X}^{\partial} (u \mid Y_i, X_i; \sigma_{Y^*} ) \, du \Big| \\
    &\leq \Big|\int_{\mathcal{U}} \big(\rho_\tau\left( \mathcal{E}_{i\tau}^* - u\right) - \rho_\tau( \mathcal{E}_{i\tau}^* )\big) \cdot f_{U_{Y^*} | Y, X}^{\partial} (u \mid Y_i, X_i; \sigma_{Y^*} ) \, du \Big| + \rho_\tau( \mathcal{E}_{i\tau}^* ) \Big|\int_{\mathcal{U}} f_{U_{Y^*} | Y, X}^{\partial} (u \mid Y_i, X_i; \sigma_{Y^*} ) \, du \Big| \\
    &\leq \int_{\mathcal{U}}|u|\cdot \big|f_{U_{Y^*} | Y, X}^{\partial} (u \mid Y_i, X_i; \sigma_{Y^*} ) \big| \, du =  \int_{\mathcal{U}}\big|u f_{U_{Y^*} | Y, X}^{\partial} (u \mid Y_i, X_i; \sigma_{Y^*} ) \big| \, du
\end{align*}since $ \int_{\mathcal{U}} f_{U_{Y^*} | Y, X}^{\partial} (u \mid Y_i, X_i; \sigma_{Y^*} ) \, du = \frac{\partial}{\partial \sigma } \int_{\mathcal{U}} f_{U_{Y^*} | Y, X}(u \mid Y_i, X_i; \sigma_{Y^*} ) \, du = \frac{\partial}{\partial \sigma } 1 = 0 $ holds under \Cref{ass:dominance_Xf}(b) by the dominance convergence theorem. From the foregoing and \Cref{ass:dominance_Xf}(e),
\begin{align*}
    \E\Big[\Big| \int_{\mathcal{U}} \rho_\tau\left( Y_i - u - X_i'\beta_{Y^*}(\tau) \right) f_{U_{Y^*} | Y, X}^{\partial} (u \mid Y_i, X_i; \sigma_{Y^*} ) \, du \Big|^2 \Big] &\leq  \E\Big[\Big(\int_{\mathcal{U}}\big|u f_{U_{Y^*} | Y, X}^{\partial} (u \mid Y_i, X_i; \sigma_{Y^*} ) \big| \, du \Big)^2 \Big]\\
    &\leq C.
\end{align*}Considering both bounds, \( \E \big[\big\lVert \mathbb{C}_{in}(\tau) \big\rVert^2 \big] \leq 2C < \infty \) whence \( \E \big[\big\lVert \mathbb{C}_{iY^*}^L \big\rVert^2 \big] \leq 2CL < \infty \) with $ \mathbb{C}_{iY^*}^L:= \begin{bmatrix}
\mathbb{C}_{in}(\tau_{1})\\
\mathbb{C}_{in}(\tau_{2})\\
\vdots\\
\mathbb{C}_{in}(\tau_{L})
\end{bmatrix}  $.

The conclusion follows from the Multivariate Lindeberg-L\'evy Central Limit Theorem (CLT) and the Continuous Mapping Theorem (CMT).
\end{proof}

\subsection{Proof of Corollary \ref{cor:anorm_Ystar_given_X}} \label{app:proof-cor1}

\begin{proof}
        The estimator of \( \F_{Y^* \mid X}(y \mid x) \) is given by \( \displaystyle \widehat{\F}_{Y^* \mid X}(y \mid x) = \int_0^1 \indicator{ x'\widehat{\beta}_{Y^*}(\tilde{\tau}) \leq y } \, d\tilde{\tau} \). Thanks to the strict monotonicity of the map $\tau \mapsto x'\beta_{Y^*}(\tau)$ and its sample counterpart analogously, there exist unique values $ \tau_y \in (0,1) $ and $ \hat{\tau}_y $ such that $ x'\beta_{Y^*}(\tau_y) = y $ and $ x'\widehat{\beta}_{Y^*}(\hat{\tau}_y) = y $. Consider the function $ \hat{h}(\tau):= y - x'\widehat{\beta}_{Y^*}(\tau) $. By \Cref{ass:splines} and the MVT,
\begin{align*}
    0&=\hat{h}(\hat{\tau}_y)\\
    &= \hat{h}(\tau_y) + \hat{h}^\partial(\bar{\tau}_y)(\hat{\tau}_y - \tau_y)\\
    &= y - x'\widehat{\beta}_{Y^*}(\tau_y) -  x'\widehat{\beta}_{Y^*}^\partial(\bar{\tau}_y)(\hat{\tau}_y - \tau_y)\\
    &= -x'\big( \widehat{\beta}_{Y^*}(\tau_y) - \beta_{Y^*}(\tau_y) \big) -  x'\widehat{\beta}_{Y^*}^\partial(\bar{\tau}_y)(\hat{\tau}_y - \tau_y).
\end{align*}Notice that under \Cref{ass:qr}, $ \tau_y = \F_{Y^* \mid X}(y \mid x) $ and $ \hat{\tau}_y = \widehat{\F}_{Y^* \mid X}(y \mid x) $ by the definition of the quantile function and its estimator. Thus by \Cref{ass:dominance_Xf}(d), the Continuous Mapping Theorem (CMT) and \Cref{theorem:consistency},
\begin{align*}
    \sqrt{n}\big( \widehat{\F}_{Y^* \mid X}(y \mid x) - \F_{Y^* \mid X}(y \mid x) \big)
    &= -\frac{1}{x'\widehat{\beta}_{Y^*}^\partial(\bar{\tau}_y)}x'\sqrt{n}\big( \widehat{\beta}_{Y^*}(\tau_y) - \beta_{Y^*}(\tau_y) \big)\\
    &= -f_{Y^* \mid X}(y \mid x)x'\sqrt{n}\big( \widehat{\beta}_{Y^*}(\tau_y) - \beta_{Y^*}(\tau_y) \big) + o_p(1).
\end{align*}

It follows from the foregoing, \Cref{ass:splines}, and \eqref{eqn:beta_expand_L} that
\begin{align*}
    \sqrt{n}&\big( \widehat{\F}_{Y^* \mid X}(y \mid x) - \F_{Y^* \mid X}(y \mid x) \big)
    = -f_{Y^* \mid X}(y \mid x)\sum_{\ell=1}^{L} \omega_{\ell,\beta}(\tau_y)x'\sqrt{n}\big( \widehat{\beta}_{Y^*}(\tau_\ell) - \beta_{Y^*}(\tau_\ell) \big) + o_p(1)\\
    &= \underbrace{-f_{Y^* \mid X}(y \mid x)\Big[\omega_{1,\beta}(\tau_y)x', \ldots, \omega_{L,\beta}(\tau_y)x', \, 0 \Big]}_{{\mathbb{M}_{FY}^L}(y,x)'}\sqrt{n} \begin{bmatrix}
        \widehat{\bm\beta}_{Y^*}^L - {\bm\beta}_{Y^*}^L\\
        \widehat{\sigma}_{Y^*} - \sigma_{Y^*}
    \end{bmatrix} + o_p(1).
\end{align*}The conclusion follows from the CMT and \Cref{theorem:anorm_beta} with \( \sigma(\F_{Y^* \mid X}(y \mid x)) := {\mathbb{M}_{FY}^L}(y,x)'\Omega_{Y^*}^L\mathbb{M}_{FY}^L(y,x) \).

\end{proof}

\subsection{Proof of Proposition \ref{prop:anorm_copula}} \label{app:anorm_copula}

\begin{proof}
    Consider the simulated maximum likelihood estimator
\begin{align*}
  \widehat{\delta} = \arg \max_{\delta} \sum_{i=1}^n \log\big(\widehat{L}_i^{(S)}(\delta)\big)
\end{align*}where

\begin{align*}
  \widehat{L}_i^{(S)}(\delta) &= \int_{\mathcal{U} \times \mathcal{U} } \Big[ c_{Y^*T^*\mid X}\big(\widehat{\F}_{Y^*\mid X}(Y_i- u\mid X_i), \widehat{\F}_{T^*\mid X}(T_i- v\mid X_i); \delta\big)   \times \widehat{f}_{Y^*\mid X}(Y_i - u | X_i) \widehat{f}_{T^*\mid X}(T_i- v | X_i)\\
  &\hspace{50pt} \times \widehat{f}_{U_{Y^*} | Y, X}^{(S)}(u \mid Y_i, X_i; \widehat{\sigma}_{Y^*} )\widehat{f}_{U_{T^*} | Y, X}^{(S)}(v \mid Y_i, X_i; \widehat{\sigma}_{T^*} )\Big] \, du dv.
\end{align*} Similarly, define
\begin{align*}
  \widehat{L}_i(\delta) &:= \int_{\mathcal{U} \times \mathcal{U} } \Big[ c_{Y^*T^*\mid X}\big(\widehat{\F}_{Y^*\mid X}(Y_i- u\mid X_i), \widehat{\F}_{T^*\mid X}(T_i- v\mid X_i); \delta\big)   \times \widehat{f}_{Y^*\mid X}(Y_i - u | X_i) \widehat{f}_{T^*\mid X}(T_i- v | X_i)\\
  &\hspace{50pt} \times \widehat{f}_{U_{Y^*} | Y, X}(u \mid Y_i, X_i; \widehat{\sigma}_{Y^*} )\widehat{f}_{U_{T^*} | Y, X}(v \mid Y_i, X_i; \widehat{\sigma}_{T^*} )\Big] \, du dv \quad \text{ and}\\
  L_i(\delta) &:= \int_{\mathcal{U} \times \mathcal{U} } \Big[ c_{Y^*T^*\mid X}\big(\F_{Y^*\mid X}(Y_i- u\mid X_i), \F_{T^*\mid X}(T_i- v\mid X_i); \delta\big)   \times f_{Y^*\mid X}(Y_i - u | X_i) f_{T^*\mid X}(T_i- v | X_i)\\
  &\hspace{50pt} \times f_{U_{Y^*} | Y, X}(u \mid Y_i, X_i; \sigma_{Y^*} )f_{U_{T^*} | Y, X}(v \mid Y_i, X_i; \sigma_{T^*} )\Big] \, du dv.
\end{align*}

\noindent Let
$$
 \widehat{\mathcal S}_i^{(S)}(\delta):=\nabla_\delta\log \widehat{L}_i^{(S)}(\delta) = \frac{\nabla_\delta \widehat{L}_i^{(S)}(\delta)}{\widehat{L}_i^{(S)}(\delta)},
\qquad
\mathcal H_i(\delta):=\nabla_\delta^2\log L_i(\delta)
=\frac{\nabla_\delta^2 L_i(\delta)}{L_i(\delta)}-\frac{\nabla_\delta L_i(\delta)\,\nabla_\delta L_i(\delta)'}{L_i(\delta)^2},
$$
\begin{align*}
    \nabla_\delta {L}_i(\delta)\;&=\;\iint c_\delta(\cdot;\delta)\;
 f_{Y^*\mid X}\, f_{T^*\mid X}\, f_{U_{Y^*}|Y,X}\, f_{U_{T^*}|Y,X}\,du\,dv, \quad \text{ and}\\
\nabla_\delta^2 {L}_i(\delta)\;&=\;\iint c_{\delta\delta}(\cdot;\delta)\;
 f_{Y^*\mid X}\, f_{T^*\mid X}\, f_{U_{Y^*}|Y,X}\, f_{U_{T^*}|Y,X}\,du\,dv
\end{align*}
where $ c_{\delta}(\cdot,\cdot;\delta):= \tfrac{\partial}{\partial \delta} c_{Y^*T^*\mid X}(\cdot,\cdot;\delta) $ and $ c_{\delta\delta}(\cdot,\cdot;\delta):= \tfrac{\partial^2}{\partial \delta \partial \delta'} c_{Y^*T^*\mid X}(\cdot,\cdot;\delta) $.

Granted the differentiability of $ c_{Y^*T^*\mid X}(\cdot, \cdot ;\delta) $ (\Cref{ass:cop}), and that $\widehat{\delta}$ is the maximizer, the following first-order condition holds:
\begin{align*}
    0&= \sum_{i=1}^n \widehat{\mathcal S}_i^{(S)}(\widehat{\delta})\\
    &= \sum_{i=1}^n \widehat{\mathcal S}_i(\widehat{\delta}) + o_p(n^{-1/2}) \\
    &= \sum_{i=1}^n \mathcal S_i(\widehat{\delta}) + \sum_{i=1}^n \big( \widehat{\mathcal S}_i(\widehat{\delta}) - \mathcal S_i(\widehat{\delta}) \big) + o_p(n^{-1/2})\\
    &=: \sum_{i=1}^n \mathcal S_i(\widehat{\delta}) + \sum_{i=1}^n \Delta_i(\widehat{\delta};\widehat{\cdot}-\cdot) + o_p(n^{-1/2})
\end{align*}where $ \mathcal S_i(\delta):=\nabla_\delta\log L_i(\delta) $. The second equality holds under \Cref{ass:MCMC_conv,ass:dominance_Xf,ass:cop} and the condition $n/S=o(1)$.

Under \Cref{ass:cop}, the copula density function is twice continuously differentiable with respect to $\delta$. Plugging in a first–order expansion of $ \mathcal S_i(\widehat{\delta}) $ around $\delta$ into the first-order condition then gives
\begin{align*}
    o_p(n^{-1/2}) = \frac1n\sum_{i=1}^n \mathcal{S}_i(\delta)
+\Big(\frac1n\sum_{i=1}^n \mathcal{H}_i(\delta)\Big)(\widehat\delta-\delta)
+\frac1n\sum_{i=1}^n \Delta_i(\widehat{\delta};\widehat{\cdot}-\cdot)
\end{align*}
where $ \E[\mathcal{S}_i(\delta)] = 0 $. Therefore,
\begin{equation}\label{eqn:delta_IF_1}
    \sqrt n(\widehat\delta-\delta) =
-\Big[\frac1n\sum_{i=1}^n \mathcal{H}_i(\delta)\Big]^{-1}
\Bigg\{
\frac1{\sqrt n}\sum_{i=1}^n \mathcal{S}_i(\delta)
+\frac1{\sqrt n}\sum_{i=1}^n \Delta_i(\widehat{\delta};\widehat{\cdot}-\cdot)
\Bigg\}
+o_p(1)
\end{equation} where $\Delta_i(\widehat{\delta};\widehat{\cdot}-\cdot)$ has the following weighted sum representation:
$$
\begin{aligned}
\Delta_i(\widehat{\delta};\widehat{\cdot}-\cdot) :&= \widehat{\mathcal S}_i(\widehat{\delta}) - \mathcal S_i(\widehat{\delta})\\
&= \nabla_\delta \big(\log \widehat{L}_i(\widehat{\delta}) - \log L_i(\widehat{\delta})\big)\\
&= \widehat{L}_i(\widehat{\delta})^{-1}\nabla_\delta\big( \widehat{L}_i(\widehat{\delta}) - L_i(\widehat{\delta}) \big) - \big( \widehat{L}_i(\widehat{\delta})L_i(\widehat{\delta}) \big)^{-1}\nabla_\delta L_i(\widehat{\delta})\big( \widehat{L}_i(\widehat{\delta}) - L_i(\widehat{\delta}) \big)\\
&=:\int\Big(\int W_i^{(F_Y)}(u,v;\widehat{\delta})\,dv\Big)\,\big(\widehat \F_{Y^*\mid X}- \F_{Y^*\mid X}\big)(Y_i-u\mid X_i)\,du\\
&\quad+\int \Big(\int W_i^{(f_Y)}(u,v;\widehat{\delta})\,dv\Big)\,\big(\widehat f_{Y^*\mid X}-f_{Y^*\mid X}\big)(Y_i-u\mid X_i)\,du\\
&\quad+\int \Big( \int W_i^{(f_{U_Y})}(u,v;\widehat{\delta})\,dv\Big)\,\big(\widehat f_{U_{Y^*}\mid Y,X}-f_{U_{Y^*}\mid Y,X}\big)(u\mid Y_i,X_i)\,du\\
&\quad+\int\Big(\int W_i^{(F_T)}(u,v;\widehat{\delta})\,du\Big)\,\big(\widehat \F_{T^*\mid X}- \F_{T^*\mid X}\big)(T_i-v\mid X_i)\,dv\\
&\quad+\int \Big(\int W_i^{(f_T)}(u,v;\widehat{\delta})\,du\Big)\,\big(\widehat f_{T^*\mid X}-f_{T^*\mid X}\big)(T_i-v\mid X_i)\,dv\\
&\quad+\int \Big(\int W_i^{(f_{U_T})}(u,v;\widehat{\delta})\,du\Big)\,\big(\widehat f_{U_{T^*}\mid Y,X}-f_{U_{T^*}\mid Y,X}\big)(v\mid Y_i,X_i)\,dv.
\end{aligned}
$$

First, from the proof of \Cref{cor:anorm_Ystar_given_X},
\begin{align*}
    \sqrt{n}\big(\widehat \F_{Y^*\mid X}- \F_{Y^*\mid X}\big)(Y_i-u\mid X_i) = {\mathbb{M}_{FY}^L}(Y_i-u,X_i)'\sqrt{n} \begin{bmatrix}
        \widehat{\bm\beta}_{Y^*}^L - {\bm\beta}_{Y^*}^L\\
        \widehat{\sigma}_{Y^*} - \sigma_{Y^*}
    \end{bmatrix} + o_p(1).
\end{align*}

Second, from \Cref{lem:R_AB}(a) and the conditions of \Cref{theorem:consistency},
\begin{align*}
    \sqrt{n}\big(\widehat f_{Y^*\mid X} & - f_{Y^*\mid X}\big)(Y_i-u\mid X_i) = \sum_{\ell=1}^{L}\widehat{\mathbb{R}}_{\ell,f_{Y^*\mid X}}(Y_i-u,X_i)'\,
\sqrt{n}\big(\widehat{\beta}_{Y^*}(\tau_{\ell})-\beta_{Y^*}(\tau_{\ell})\big)\\
    &= \begin{bmatrix}
        \widehat{\mathbb{R}}_{1,f_{Y^*\mid X}}(Y_i-u,X_i)' & \ldots & \widehat{\mathbb{R}}_{L,f_{Y^*\mid X}}(Y_i-u,X_i)' & \, 0
    \end{bmatrix} \sqrt{n} \begin{bmatrix}
        \widehat{\bm\beta}_{Y^*}^L - {\bm\beta}_{Y^*}^L\\
        \widehat{\sigma}_{Y^*} - \sigma_{Y^*}
    \end{bmatrix} \\
    &=: \mathbb{M}_{f_{Y^*\mid X}}^L(Y_i-u,X_i)\sqrt{n} \begin{bmatrix}
        \widehat{\bm\beta}_{Y^*}^L - {\bm\beta}_{Y^*}^L\\
        \widehat{\sigma}_{Y^*} - \sigma_{Y^*}
    \end{bmatrix} + o_p(1).
\end{align*}

Lastly, under \Cref{ass:dominance_Xf}(e), the conditions of \Cref{lem:R_AB}(b), and the conditions of \Cref{theorem:consistency},
\begin{align*}
    &\sqrt{n}\big( \widehat{f}_{U_{Y^*} | Y, X}(u \mid Y_i, X_i; \widehat{\sigma}_{Y^*} ) - f_{U_{Y^*} | Y, X}(u \mid Y_i, X_i; \sigma_{Y^*} ) \big)\\
    &= \sqrt{n}\big( \widehat{f}_{U_{Y^*} | Y, X} - f_{U_{Y^*} | Y, X}\big)(u \mid Y_i, X_i; \sigma_{Y^*} ) + \sqrt{n}\big( \widehat{f}_{U_{Y^*} | Y, X}(u \mid Y_i, X_i; \widehat{\sigma}_{Y^*} ) - \widehat{f}_{U_{Y^*} | Y, X}(u \mid Y_i, X_i; \sigma_{Y^*} ) \big) \\
    &= \sum_{\ell=1}^{L}\widehat{\mathbb{R}}_{\ell,f_{U_{Y^*} | Y, X}}(u,Y_i,X_i;\sigma_{Y^*})'\,
\sqrt{n}\big(\widehat{\beta}_{Y^*}(\tau_{\ell})-\beta_{Y^*}(\tau_{\ell})\big) + \widehat{f}_{U_{Y^*} | Y, X}^\partial(u \mid Y_i, X_i; \bar{\sigma}_{Y^*} )\sqrt{n}(\widehat{\sigma}_{Y^*} -  \sigma_{Y^*} )\\
    &= \begin{bmatrix}
        \widehat{\mathbb{R}}_{1,f_{U_{Y^*} | Y, X}}(u,Y_i,X_i;\sigma_{Y^*})', \ldots, \widehat{\mathbb{R}}_{L,f_{U_{Y^*} | Y, X}}(u,Y_i,X_i;\sigma_{Y^*})' &
        \widehat{f}_{U_{Y^*} | Y, X}^\partial(u \mid Y_i, X_i; \bar{\sigma}_{Y^*} )
    \end{bmatrix}\sqrt{n} \begin{bmatrix}
        \widehat{\bm\beta}_{Y^*}^L - {\bm\beta}_{Y^*}^L\\
        \widehat{\sigma}_{Y^*} - \sigma_{Y^*}
    \end{bmatrix}\\
    &=: {\mathbb{M}_{f_{U_{Y^*} | Y, X}}^L(u;Y_i,X_i)}'\sqrt{n} \begin{bmatrix}
        \widehat{\bm\beta}_{Y^*}^L - {\bm\beta}_{Y^*}^L\\
        \widehat{\sigma}_{Y^*} - \sigma_{Y^*}
    \end{bmatrix} + o_p(1).
\end{align*} Putting together the foregoing, and applying arguments with respect to $Y^*$ analogously for $T^*$,

\begin{align*}
    \sqrt{n}\Delta_i(\widehat{\delta};\widehat{\cdot}-\cdot)
    &=\bigg\{\int\Big[\Big(\int W_i^{(F_Y)}(u,v;\widehat{\delta})\,dv\Big){\mathbb{M}_{FY}^L} (Y_i-u,X_i)'\\
    &\quad +  \Big(\int W_i^{(f_Y)}(u,v;\widehat{\delta})\,dv\Big){\mathbb{M}_{f_{Y^*\mid X}}^L} (Y_i-u,X_i)'\\
    &\quad + \Big( \int W_i^{(f_{U_Y})}(u,v;\widehat{\delta})\,dv\Big){\mathbb{M}_{f_{U_{Y^*} | Y, X}}^L(u;Y_i,X_i)}' \,\Big]du\bigg\}\sqrt{n} \begin{bmatrix}
        \widehat{\bm\beta}_{Y^*}^L - {\bm\beta}_{Y^*}^L\\
        \widehat{\sigma}_{Y^*} - \sigma_{Y^*}
    \end{bmatrix}   \\
&+\bigg\{\int\bigg[\Big(\int W_i^{(F_T)}(u,v;\widehat{\delta})\,du\Big)\mathbb{M}_{FT}^{\,L}(T_i-v,X_i)' \\
&+  \Big(\int W_i^{(f_T)}(u,v;\widehat{\delta})\,du\Big)\mathbb{M}_{f_{T^*\mid X}}^L(T_i-v,X_i)'\\
    &\quad + \Big( \int W_i^{(f_{U_T})}(u,v;\widehat{\delta})\,du\Big)\mathbb{M}_{f_{U_{T^*}|T,X}}^L(v;T_i,X_i)' \,\bigg]dv\bigg\}\sqrt{n} \begin{bmatrix}
        \widehat{\bm\beta}_{T^*}^L - {\bm\beta}_{T^*}^L\\
        \widehat{\sigma}_{T^*} - \sigma_{T^*}
    \end{bmatrix} + o_p(1)\\
    &=: \mathbb{M}_{i\Delta Y}^{\,L}(Y_i,X_i;\widehat{\delta})'\sqrt{n} \begin{bmatrix}
        \widehat{\bm\beta}_{Y^*}^L - {\bm\beta}_{Y^*}^L\\
        \widehat{\sigma}_{Y^*} - \sigma_{Y^*}
    \end{bmatrix} + \mathbb{M}_{i\Delta T}^{\,L}(Y_i,X_i;\widehat{\delta})'\sqrt{n} \begin{bmatrix}
        \widehat{\bm\beta}_{T^*}^L - {\bm\beta}_{T^*}^L\\
        \widehat{\sigma}_{T^*} - \sigma_{T^*}
    \end{bmatrix} + o_p(1).
\end{align*}

\begin{comment}
    From the above and \Cref{theorem:anorm_beta},

\begin{align*}
    \sqrt{n}\Delta_i(\widehat{\delta};\widehat{\cdot}-\cdot) &= \mathrm{Diag}\Big[ \mathbb{M}_{i\Delta Y}^{\,L}, \ \mathbb{M}_{i\Delta T}^{\,L}  \Big]\times \sqrt{n} \begin{bmatrix}
        \widehat{\bm\beta}_{T^*}^L - {\bm\beta}_{T^*}^L\\
        \widehat{\sigma}_{T^*} - \sigma_{T^*}\\
        \widehat{\bm\beta}_{T^*}^L - {\bm\beta}_{T^*}^L\\
        \widehat{\sigma}_{T^*} - \sigma_{T^*}
    \end{bmatrix} + o_p(1)\\
    &= \mathrm{Diag}\Big[ \mathbb{M}_{i\Delta Y}^{\,L}\times \big({\mathbb{H}_{Y^*}^L}'\mathbb{H}_{Y^*}^L\big)^{-1}{\mathbb{H}_{Y^*}^L}', \ \mathbb{M}_{i\Delta T}^{\,L}\times \big({\mathbb{H}_{T^*}^L}'\mathbb{H}_{T^*}^L\big)^{-1}{\mathbb{H}_{T^*}^L}'  \Big]
    \begin{bmatrix}
    \sqrt{n}\mathbb{C}_{n,Y^*}^L \\
    \sqrt{n}\mathbb{C}_{n,T^*}^L
    \end{bmatrix}
     + o_p(1)\\
     &=: \mathbb{M}_{i\Delta \cdot}^{\,L}\begin{bmatrix}
    \sqrt{n}\mathbb{C}_{n,Y^*}^L \\
    \sqrt{n}\mathbb{C}_{n,T^*}^L
    \end{bmatrix} + o_p(1).
\end{align*}
\end{comment}

Plugging the above into \eqref{eqn:delta_IF_1} and applying the CMT,
\begin{align*}
    \sqrt n(\widehat\delta-\delta) &=
-\mathcal{H}_{\delta}^{-1}
\frac1{\sqrt n}\sum_{i=1}^n \mathcal{S}_i(\delta) -\mathcal{H}_{\delta}^{-1}
\frac1{n}\sum_{i=1}^n \mathbb{M}_{i\Delta Y}^{\,L}(Y_i,X_i;\widehat{\delta})'\sqrt{n} \begin{bmatrix}
        \widehat{\bm\beta}_{Y^*}^L - {\bm\beta}_{Y^*}^L\\
        \widehat{\sigma}_{Y^*} - \sigma_{Y^*}
    \end{bmatrix} \\
    &-\mathcal{H}_{\delta}^{-1}
\frac1{n}\sum_{i=1}^n \mathbb{M}_{i\Delta T}^{\,L}(Y_i,X_i;\widehat{\delta})'\sqrt{n} \begin{bmatrix}
        \widehat{\bm\beta}_{T^*}^L - {\bm\beta}_{T^*}^L\\
        \widehat{\sigma}_{T^*} - \sigma_{T^*}
    \end{bmatrix}
+o_p(1)\\
&=
-\mathcal{H}_{\delta}^{-1}
\frac1{\sqrt n}\sum_{i=1}^n \mathcal{S}_i(\delta) -\mathcal{H}_{\delta}^{-1}
\mathbb{M}_{\Delta Y}^{L\, \prime}\sqrt{n} \begin{bmatrix}
        \widehat{\bm\beta}_{Y^*}^L - {\bm\beta}_{Y^*}^L\\
        \widehat{\sigma}_{Y^*} - \sigma_{Y^*}
    \end{bmatrix} -\mathcal{H}_{\delta}^{-1}
\mathbb{M}_{\Delta T}^{L\, \prime}\sqrt{n} \begin{bmatrix}
        \widehat{\bm\beta}_{T^*}^L - {\bm\beta}_{T^*}^L\\
        \widehat{\sigma}_{T^*} - \sigma_{T^*}
    \end{bmatrix}
+o_p(1) \\
&=: \frac{1}{\sqrt{n}} \sum_{i=1}^n \psi_i^{\delta} + o_p(1) \xrightarrow{d} \N(0,\Sigma(\delta))
\end{align*}
where $\Sigma(\delta) = \E\big[\psi^{\delta}{\psi^{\delta}}'\big]$, $ \displaystyle \mathcal{H}_{\delta}:= \plim_{n\rightarrow \infty}\frac1n\sum_{i=1}^n \mathcal{H}_i(\delta) $ and $ \displaystyle \mathbb{M}_{\Delta \, j}^{\,L}:= \plim_{n\rightarrow \infty} \frac1{n}\sum_{i=1}^n \mathbb{M}_{i\Delta \, j}^{\,L}(Y_i,X_i;\widehat{\delta}), \ j\in \{Y,T\} $. Since a linear combination of asymptotically linear quantities is asymptotically linear, the concluding part follows from the CMT, \Cref{theorem:anorm_beta}, in addition to the condition $ \E[\lVert \mathcal{S}(\delta) \rVert^2] < \infty $ which is guaranteed under \Cref{ass:dominance_Xf,ass:cop}.
\end{proof}

\subsection{Proof of Theorem \ref{theorem:anorm_pars}}\label{sec:proof_anorm_pars}
We provide the proofs of the first result in \Cref{theorem:anorm_pars} and of bootstrap validity in this subsection. The proofs of the remaining parts are provided in the Supplementary Appendix.

\subsubsection{Proof of Theorem \ref{theorem:anorm_pars}(a)}

The following lemma provides a proof of part (a) of \Cref{theorem:anorm_pars}.

\begin{lemma}\label{lem:FYT_anorm} Suppose \Cref{ass:me,ass:additional-qr,ass:sampling,ass:dominance_Xf,ass:qr,ass:splines,ass:MCMC_conv,ass:unobs,ass:compact_par_space,ass:cop} hold, then
    \( \displaystyle \sqrt{n}\big( \widehat{\F}_{Y^*T^*\mid X}(y,t\mid x) - \F_{Y^*T^*\mid X}(y,t\mid x) \big) \xrightarrow{d} \mathcal{N}\big(0,\, \sigma(\F_{Y^*T^*\mid X}(y,t\mid x))\big). \)
\end{lemma}

\begin{proof}
    Recall
\( \F_{Y^*T^*\mid X}(y,t\mid x) = C_{Y^*T^*\mid X}\big(\F_{Y^*\mid X}(y\mid x),\F_{T^*\mid X}(t\mid x)\mid x\big) = C_{Y^*T^*\mid X}\big(\F_{Y^*\mid X}(y\mid x),\F_{T^*\mid X}(t\mid x);\, \delta\big) \) and the estimator is given by
\( \widehat{\F}_{Y^*T^*\mid X}(y,t\mid x) = C_{Y^*T^*\mid X}\big(\widehat{\F}_{Y^*\mid X}(y\mid x),\, \widehat{\F}_{T^*\mid X}(t\mid x);\, \widehat{\delta}\big) \). Consider the following decomposition:
\begin{align*}
    \sqrt{n}&\big( \widehat{\F}_{Y^*T^*\mid X}(y,t\mid x) - \F_{Y^*T^*\mid X}(y,t\mid x) \big)\\
    &= \sqrt{n}\Big(C_{Y^*T^*\mid X}\big(\widehat{\F}_{Y^*\mid X}(y\mid x),\, \widehat{\F}_{T^*\mid X}(t\mid x);\, \widehat{\delta}\big) - C_{Y^*T^*\mid X}\big(\F_{Y^*\mid X}(y\mid x),\F_{T^*\mid X}(t\mid x);\, \delta\big)\Big) \\
    &= \sqrt{n}\Big(C_{Y^*T^*\mid X}\big(\widehat{\F}_{Y^*\mid X}(y\mid x),\, \widehat{\F}_{T^*\mid X}(t\mid x);\, \widehat{\delta}\big) - C_{Y^*T^*\mid X}\big(\F_{Y^*\mid X}(y\mid x),\, \widehat{\F}_{T^*\mid X}(t\mid x);\, \widehat{\delta}\big)\Big) \\
    &+\sqrt{n}\Big(C_{Y^*T^*\mid X}\big(\F_{Y^*\mid X}(y\mid x),\, \widehat{\F}_{T^*\mid X}(t\mid x);\, \widehat{\delta}\big) - C_{Y^*T^*\mid X}\big(\F_{Y^*\mid X}(y\mid x),\, \F_{T^*\mid X}(t\mid x);\, \widehat{\delta}\big)\Big) \\
    &+\sqrt{n}\Big(C_{Y^*T^*\mid X}\big(\F_{Y^*\mid X}(y\mid x),\, \F_{T^*\mid X}(t\mid x);\, \widehat{\delta}\big) - C_{Y^*T^*\mid X}\big(\F_{Y^*\mid X}(y\mid x),\, \F_{T^*\mid X}(t\mid x);\, \delta\big)\Big).
\end{align*}
Let $ \tau_y:= \F_{Y^*\mid X}(y\mid x) $ and $ \tau_t: = \F_{T^*\mid X}(t\mid x)$. Since (1) $C_{Y^*T^*\mid X}(\tau_y,\tau_t;\delta)$ is continuously differentiable in $(\tau_y,\tau_t,\delta')' \in \mathcal{T}^2\times\Gamma_\delta $, with partials $C_1=\partial C/\partial u$, $C_2=\partial C/\partial v$, and $ C_\delta$ (\Cref{ass:cop}), it follows from the above decomposition and the CMT that
$$
\begin{aligned}
\sqrt{n}&\big(\widehat \F_{Y^*T^*\mid X}(y,t\mid x)- \F_{Y^*T^*\mid X}(y,t\mid x)\big)\\
=& C_1(\tau_y,\tau_t;\delta)\sqrt{n}\big(\widehat \F_{Y^*\mid X}(y\mid x)- \F_{Y^*\mid X}(y\mid x)\big) + C_2(\tau_y,\tau_t;\delta)\sqrt{n}\big(\widehat \F_{T^*\mid X}(t\mid x)- \F_{T^*\mid X}(t\mid x)\big) \\
& + C_\delta(\tau_y,\tau_t;\delta)\big)'\,\sqrt{n}(\widehat\delta-\delta)\;+\;o_p(1) \\
=:& \frac{1}{\sqrt{n}} \sum_{i=1}^n \psi_i^{(a)} + o_p(1) \xrightarrow{d} \N(0,\sigma(F_{Y^*T^*|X}(y,t|x)))
\end{aligned}
$$where \(  \sigma(F_{Y^*T^*|X}(y,t|x)) = \E\big[(\psi^{(a)})^2\big] \). Since a linear combination of asymptotically linear quantities is asymptotically linear, the conclusion follows from \Cref{cor:anorm_Ystar_given_X} (applied analogously to $T$), \Cref{prop:anorm_copula}, and the CMT.
\end{proof}

\subsubsection{Validity of the Bootstrap}

Given our previous results establishing the limiting distributions of the QR parameters, measurement error parameters, copula parameters, and target parameters, the validity of the bootstrap follows from standard arguments that closely parallel the arguments provided above.  For brevity, we only sketch these arguments here.  To start with, we consider the validity of the bootstrap for the QR parameters and measurement error parameters.  The bootstrap version of these estimators is defined by replacing the score equations (Equations \ref{eqn:sample-qr-score} and \ref{eqn:sample-me-score}) with their weighted counterparts, where each observation is multiplied by bootstrap weights $w_i$ drawn independently of the data with mean one and variance one (for the empirical bootstrap, these are multinomial weights). Then, the validity of the bootstrap holds by linearizing the weighted score equations around the original estimates (which follows using an analogous argument to the one provided above), and then employing the bootstrap CLT (\citet{praestgaard-wellner-1993}).  Finally, the validity of the bootstrap for the target parameters in \Cref{theorem:anorm_pars} follows using similar arguments to the ones provided in the proof of \Cref{theorem:anorm_pars} in conjunction with the bootstrap delta method (\citet{vaart-wellner-1996}).

\section{Supporting Lemmas}

\begin{lemma}\label{lem:R_AB}
    Under \Cref{ass:splines},
    \begin{align*}
    (a)\ &\big( \widehat{f}_{Y^*\mid X}(y \mid x ) - f_{Y^*\mid X}(y \mid x ) \big) = \sum_{\ell=1}^{L}\widehat{\mathbb{R}}_{\ell,f_{Y^*\mid X}} (y,x)'\,
\big(\widehat{\beta}_{Y^*}(\tau_{\ell})-\beta_{Y^*}(\tau_{\ell})\big) \text{ and} \\
    (b)\ &\big( \widehat{f}_{U_{Y^*} | Y, X}(u \mid Y_i, X_i; \sigma_{Y^*} ) - f_{U_{Y^*} | Y, X}(u \mid Y_i, X_i; \sigma_{Y^*} ) \big) = \sum_{\ell=1}^{L}\widehat{\mathbb{R}}_{\ell,f_{U_{Y^*} | Y, X}}(u,Y_i,X_i;\sigma_{Y^*})'\,
\big(\widehat{\beta}_{Y^*}(\tau_{\ell})-\beta_{Y^*}(\tau_{\ell})\big)
\end{align*}for some random vectors $ \widehat{\mathbb{R}}_{\ell,f_{Y^*\mid X}}(y,x) $ and $ \widehat{\mathbb{R}}_{\ell,f_{U_{Y^*} | Y, X}}(u,\sigma_{Y^*},Y_i,X_i) $, $ \ell \in \{ 1, \ldots, L \} $.
\end{lemma}

\begin{proof}

\noindent \textbf{Part (a):}

Recall \( f_{Y^*\mid X}(y \mid x ) = \big[ x'\beta_{Y^*}^{\partial}\big(  \F_{Y^*\mid X}(y \mid x ) \big) \big]^{-1} \) with the estimator \( \widehat{f}_{Y^*\mid X}(y \mid x ) = \big[ x'\widehat{\beta}_{Y^*}^{\partial}\big( \widehat{\F}_{Y^*\mid X}(y \mid x ) \big) \big]^{-1} \). Let $ \tau_y: = \F_{Y^*\mid X}(y \mid x ) $ and $ \widehat{\tau}_y: = \widehat{\F}_{Y^*\mid X}(y \mid x ) $, then
\begin{align*}
    \big( \widehat{f}_{Y^*\mid X}(y \mid x ) - f_{Y^*\mid X}(y \mid x ) \big) &= \big[ x'\widehat{\beta}_{Y^*}^{\partial}\left(\widehat{\tau}_y\right) \big]^{-1} - \big[ x'\beta_{Y^*}^{\partial}\left(\tau_y\right) \big]^{-1} \\
   &=-\big[\widehat{\beta}_{Y^*}^{\partial}\big(\widehat{\tau}_y\big)' xx'\beta_{Y^*}^{\partial}\big(\tau_y\big)\big]^{-1}\\
   & \qquad \times x'\big(\widehat{\beta}_{Y^*}^{\partial}(\widehat{\tau}_y)
          - \beta_{Y^*}^{\partial}(\tau_y)\big).
\end{align*}Thanks to \Cref{ass:splines}, the process $ \beta_{Y^*}(\cdot) $ is piecewise linear. Thus, on each open interval between knots, $\beta_{Y^*}(\cdot)$ is linear and the second derivative is zero. At each knot $ \tau_\ell, \ell \in \{1,\ldots,L \} $, the first derivative has a jump and the second derivative does not exist (unless the slopes match there). Since $L$ is finite, the set of knots has Lebesgue measure zero and thus, $ \beta_{Y^*}^{\partial\partial}(\tau):= \frac{\partial^2 \beta_{Y^*}(\tau)}{\partial \tau^2} = 0 $ for Lebesgue-almost everywhere $\tau \in \mathcal{T} $. Further, under \Cref{ass:splines}, the indices in the set are evenly spaced, i.e., $ \tau_\ell = \tau_1 + (\ell-1)\epsilon, \, \epsilon:=\tau_2-\tau_1 $, and
\begin{align}\label{eqn:beta_expand_L}
    \beta_{Y^*}(\tau) &= \sum_{\ell = 1}^{L-1} \indicator{ \tau \in (\tau_{\ell},\, \tau_{\ell+1} ) } \Big( \beta_{Y^*}(\tau_\ell) + \frac{\tau - \tau_\ell }{\epsilon} \big( \beta_{Y^*}(\tau_{\ell+1}) - \beta_{Y^*}(\tau_\ell) \big) \Big) + \sum_{\ell = 1}^{L}\indicator{\tau = \tau_\ell}\beta_{Y^*}(\tau_\ell) \nonumber \\
&=\sum_{\ell=1}^{L}
\Big[
\indicator{\ell\le L-1}\,\indicator{\tau\in[\tau_\ell,\tau_{\ell+1})}\Big(1-\tfrac{\tau-\tau_\ell}{\epsilon}\Big)
\;+\;
\indicator{\ell\ge 2}\,\indicator{\tau\in(\tau_{\ell-1},\tau_\ell)}\tfrac{\tau-\tau_{\ell-1}}{\epsilon}
\Big]\,
\beta_{Y^*}(\tau_\ell) \nonumber \\
&=:\sum_{\ell=1}^{L}\omega_{\ell,\beta}(\tau)\,
\beta_{Y^*}(\tau_\ell).
\end{align} Hence
\begin{align*}
    \frac{\partial}{\partial \tau}\beta_{Y^*}(\tau) = \sum_{\ell = 1}^{L} \omega_{\ell,\beta}^\partial(\tau) \beta_{Y^*}(\tau_\ell)
\end{align*}where \( \omega_{\ell,\beta}^\partial(\tau) :=
\epsilon^{-1}\Big(\indicator{\ell\ge2}\,\indicator{\tau \in[\tau_{\ell-1},\tau_{\ell})}
- \indicator{\ell\le L-1}\,\indicator{\tau \in (\tau_{\ell},\tau_{\ell+1})}\Big). \)

Thus,
\begin{align*}
    \widehat{\beta}_{Y^*}^{\partial}(\widehat{\tau}_y)
          - \beta_{Y^*}^{\partial}(\tau_y)
          &= \big( \widehat{\beta}_{Y^*}^{\partial}(\widehat{\tau}_y)
          - \beta_{Y^*}^{\partial}(\widehat{\tau}_y) \big) + \big( \beta_{Y^*}^{\partial}(\widehat{\tau}_y)
          - \beta_{Y^*}^{\partial}(\tau_y) \big) \\
          &= \frac{\partial}{\partial \tau} \big( \widehat{\beta}_{Y^*}(\widehat{\tau}_y)
          - \beta_{Y^*}(\widehat{\tau}_y) \big) + \beta_{Y^*}^{\partial\partial}(\bar{\tau}_y)(\widehat{\tau}_y - \tau_y) \\
          & = \frac{\partial}{\partial \tau} \big( \widehat{\beta}_{Y^*}(\widehat{\tau}_y)
          - \beta_{Y^*}(\widehat{\tau}_y) \big)\\
          &=\sum_{\ell=1}^{L}\omega_{\ell,\beta}^\partial(\widehat{\tau}_y)\,
\big(\widehat{\beta}_{Y^*}(\tau_{\ell})-\beta_{Y^*}(\tau_{\ell})\big).
\end{align*} Putting terms together,
\begin{align}\label{eqn:f_expand}
    \big( \widehat{f}_{Y^*\mid X}(y \mid x ) - f_{Y^*\mid X}(y \mid x ) \big) &=-\big[\widehat{\beta}_{Y^*}^{\partial}\big(\widehat{\tau}_y\big)' xx'\beta_{Y^*}^{\partial}\big(\tau_y\big)\big]^{-1}x'\sum_{\ell=1}^{L}\omega_{\ell,\beta}^\partial(\widehat{\tau}_y)\,
\big(\widehat{\beta}_{Y^*}(\tau_{\ell})-\beta_{Y^*}(\tau_{\ell})\big) \nonumber \\
&=:\sum_{\ell=1}^{L}\widehat{\mathbb{R}}_{\ell,f_{Y^*\mid X}}(y,x)'\,
\big(\widehat{\beta}_{Y^*}(\tau_{\ell})-\beta_{Y^*}(\tau_{\ell})\big).
\end{align}

\noindent \textbf{Part (b):}

From part (a) above,
\begin{align*}
    \big( \widehat{f}_{U_{Y^*} | Y, X}(u \mid Y_i, X_i; \sigma_{Y^*} ) &- f_{U_{Y^*} | Y, X}(u \mid Y_i, X_i; \sigma_{Y^*} ) \big)\\
    &= \frac{f_{U_{Y^*}}(u ; \sigma_{Y^*})}{\int_{\mathcal{U}} f_{Y \mid U_{Y^*}, X}(Y_i \mid \tilde{u}, X_i)\,
   f_{U_{Y^*} \mid X}(\tilde{u} ; \sigma_{Y^*})\, d\tilde{u}}\big( \widehat{f}_{Y^*\mid X}(Y_i \mid X_i ) - f_{Y^*\mid X}(Y_i \mid X_i ) \big)\\
   &=\frac{ f_{U_{Y^*}}(u ; \sigma_{Y^*})}{\int_{\mathcal{U}} f_{Y \mid U_{Y^*}, X}(Y_i \mid \tilde{u}, X_i)\,
   f_{U_{Y^*} \mid X}(\tilde{u} ; \sigma_{Y^*})\, d\tilde{u}}  \times \sum_{\ell=1}^{L}\widehat{\mathbb{R}}_{\ell,f_{Y^*\mid X}}(Y_i,X_i)'\,
\big(\widehat{\beta}_{Y^*}(\tau_{\ell})-\beta_{Y^*}(\tau_{\ell})\big)\\
    &=: \sum_{\ell=1}^{L}\widehat{\mathbb{R}}_{\ell,f_{U_{Y^*} | Y, X}}(u,Y_i,X_i;\sigma_{Y^*})'\,
\big(\widehat{\beta}_{Y^*}(\tau_{\ell})-\beta_{Y^*}(\tau_{\ell})\big).
\end{align*}
\end{proof}

\section{Monte Carlo Simulations} \label{app:simulations}

In this section, we provide some Monte Carlo simulations to demonstrate the performance of our proposed approach.  For the simulations below, we generate the outcome and the treatment according to
\begin{align*}
  Y_i &= \beta_{0Y^*}(V_{iY^*}) + \beta_{1Y^*}(V_{iY^*}) X_i + U_{iY^*} \\
  T_i &= \beta_{0T^*}(V_{iT^*}) + \beta_{1T^*}(V_{iT^*}) X_i + U_{iT^*}
\end{align*}
where, for $\tau \in [0,1]$, we set
\begin{align*}
  \beta_{0Y^*}(\tau) &= 1 + 3\tau - \tau^2 \\
  \beta_{1Y^*}(\tau) &= \exp(\tau) \\
  \beta_{0T^*}(\tau) &= 1 + 2\tau \\
  \beta_{1T^*}(\tau) &= 1 + \tau
\end{align*}
We set $V_{Y^*}$ and $V_{T^*}$ to be draws from a parametric copula $C_{V_{Y^*}V_{T^*}\mid X}$ which we vary across simulations.  In particular, we vary this copula between a Gaussian copula with parameter $\rho = 0.5$ and a Clayton copula with parameter $\delta=1.5$.

$U_{Y^*}$ and $U_{T^*}$ are $i.i.d.$ draws from $\N(0, \sigma^2_j)$ for $j \in \{Y^*,T^*\}$, and we vary $\sigma^2_j$ across simulations.  Larger values of $\sigma^2_j$ indicate more measurement error, and, in particular, we consider the cases where $\sigma^2_j$ is equal to $1$, $0.25$, $0.01$, and $0$, which roughly correspond to a relatively large amount of measurement error, a relatively small amount of measurement error, a very small amount of measurement error, and no measurement error at all.  Finally, we set $\ln(X) \sim \N(0, 0.5^2)$.

Table \ref{tab:mc_sims} contains the results from the Monte Carlo simulations.  The first set of results compares the performance of our first step estimators (which account for measurement error in the outcome) with quantile regression that ignores measurement error (these results are in the columns labeled ``QR Coef.'').  For this part, we estimate the quantile regression parameters for the outcome across a grid of 10 equally spaced values of $\tau$ from 0.02 to 0.98.  The numbers in those columns correspond to an aggregated version of root mean squared error---in each Monte Carlo simulation, we compute the squared error of the estimate of $\beta_{0Y^*}(\tau)$ and $\beta_{1Y^*}(\tau)$, average these across all 20 parameter estimates, and then calculate the square root of the mean of this quantity across Monte Carlo simulations.  For the first step quantile regression estimates, we find that our approach performs substantially better than ignoring measurement error in the case when the measurement error is large, each approach performs similarly in the case with small measurement error, and ignoring measurement error tends to perform better when the amount of measurement error is very small or non-existent.  Additionally, both approaches perform moderately better with larger sample sizes.  And the results are qualitatively similar for the Gaussian copula and the Clayton copula that we consider.

\begin{table}[th]
\centering
\caption{Monte Carlo Simulations}
\label{tab:mc_sims}
\resizebox{\linewidth}{!}{
\begin{tabular}[t]{lrrrrrrr}
\toprule
\multicolumn{1}{c}{ } & \multicolumn{2}{c}{QR Coef.} & \multicolumn{2}{c}{Cop. Param} & \multicolumn{3}{c}{Transition Mat.} \\
\cmidrule(l{3pt}r{3pt}){2-3} \cmidrule(l{3pt}r{3pt}){4-5} \cmidrule(l{3pt}r{3pt}){6-8}
ME s.d. & ME & No ME & ME & No ME & ME & No ME & Obs.\\
\midrule
\addlinespace[0.3em]
\multicolumn{8}{l}{\textbf{$n=250$, Gaussian Copula ($\rho=0.5$)}}\\
\hspace{1em}1 & 0.418 & 0.580 & 0.163 & 0.240 & 0.046 & 0.071 & 0.086\\
\hspace{1em}0.5 & 0.296 & 0.308 & 0.124 & 0.101 & 0.043 & 0.027 & 0.050\\
\hspace{1em}0.1 & 0.307 & 0.217 & 0.162 & 0.050 & 0.056 & 0.017 & 0.043\\
\hspace{1em}0 & 0.306 & 0.212 & 0.161 & 0.051 & 0.059 & 0.016 & 0.043\\
\addlinespace[0.3em]
\multicolumn{8}{l}{\textbf{$n=1000$, Gaussian Copula ($\rho=0.5$)}}\\
\hspace{1em}1 & 0.256 & 0.493 & 0.130 & 0.220 & 0.044 & 0.076 & 0.083\\
\hspace{1em}0.5 & 0.185 & 0.213 & 0.098 & 0.079 & 0.032 & 0.027 & 0.037\\
\hspace{1em}0.1 & 0.204 & 0.108 & 0.132 & 0.026 & 0.048 & 0.012 & 0.024\\
\hspace{1em}0 & 0.204 & 0.105 & 0.135 & 0.024 & 0.050 & 0.009 & 0.022\\
\addlinespace[0.3em]
\multicolumn{8}{l}{\textbf{$n=250$, Clayton Copula ($\delta=1.5$)}}\\
\hspace{1em}1 & 0.472 & 0.578 & 0.764 & 1.043 & 0.068 & 0.098 & 0.106\\
\hspace{1em}0.5 & 0.302 & 0.313 & 0.343 & 0.694 & 0.028 & 0.059 & 0.065\\
\hspace{1em}0.1 & 0.306 & 0.213 & 0.503 & 0.268 & 0.043 & 0.023 & 0.043\\
\hspace{1em}0 & 0.305 & 0.209 & 0.569 & 0.218 & 0.049 & 0.017 & 0.042\\
\addlinespace[0.3em]
\multicolumn{8}{l}{\textbf{$n=1000$, Clayton Copula ($\delta=1.5$)}}\\
\hspace{1em}1 & 0.253 & 0.489 & 0.693 & 1.016 & 0.054 & 0.093 & 0.097\\
\hspace{1em}0.5 & 0.185 & 0.212 & 0.252 & 0.632 & 0.025 & 0.045 & 0.047\\
\hspace{1em}0.1 & 0.205 & 0.107 & 0.524 & 0.178 & 0.044 & 0.013 & 0.022\\
\hspace{1em}0 & 0.205 & 0.106 & 0.561 & 0.119 & 0.048 & 0.012 & 0.023\\
\bottomrule
\end{tabular}}
\begin{justify}
{\footnotesize \textit{Notes:} The table reports root mean squared error for (i) estimates of quantile regression parameters (columns labeled ``QR Coef''), (ii) estimates of the conditional copula parameter (columns labeled ``Cop.\,Param''), and (iii) estimates of the transition matrix (columns labeled ``Transition Mat.'').  The column labeled ``ME s.d.'' refers to the standard deviation of the measurement error for $Y$ and $T$ (which are set to be equal to each other).  Columns labeled ``ME'' use the approach suggested in the paper; columns labeled ``No ME'' ignore measurement error but otherwise follow the same estimation strategy consisting of first step quantile regressions followed by estimating the conditional copula; the column labeled ``Obs.'' directly uses the observed data to calculate the unconditional transition matrix.  The rows differ by (i) the amount of measurement error, (ii) the number of observations, and (iii) the particular conditional copula used.  The results are based on 1000 Monte Carlo simulations.}
\end{justify}
\end{table}

\FloatBarrier

The second set of results is for estimates of the copula parameter (these results are in the columns labeled ``Cop.\,Param'').  As before, our approach outperforms ignoring measurement error in the case of a large amount of measurement error and underperforms ignoring measurement error in the cases with a very small amount of measurement error or non-existent measurement error.  However, in the middle case (recall, this is the case with a small but non-negligible amount of measurement error), our approach tends to perform moderately better than ignoring measurement error across simulations, especially for the case with a Clayton copula.  When measurement error is ignored, estimates of the conditional copula parameter tend to be downward biased for the dependence between $Y^*$ and $T^*$ due to measurement error.

The next set of results is for estimates of unconditional transition matrices (these results are in the columns labeled ``Transition Mat.'').  For the transition matrix, we report an aggregated version of the root mean squared error based on computing the average squared error across cells in the transition matrix for a particular simulation and then taking the square root of the average of these across simulations.  For the transition matrix, we also report estimates of the root mean squared error of just taking the raw data and computing a transition matrix directly.  As for the copula parameter, our approach tends to perform better than ignoring measurement error for estimating transition matrices, except in cases where the measurement error is extremely small.  This can be understood in light of the results on estimating the QR coefficients and copula parameters themselves.  In the middle case with a small but non-negligible amount of measurement error, both approaches tend to perform about equally for estimating the QR parameters, but our approach tends to perform better for estimating the copula parameter.  Since the transition matrix parameter depends on both of these, our approach tends to perform somewhat better in this case.

In \Cref{app:mc_sims_additional} in the Supplementary Appendix, we provide additional Monte Carlo simulations with Laplace measurement error and with correlated measurement error.

\end{document}